\documentclass{cernyrep}
\usepackage{texnames}
\usepackage[T1]{fontenc}
\begin{document}
\title{Neutrino Physics}

\author{Zhi-zhong Xing}

\institute{Institute of High Energy Physics and Theoretical Physics
Center for Science Facilities, \\ Chinese Academy of Sciences, Beijing, China}

\maketitle 

\begin{abstract}
I give a theoretical overview of some basic properties of massive
neutrinos in these lectures. Particular attention is paid to the
origin of neutrino masses, the pattern of lepton flavor mixing, the
feature of leptonic CP violation and the electromagnetic properties
of massive neutrinos. I highlight the TeV seesaw mechanisms as a
possible bridge between neutrino physics and collider physics in the
era characterized by the Large Hadron Collider.
\end{abstract}

\section{Finite Neutrino Masses}

It is well known that the mass of an elementary particle represents
its inertial energy when it exists at rest. Hence a massless
particle has no way to exist at rest --- instead, it must always
move at the speed of light. A massive fermion (either lepton or
quark) must exist in both left-handed and right-handed states, since
the field operators responsible for the non-vanishing mass of a
fermion have to be bilinear products of the spinor fields which flip
the fermion's handedness or chirality.

The standard model (SM) of electroweak interactions contains three
neutrinos $(\nu^{}_e, \nu^{}_\mu, \nu^{}_\tau)$ which are purely
left-handed and massless. In the SM the masslessness of the photon
is guaranteed by the electromagnetic $U(1)^{}_{\rm Q}$ gauge
symmetry. Although the masslessness of three neutrinos corresponds
to the lepton number conservation
\footnote{It is actually the $B$$-$$L$ symmetry that makes neutrinos
exactly massless in the SM, where $B$ = baryon number and $L$ =
lepton number. The reason is simply that a neutrino and an
antineutrino have different values of $B$$-$$L$. Thus the naive
argument for massless neutrinos is valid to all orders in
perturbation and non-perturbation theories, if $B$$-$$L$ is an exact
symmetry.},
the latter is an accidental symmetry rather than a fundamental
symmetry of the SM. Hence many physicists strongly believed that
neutrinos should be massive even long before some incontrovertible
experimental evidence for massive neutrinos were accumulated. A good
reason for this belief is that neutrinos are more natural to be
massive than to be massless in some grand unified theories, such as
the $\rm SO(10)$ theory, which try to unify electromagnetic, weak
and strong interactions as well as leptons and quarks.

If neutrinos are massive and their masses are non-degenerate, it
will in general be impossible to find a flavor basis in which the
coincidence between flavor and mass eigenstates holds both for
charged leptons $(e, \mu, \tau)$ and for neutrinos $(\nu^{}_e,
\nu^{}_\mu, \nu^{}_\tau)$. In other words, the phenomenon of flavor
mixing is naturally expected to appear between three charged leptons
and three massive neutrinos, just like the phenomenon of flavor
mixing between three up-type quarks $(u, c, t)$ and three down-type
quarks $(d, s, b)$. If there exist irremovable complex phases in the
Yukawa interactions, CP violation will naturally appear both in the
quark sector and in the lepton sector.

\subsection{Some preliminaries}

To write out the mass term for three known neutrinos, let us make a
minimal extension of the SM by introducing three right-handed
neutrinos. Then we totally have six neutrino fields
\footnote{The left- and right-handed components of a fermion field
$\psi (x)$ are denoted as $\psi^{}_{\rm L}(x) = P^{}_{\rm L} \psi
(x)$ and $\psi^{}_{\rm R}(x) = P^{}_{\rm R} \psi (x)$, respectively,
where $P^{}_{\rm L} \equiv (1-\gamma^{}_5)/2$ and $P^{}_{\rm R}
\equiv (1 +\gamma^{}_5)/2$ are the chiral projection operators.
Note, however, that $\nu^{}_{\rm L} = P^{}_{\rm L} \nu^{}_{\rm L}$
and $N^{}_{\rm R} = P^{}_{\rm R} N^{}_{\rm R}$ are in general
independent of each other.}:
\begin{eqnarray}
\nu^{}_{\rm L} = \left( \begin{matrix} \nu^{}_{e \rm L} \cr
\nu^{}_{\mu \rm L} \cr \nu^{}_{\tau \rm L} \cr \end{matrix} \right)
\; , ~~~~ N^{}_{\rm R} = \left( \begin{matrix} N^{}_{1 \rm R} \cr
N^{}_{2 \rm R} \cr N^{}_{3 \rm R} \cr \end{matrix} \right) \; ,
\end{eqnarray}
where only the left-handed fields take part in the electroweak
interactions. The charge-conjugate counterparts of $\nu^{}_{\rm L}$
and $N^{}_{\rm R}$ are defined as
\begin{eqnarray}
\left( \nu^{}_{\rm L} \right)^c \equiv {\cal C}
\overline{\nu^{}_{\rm L}}^T \; , ~~~~ \left( N^{}_{\rm R} \right)^c
\equiv {\cal C} \overline{N^{}_{\rm R}}^T \; ;
\end{eqnarray}
and accordingly,
\begin{eqnarray}
\overline{\left( \nu^{}_{\rm L} \right)^c} = (\nu^{}_{\rm L})^T
{\cal C} \; , ~~~~ \overline{\left( N^{}_{\rm R} \right)^c} =
(N^{}_{\rm R})^T {\cal C} \; ,
\end{eqnarray}
where $\cal C$ denotes the charge-conjugation matrix and satisfies
the conditions
\begin{eqnarray}
{\cal C}\gamma^T_\mu {\cal C}^{-1} = -\gamma^{}_\mu \; , ~~~~ {\cal
C}\gamma^T_5 {\cal C}^{-1} = \gamma^{}_5 \; , ~~~~ {\cal C}^{-1} =
{\cal C}^\dagger = {\cal C}^T = -{\cal C} \; .
\end{eqnarray}
It is easy to check that $P^{}_{\rm L}(N^{}_{\rm R})^c = (N^{}_{\rm
R})^c$ and $P^{}_{\rm R}(\nu^{}_{\rm L})^c = (\nu^{}_{\rm L})^c$
hold; namely, $(\nu^{}_{\rm L})^c = (\nu^c)^{}_{\rm R}$ and
$(N^{}_{\rm R})^c = (N^c)^{}_{\rm L}$ hold. Hence $(\nu^{}_{\rm
L})^c$ and $(N^{}_{\rm R})^c$ are right- and left-handed fields,
respectively. One may then use the neutrino fields $\nu^{}_{\rm L}$,
$N^{}_{\rm R}$ and their charge-conjugate partners to write out the
gauge-invariant and Lorentz-invariant neutrino mass terms.

In the SM the weak charged-current interactions of three active
neutrinos are given by
\begin{equation}
{\cal L}^{}_{\rm cc} = \frac{g}{\sqrt{2}} \overline{\left(e~ \mu~
\tau\right)^{}_{\rm L}} ~\gamma^\mu \left( \begin{matrix} \nu^{}_e
\cr \nu^{}_\mu \cr \nu^{}_\tau \end{matrix} \right)^{}_{\rm L}
W^-_\mu + {\rm h.c.} \; .
\end{equation}
Without loss of generality, we choose the basis in which the mass
eigenstates of three charged leptons are identified with their
flavor eigenstates. If neutrinos have non-zero and non-degenerate
masses, their flavor and mass eigenstates are in general not
identical in the chosen basis. This mismatch signifies lepton flavor
mixing.

\subsection{Dirac neutrino masses}

A Dirac neutrino is described by a four-component Dirac spinor
$\nu^{} = \nu^{}_{\rm L} + N^{}_{\rm R}$, whose left-handed and
right-handed components are just $\nu^{}_{\rm L}$ and $N^{}_{\rm
R}$. The Dirac neutrino mass term comes from the Yukawa interactions
\begin{equation}
-{\cal L}^{}_{\rm Dirac} = \overline{\ell^{}_{\rm L}} Y^{}_\nu
\tilde{H} N^{}_{\rm R} + {\rm h.c.} \; ,
\end{equation}
where $\tilde{H} \equiv i\sigma^{~}_2 H^*$ with $H$ being the SM
Higgs doublet, and $\ell^{}_{\rm L}$ denotes the left-handed lepton
doublet. After spontaneous gauge symmetry breaking (i.e.,
$SU(2)^{}_{\rm L} \times U(1)^{}_{\rm Y} \to U(1)^{}_{\rm Q}$), we
obtain
\begin{equation}
-{\cal L}^\prime_{\rm Dirac} = \overline{\nu^{}_{\rm L}} M^{}_{\rm
D} N^{}_{\rm R} + {\rm h.c.} \; ,
\end{equation}
where $M^{}_{\rm D}= Y^{}_\nu \langle H\rangle$ with $\langle
H\rangle \simeq 174 ~ {\rm GeV}$ being the vacuum expectation value
of $H$. This mass matrix can be diagonalized by a bi-unitary
transformation: $V^\dagger M^{}_{\rm D} U = \widehat{M}^{}_\nu
\equiv {\rm Diag}\{m^{}_1, m^{}_2, m^{}_3 \}$ with $m^{}_i$ being
the neutrino masses (for $i=1, 2, 3$). After this diagonalization,
\begin{equation}
-{\cal L}^{\prime}_{\rm Dirac} = \overline{\nu^\prime_{\rm L}}
\widehat{M}^{}_\nu N^\prime_{\rm R} + {\rm h.c.} \; ,
\end{equation}
where $\nu^\prime_{\rm L} = V^\dagger \nu^{}_{\rm L}$ and
$N^\prime_{\rm R} = U^\dagger N^{}_{\rm R}$. Then the four-component
Dirac spinor
\begin{equation}
\nu^\prime = \nu^\prime_{\rm L} + N^\prime_{\rm R} = \left(
\begin{matrix} \nu^{}_1 \cr \nu^{}_2 \cr \nu^{}_3 \cr \end{matrix}
\right) \; ,
\end{equation}
which automatically satisfies $P^{}_{\rm L} \nu^\prime =
\nu^\prime_{\rm L}$ and $P^{}_{\rm R} \nu^\prime = N^\prime_{\rm
R}$, describes the mass eigenstates of three Dirac neutrinos. In
other words,
\begin{equation}
-{\cal L}^{\prime}_{\rm Dirac} = \overline{\nu^\prime}
\widehat{M}^{}_\nu \nu^\prime = \sum^3_{i=1} m^{}_i
\overline{\nu^{}_i} \nu^{}_i \; .
\end{equation}
The kinetic term of Dirac neutrinos reads
\begin{eqnarray}
{\cal L}^{}_{\rm kinetic} = i \overline{\nu^{}_{\rm L}}
\gamma^{}_\mu \partial^\mu \nu^{}_{\rm L} + i \overline{N^{}_{\rm
R}} \gamma^{}_\mu \partial^\mu N^{}_{\rm R} = i
\overline{\nu^\prime} \gamma^{}_\mu
\partial^\mu \nu^\prime = i \sum^3_{k=1} \overline{\nu^{}_k}
\gamma^{}_\mu \partial^\mu \nu^{}_k \; ,
\end{eqnarray}
where $V^\dagger V = V V^\dagger = {\bf 1}$ and $U^\dagger U = U
U^\dagger = {\bf 1}$ have been used.

Now we rewrite the weak charged-current interactions of three
neutrinos in Eq. (5) in terms of their mass eigenstates
$\nu^\prime_{\rm L} = V^\dagger \nu^{}_{\rm L}$ in the chosen basis
where the flavor and mass eigenstates of three charged leptons are
identical:
\begin{equation}
{\cal L}^{}_{\rm cc} = \frac{g}{\sqrt{2}} \overline{\left(e~ \mu~
\tau\right)^{}_{\rm L}} ~\gamma^\mu V \left( \begin{matrix} \nu^{}_1
\cr \nu^{}_2 \cr \nu^{}_3 \end{matrix} \right)^{}_{\rm L} W^-_\mu +
{\rm h.c.} \; .
\end{equation}
The $3\times 3$ unitary matrix $V$, which actually links the
neutrino mass eigenstates $(\nu^{}_1, \nu^{}_2, \nu^{}_3)$ to the
neutrino flavor eigenstates $(\nu^{}_e, \nu^{}_\mu, \nu^{}_\tau)$,
just measures the phenomenon of neutrino mixing.

A salient feature of massive Dirac neutrinos is lepton number
conservation. To see why massive Dirac neutrinos are
lepton-number-conserving, we make the global phase transformations
\begin{eqnarray}
l (x) \rightarrow e^{i\Phi} l (x) \; , ~~~~ \nu^\prime_{\rm L} (x)
\rightarrow e^{i\Phi} \nu^\prime_{\rm L} (x) \; , ~~~~ N^\prime_{\rm
R} (x) \rightarrow e^{i\Phi} N^\prime_{\rm R} (x) \; ,
\end{eqnarray}
where $l$ denotes the column vector of $e$, $\mu$ and $\tau$ fields,
and $\Phi$ is an arbitrary spacetime-independent phase. As the mass
term ${\cal L}^\prime_{\rm Dirac}$, the kinetic term ${\cal
L}^{}_{\rm kinetic}$ and the charged-current interaction term ${\cal
L}^{}_{\rm cc}$ are all invariant under these transformations, the
lepton number must be conserved for massive Dirac neutrinos. It is
evident that lepton flavors are violated, unless $M^{}_{\rm D}$ is
diagonal or equivalently $V$ is the identity matrix. In other words,
lepton flavor mixing leads to lepton flavor violation, or vice
versa.

For example, the decay mode $\pi^- \rightarrow \mu^- +
\overline{\nu}^{}_\mu$ preserves both the lepton number and lepton
flavors. In contrast, $\mu^+ \rightarrow e^+ + \gamma$ preserves the
lepton number but violates the lepton flavors. The observed
phenomena of neutrino oscillations verify the existence of neutrino
flavor violation. Note that the $0\nu 2\beta$ decay $(A,Z)
\rightarrow (A,Z+2) + 2e^-$ violates the lepton number. This process
cannot take place if neutrinos are massive Dirac particles, but it
may naturally happen if neutrinos are massive Majorana particles.

\subsection{Majorana neutrino masses}

The left-handed neutrino field $\nu^{}_{\rm L}$ and its
charge-conjugate counterpart $(\nu^{}_{\rm L})^c$ can in principle
form a neutrino mass term, as $(\nu^{}_{\rm L})^c$ is actually
right-handed. But this Majorana mass term is forbidden by the
$SU(2)^{}_{\rm L} \times U(1)^{}_{\rm Y}$ gauge symmetry in the SM,
which contains only one $SU(2)^{}_{\rm L}$ Higgs doublet and
preserves lepton number conservation. We shall show later that the
introduction of an $SU(2)^{}_{\rm L}$ Higgs triplet into the SM can
accommodate such a neutrino mass term with gauge invariance. Here we
ignore the details of the Higgs triplet models and focus on the
Majorana neutrino mass term itself:
\begin{equation}
-{\cal L}^\prime_{\rm Majorana} = \frac{1}{2} \overline{\nu^{}_{\rm
L}} M^{}_{\rm L} (\nu^{}_{\rm L} )^c + {\rm h.c.} \; .
\end{equation}
Note that the mass matrix $M^{}_{\rm L}$ must be symmetric. Because
the mass term is a Lorentz scalar whose transpose keeps unchanged,
we have
\begin{eqnarray}
\overline{\nu^{}_L} M^{}_{\rm L} (\nu^{}_L )^c = \left[
\overline{\nu^{}_L} M^{}_{\rm L} (\nu^{}_L )^c \right]^T = -
\overline{\nu^{}_L} {\cal C}^T M^T_{\rm L} \overline{\nu^{}_L}^T =
\overline{\nu^{}_L} M^T_{\rm L} (\nu^{}_L )^c \; ,
\end{eqnarray}
where a minus sign appears when interchanging two fermion field
operators, and ${\cal C}^T = -{\cal C}$ has been used. Hence
$M^T_{\rm L} = M^{}_{\rm L}$ holds. This symmetric mass matrix can
be diagonalized by the transformation $V^\dagger M^{}_{\rm L} V^*
=\widehat{M}^{}_\nu \equiv {\rm Diag}\{m^{}_1, m^{}_2, m^{}_3 \}$,
where $V$ is a unitary matrix. After this, Eq. (14) becomes
\begin{equation}
-{\cal L}^{\prime}_{\rm Majorana} = \frac{1}{2}
\overline{\nu^\prime_{\rm L}} \widehat{M}^{}_\nu (\nu^\prime_{\rm L}
)^c + {\rm h.c.} \; ,
\end{equation}
where $\nu^\prime_{\rm L} = V^\dagger \nu^{}_{\rm L}$ and
$(\nu^\prime_{\rm L})^c = {\cal C}\overline{\nu^\prime_{\rm L}}^T$.
Then the Majorana field
\begin{equation}
\nu^\prime = \nu^\prime_{\rm L} + (\nu^\prime_{\rm L} )^c = \left(
\begin{matrix} \nu^{}_1 \cr \nu^{}_2 \cr \nu^{}_3 \cr \end{matrix}
\right) \; ,
\end{equation}
which certainly satisfies the Majorana condition $(\nu^\prime)^c =
\nu^\prime$, describes the mass eigenstates of three Majorana
neutrinos. In other words,
\begin{equation}
-{\cal L}^{\prime}_{\rm Majorana} = \frac{1}{2}
\overline{\nu^\prime} \widehat{M}^{}_\nu \nu^\prime = \frac{1}{2}
\sum^3_{i=1} m^{}_i \overline{\nu^{}_i} \nu^{}_i \; .
\end{equation}
The kinetic term of Majorana neutrinos reads
\begin{eqnarray}
{\cal L}^{}_{\rm kinetic} = i \overline{\nu^{}_{\rm L}}
\gamma^{}_\mu \partial^\mu \nu^{}_{\rm L} = i
\overline{\nu^\prime_{\rm L}} \gamma^{}_\mu \partial^\mu
\nu^\prime_{\rm L} = \frac{i}{2} \overline{\nu^\prime} \gamma^{}_\mu
\partial^\mu \nu^\prime = \frac{i}{2} \sum^3_{k=1}
\overline{\nu^{}_k} \gamma^{}_\mu
\partial^\mu \nu^{}_k \; ,
\end{eqnarray}
where we have used a generic relationship $\overline{(\psi^{}_{\rm
L})^c} \gamma^{}_\mu \partial^\mu (\psi^{}_{\rm L})^c =
\overline{\psi^{}_{\rm L}} \gamma^{}_\mu \partial^\mu \psi^{}_{\rm
L}$. This relationship can easily be proved by taking account of
$\partial^\mu \left[\overline{(\psi^{}_{\rm L})^c} \gamma^{}_\mu
(\psi^{}_{\rm L})^c\right] =0$; i.e., we have
\begin{eqnarray}
\overline{(\psi^{}_{\rm L})^c} \gamma^{}_\mu \partial^\mu
(\psi^{}_{\rm L})^c = - \partial^\mu \overline{(\psi^{}_{\rm L})^c}
\gamma^{}_\mu (\psi^{}_{\rm L})^c = - \left[
\partial^\mu \overline{(\psi^{}_{\rm L})^c} \gamma^{}_\mu
(\psi^{}_{\rm L})^c \right]^T \nonumber \\
= \left({\cal
C}\overline{\psi^{}_{\rm L}}^T\right)^T \gamma^T_\mu
\partial^\mu \left[\left(\psi^{}_{\rm L}\right)^T {\cal C}\right]^T
= \overline{\psi^{}_{\rm L}} \gamma^{}_\mu \partial^\mu \psi^{}_{\rm
L} , \; ~
\end{eqnarray}
where ${\cal C}^T \gamma^T_\mu {\cal C}^T = \gamma^{}_\mu$, which
may be read off from Eq. (4), has been used.

It is worth pointing out that the factor $1/2$ in ${\cal
L}^\prime_{\rm Majorana}$ allows us to get the Dirac equation of
massive Majorana neutrinos analogous to that of massive Dirac
neutrinos. To see this point more clearly, let us consider the
Lagrangian of free Majorana neutrinos (i.e., their kinetic and mass
terms):
\begin{eqnarray}
{\cal L}^{}_\nu = i \overline{\nu^{}_{\rm L}} \gamma^{}_\mu
\partial^\mu \nu^{}_{\rm L} - \left[ \frac{1}{2}
\overline{\nu^{}_{\rm L}} M^{}_{\rm L} (\nu^{}_{\rm L} )^c + {\rm
h.c.} \right] = i \overline{\nu^\prime_{\rm L}} \gamma^{}_\mu
\partial^\mu \nu^\prime_{\rm L} - \left[ \frac{1}{2}
\overline{\nu^\prime_{\rm L}} \widehat{M}^{}_\nu (\nu^\prime_{\rm L}
)^c + {\rm h.c.} \right]
\nonumber \\
= \frac{1}{2} \left( i \overline{\nu^\prime} \gamma^{}_\mu
\partial^\mu \nu^\prime - \overline{\nu^\prime} \widehat{M}^{}_\nu
\nu^\prime \right) = -\frac{1}{2} \left( i
\partial^\mu \overline{\nu^\prime} \gamma^{}_\mu \nu^\prime +
\overline{\nu^\prime} \widehat{M}^{}_\nu \nu^\prime \right) \; ,
~~~~~~~~~~~~~~~~~~~~~~~~~~~~~~
\end{eqnarray}
where $\partial^\mu (\overline{\nu^\prime} \gamma^{}_\mu \nu^\prime)
=0$ has been used. Then we substitute ${\cal L}^{}_\nu$ into the
Euler-Lagrange equation
\begin{equation}
\partial^\mu \frac{\partial {\cal L}^{}_\nu}{\partial \left(\partial^\mu
\overline{\nu^\prime} \right)} - \frac{\partial {\cal
L}^{}_\nu}{\partial \overline{\nu^\prime}} = 0 \; ~~~~
\end{equation}
and obtain the Dirac equation
\begin{eqnarray}
i \gamma^{}_\mu \partial^\mu \nu^\prime - \widehat{M}^{}_\nu
\nu^\prime = 0 \; .
\end{eqnarray}
More explicitly, $i \gamma^{}_\mu \partial^\mu \nu^{}_k - m^{}_k
\nu^{}_k = 0$ holds (for $k=1, 2, 3$). That is why the factor $1/2$
in ${\cal L}^\prime_{\rm Majorana}$ makes sense.

The weak charged-current interactions of three neutrinos in Eq. (5)
can now be rewritten in terms of their mass eigenstates
$\nu^\prime_{\rm L} = V^\dagger \nu^{}_{\rm L}$. In the chosen basis
where the flavor and mass eigenstates of three charged leptons are
identical, the expression of ${\cal L}^{}_{cc}$ for Majorana
neutrinos is the same as that given in Eq. (12) for Dirac neutrinos.
The unitary matrix $V$ is just the $3\times 3$ Majorana neutrino
mixing matrix, which contains two more irremovable CP-violating
phases than the $3\times 3$ Dirac neutrino mixing matrix (see
section 4 for detailed discussions).

The most salient feature of massive Majorana neutrinos is lepton
number violation. Let us make the global phase transformations
\begin{equation}
l (x) \rightarrow e^{i\Phi} l (x) \; , ~~~~ \nu^\prime_{\rm L} (x)
\rightarrow e^{i\Phi} \nu^\prime_{\rm L} (x) \; ,
\end{equation}
where $l$ stands for the column vector of $e$, $\mu$ and $\tau$
fields, and $\Phi$ is an arbitrary spacetime-independent phase. One
can immediately see that the kinetic term ${\cal L}^{}_{\rm
kinetic}$ and the charged-current interaction term ${\cal L}^{}_{\rm
cc}$ are invariant under these transformations, but the mass term
${\cal L}^\prime_{\rm Majorana}$ is not invariant because of both
$\overline{\nu^\prime_{\rm L}} \rightarrow e^{-i\Phi}
\overline{\nu^\prime_{\rm L}}$ and $(\nu^\prime_{\rm L})^c
\rightarrow e^{-i\Phi} (\nu^\prime_{\rm L})^c$. The lepton number is
therefore violated for massive Majorana neutrinos. Similar to the
case of Dirac neutrinos, the lepton flavor violation of Majorana
neutrinos is also described by $V$.

The $0\nu 2\beta$ decay $(A,Z) \rightarrow (A,Z+2) + 2e^-$ is a
clean signature of the Majorana nature of massive neutrinos. This
lepton-number-violating process can occur when there exists
neutrino-antineutrino mixing induced by the Majorana mass term
(i.e., the neutrino mass eigenstates are self-conjugate,
$\overline{\nu}^{}_i = \nu^{}_i$). The effective mass of the $0\nu
2\beta$ decay is defined as
\begin{equation}
\langle m\rangle^{}_{ee} \equiv \left| \sum_i m^{}_i V^2_{ei}
\right| \; ,
\end{equation}
where $m^{}_i$ comes from the helicity suppression factor $m^{}_i/E$
for the $\nu^{}_i$ exchange between two beta decays with $E$ being
the energy of the virtual $\nu^{}_i$ neutrino. Current experimental
data only yield an upper bound $\langle m\rangle^{}_{ee} < 0.23$ eV
(or $< 0.85$ eV as a more conservative bound) at the $2\sigma$
level.

\subsection{Hybrid neutrino mass terms}

Similar to Eq. (14), the right-handed neutrino field $N^{}_{\rm R}$
and its charge-conjugate counterpart $(N^{}_{\rm R})^c$ can also
form a Majorana mass term. Hence it is possible to write out the
following hybrid neutrino mass terms in terms of $\nu^{}_{\rm L}$,
$N^{}_{\rm R}$, $(\nu^{}_{\rm L})^c$ and $(N^{}_{\rm R})^c$ fields:
\begin{eqnarray}
-{\cal L}^{\prime}_{\rm hybrid} = \overline{\nu^{}_{\rm L}}
M^{}_{\rm D} N^{}_{\rm R} + \frac{1}{2} \overline{\nu^{}_{\rm L}}
M^{}_{\rm L} (\nu^{}_{\rm L})^c + \frac{1}{2} \overline{(N^{}_{\rm
R})^c} M^{}_{\rm R} N^{}_{\rm R} + {\rm h.c.}
\nonumber \\
= \frac{1}{2} \left [ \begin{matrix} \overline{\nu^{}_{\rm L}} &
\overline{(N^{}_{\rm R})^c} \cr \end{matrix} \right ] \left (
\begin{matrix} M^{}_{\rm L} & M^{}_{\rm D} \cr M^T_{\rm D} & M^{}_{\rm
R} \cr \end{matrix} \right ) \left [ \begin{matrix} (\nu^{}_{\rm
L})^c \cr N^{}_{\rm R} \cr \end{matrix} \right ] + {\rm h.c.} \; ,
~~~~~~~~
\end{eqnarray}
where $M^{}_{\rm L}$ and $M^{}_{\rm R}$ are symmetric mass matrices
because the corresponding mass terms are of the Majorana type, and
the relationship
\begin{eqnarray}
\overline{(N^{}_{\rm R})^c} M^T_{\rm D} (\nu^{}_{\rm L})^c =
\left[(N^{}_{\rm R})^T {\cal C} M^T_{\rm D} {\cal C}
\overline{\nu^{}_{\rm L}}^T \right]^T = \overline{\nu^{}_{\rm L}}
M^{}_{\rm D} N^{}_{\rm R} \;
\end{eqnarray}
has been used. The overall $6\times 6$ mass matrix in Eq. (26) is
also symmetric, and thus it can be diagonalized by a $6\times 6$
unitary matrix through the transformation
\begin{eqnarray}
\left( \begin{matrix} V & R \cr S & U \cr \end{matrix}
\right)^\dagger \left ( \begin{matrix} M^{}_{\rm L} & M^{}_{\rm D}
\cr M^T_{\rm D} & M^{}_{\rm R} \cr \end{matrix} \right ) \left(
\begin{matrix} V & R \cr S & U \cr \end{matrix} \right)^* =
\left( \begin{matrix} \widehat{M}^{}_\nu & {\bf 0} \cr {\bf 0} &
\widehat{M}^{}_N \cr \end{matrix} \right) \; ,
\end{eqnarray}
where we have defined $\widehat{M}^{}_\nu \equiv {\rm Diag}\{m^{}_1,
m^{}_2, m^{}_3 \}$, $\widehat{M}^{}_N \equiv {\rm Diag}\{M^{}_1,
M^{}_2, M^{}_3 \}$, and the $3\times 3$ matrices $V$, $R$, $S$ and
$U$ satisfy the unitarity conditions
\begin{eqnarray}
VV^\dagger + RR^\dagger = SS^\dagger + UU^\dagger & = & {\bf 1} \; , \nonumber \\
V^\dagger V + S^\dagger S = R^\dagger R + U^\dagger U & = & {\bf
1} \; , \nonumber \\
V S^\dagger + R U^\dagger = V^\dagger R + S^\dagger U & = & {\bf 0}
\; .
\end{eqnarray}
After this diagonalization, Eq. (26) becomes
\begin{eqnarray}
-{\cal L}^{\prime}_{\rm hybrid} = \frac{1}{2} \left [
\begin{matrix} \overline{\nu^\prime_{\rm L}} &
\overline{(N^\prime_{\rm R})^c} \cr \end{matrix} \right ] \left(
\begin{matrix} \widehat{M}^{}_\nu & {\bf 0} \cr {\bf 0} &
\widehat{M}^{}_N \cr \end{matrix} \right) \left [ \begin{matrix}
(\nu^\prime_{\rm L})^c \cr N^\prime_{\rm R} \cr \end{matrix} \right
] + {\rm h.c.} \; ,
\end{eqnarray}
where $\nu^\prime_{\rm L} = V^\dagger \nu^{}_{\rm L} + S^\dagger
(N^{}_{\rm R})^c$ and $N^\prime_{\rm R} = R^T (\nu^{}_{\rm L})^c +
U^T N^{}_{\rm R}$ together with $(\nu^\prime_{\rm L})^c = {\cal C}
\overline{\nu^\prime_{\rm L}}^T$ and $(N^\prime_{\rm R})^c = {\cal
C} \overline{N^\prime_{\rm R}}^T$. Then the Majorana field
\begin{eqnarray}
\nu^\prime = \left [ \begin{matrix} \nu^\prime_{\rm L} \cr
(N^\prime_{\rm R})^c \cr \end{matrix} \right ] + \left [
\begin{matrix} (\nu^\prime_{\rm L})^c \cr N^\prime_{\rm R} \cr \end{matrix}
\right ] = \left( \begin{matrix} \nu^{}_1 \cr \nu^{}_2 \cr \nu^{}_3
\cr N^{}_1 \cr N^{}_2 \cr N^{}_3 \cr \end{matrix} \right)
\end{eqnarray}
satisfies the Majorana condition $(\nu^\prime)^c = \nu^\prime$ and
describes the mass eigenstates of six Majorana neutrinos. In other
words,
\begin{eqnarray}
-{\cal L}^{\prime}_{\rm hybrid} = \frac{1}{2} \overline{\nu^\prime}
\left( \begin{matrix} \widehat{M}^{}_\nu & {\bf 0} \cr {\bf 0} &
\widehat{M}^{}_N \cr \end{matrix} \right) \nu^\prime = \frac{1}{2}
\sum^3_{i=1} \left( m^{}_i \overline{\nu^{}_i} \nu^{}_i + M^{}_i
\overline{N^{}_i} N^{}_i \right) \; .
\end{eqnarray}
Because of $\nu^{}_{\rm L} = V\nu^\prime_{\rm L} + R (N^\prime_{\rm
R})^c$ and $N^{}_{\rm R} = S^* (\nu^\prime_{\rm L})^c + U^*
N^\prime_{\rm R}$, we immediately have $(\nu^{}_{\rm L})^c = V^*
(\nu^\prime_{\rm L})^c + R^* N^\prime_{\rm R}$ and $(N^{}_{\rm R})^c
= S \nu^\prime_{\rm L} + U (N^\prime_{\rm R})^c$. Given the generic
relations $\overline{(\psi^{}_{\rm L})^c} \gamma^{}_\mu \partial^\mu
(\psi^{}_{\rm L})^c = \overline{\psi^{}_{\rm L}} \gamma^{}_\mu
\partial^\mu \psi^{}_{\rm L}$ and $\overline{(\psi^{}_{\rm R})^c}
\gamma^{}_\mu \partial^\mu (\psi^{}_{\rm R})^c =
\overline{\psi^{}_{\rm R}} \gamma^{}_\mu \partial^\mu \psi^{}_{\rm
R}$ for an arbitrary fermion field $\psi$, the kinetic term of
Majorana neutrinos under consideration turns out to be
\begin{eqnarray}
{\cal L}^{}_{\rm kinetic} = i \overline{\nu^{}_{\rm L}}
\gamma^{}_\mu \partial^\mu \nu^{}_{\rm L} + i \overline{N^{}_{\rm
R}} \gamma^{}_\mu \partial^\mu N^{}_{\rm R} = i
\overline{\nu^\prime_{\rm L}} \gamma^{}_\mu
\partial^\mu \nu^\prime_{\rm L} + i \overline{N^\prime_{\rm R}}
\gamma^{}_\mu \partial^\mu N^\prime_{\rm R} = \frac{i}{2}
\overline{\nu^\prime} \gamma^{}_\mu \partial^\mu \nu^\prime
\nonumber \\
= \frac{i}{2} \sum^3_{k=1} \left( \overline{\nu^{}_k} \gamma^{}_\mu
\partial^\mu \nu^{}_k + \overline{N^{}_k} \gamma^{}_\mu \partial^\mu
N^{}_k \right) \; , ~~~~~~~~~~
\end{eqnarray}
where the unitarity conditions given in Eq. (29) have been used.

The weak charged-current interactions of active neutrinos in Eq. (5)
can now be rewritten in terms of the mass eigenstates of six
Majorana neutrinos via $\nu^{}_{\rm L} = V \nu^\prime_{\rm L} + R
(N^\prime_{\rm R})^c$. In the chosen basis where the flavor and mass
eigenstates of three charged leptons are identical, we have
\begin{eqnarray}
{\cal L}^{}_{\rm cc} = \frac{g}{\sqrt{2}} \overline{\left(e~ \mu~
\tau\right)^{}_{\rm L}} ~\gamma^\mu \left[ V \left( \begin{matrix}
\nu^{}_1 \cr \nu^{}_2 \cr \nu^{}_3 \end{matrix} \right)^{}_{\rm L} +
R \left( \begin{matrix} N^{}_1 \cr N^{}_2 \cr N^{}_3
\end{matrix} \right)^{}_{\rm L} \right] W^-_\mu + {\rm h.c.} \; .
\end{eqnarray}
Note that $V$ and $R$ are responsible for the charged-current
interactions of three known neutrinos $\nu^{}_i$ and three new
neutrinos $N^{}_i$ (for $i=1, 2, 3$), respectively. Their
correlation is described by $VV^\dagger + RR^\dagger = {\bf 1}$, and
thus $V$ is not unitary unless $\nu^{}_i$ and $N^{}_i$ are
completely decoupled (i.e., $R = {\bf 0}$).

As a consequence of lepton number violation, the $0\nu 2\beta$ decay
$(A,Z) \rightarrow (A,Z+2) + 2e^-$ can now take place via the
exchanges of both $\nu^{}_i$ and $N^{}_i$ between two beta decays,
whose coupling matrix elements are $V^{}_{ei}$ and $R^{}_{ei}$
respectively. The relative contributions of $\nu^{}_i$ and $N^{}_i$
to this lepton-number-violating process depend not only on $m^{}_i$,
$M^{}_i$, $V^{}_{ei}$ and $R^{}_{ei}$ but also on the relevant
nuclear matrix elements which cannot be reliably evaluated. For a
realistic seesaw mechanism working at the TeV scale (i.e., $M^{}_i
\sim {\cal O}(1)$ TeV) or at a superhigh-energy scale, however, the
contribution of $\nu^{}_i$ to the $0\nu 2\beta$ decay is in most
cases dominant.

The hybrid neutrino mass terms in Eq. (26) provide us with the
necessary ingredients of a dynamic mechanism to interpret why three
known neutrinos have non-zero but tiny masses. The key point is that
the mass scales of $M^{}_{\rm L}$, $M^{}_{\rm D}$ and $M^{}_{\rm R}$
may have a strong hierarchy. First, $M^{}_{\rm D} \sim \langle
H\rangle \approx 174$ GeV is naturally characterized by the
electroweak symmetry breaking scale. Second, $M^{}_{\rm L} \ll
\langle H\rangle$ satisfies 't Hooft's naturalness criterion because
this Majorana mass term violates lepton number conservation. Third,
$M^{}_{\rm R} \gg \langle H\rangle$ is naturally expected since
right-handed neutrinos are $SU(2)^{}_{\rm L}$ gauge singlets and
thus their mass term is not subject to the electroweak symmetry
breaking scale. The hierarchy $M^{}_{\rm R} \gg M^{}_{\rm D} \gg
M^{}_{\rm L}$ can therefore allow us to make reliable approximations
in deriving the effective mass matrix of three active neutrinos
$(\nu^{}_e, \nu^{}_\mu, \nu^{}_\tau)$ from Eq. (28). The latter
yields
\begin{eqnarray}
R\widehat{M}^{}_N = M^{}_{\rm L} R^* + M^{}_{\rm D} U^* \; ,
\nonumber \\
S\widehat{M}^{}_\nu = M^T_{\rm D} V^* + M^{}_{\rm R} S^* \; ; ~
\end{eqnarray}
and
\begin{eqnarray}
U\widehat{M}^{}_N = M^{}_{\rm R} U^* + M^T_{\rm D} R^* \; ,
\nonumber \\
V\widehat{M}^{}_\nu = M^{}_{\rm L} V^* + M^{}_{\rm D} S^* \; . ~~
\end{eqnarray}
Given $M^{}_{\rm R} \gg M^{}_{\rm D} \gg M^{}_{\rm L}$, $R \sim S
\sim {\cal O}(M^{}_{\rm D}/M^{}_{\rm R})$ naturally holds, implying
that $U$ and $V$ are almost unitary up to the accuracy of ${\cal
O}(M^2_{\rm D}/M^2_{\rm R})$. Hence Eq. (36) leads to
\begin{eqnarray}
&& U\widehat{M}^{}_N U^T = M^{}_{\rm R} (U U^\dagger)^T + M^T_{\rm
D} (R^*U^T) \approx M^{}_{\rm R} \; ,
\nonumber \\
&& V\widehat{M}^{}_\nu V^T = M^{}_{\rm L} (V V^\dagger)^T +
M^{}_{\rm D} (S^*V^T) \approx M^{}_{\rm L} + M^{}_{\rm D} (S^*V^T)
\; .
\end{eqnarray}
$S^*V^T = M^{-1}_{\rm R} S \widehat{M}^{}_\nu V^T - M^{-1}_{\rm R}
M^T_{\rm D} (V V^\dagger)^T \approx -M^{-1}_{\rm R} M^T_{\rm D}$ can
be derived from Eq. (35). We substitute this expression into Eq.
(37) and then obtain
\begin{eqnarray}
M^{}_\nu \equiv V \widehat{M}^{}_\nu V^T \approx M^{}_{\rm L} -
M^{}_{\rm D} M^{-1}_{\rm R} M^T_{\rm D} \; .
\end{eqnarray}
This result, known as the type-(I+II) seesaw relation, is just the
effective mass matrix of three light neutrinos. The small mass scale
of $M^{}_\nu$ is attributed to the small mass scale of $M^{}_{\rm
L}$ and the large mass scale of $M^{}_{\rm R}$. There are two
particularly interesting limits: (1) If $M^{}_{\rm L}$ is absent
from Eq. (26), one will be left with the canonical or type-I seesaw
relation $M^{}_\nu \approx - M^{}_{\rm D} M^{-1}_{\rm R} M^T_{\rm
D}$; (2) If only $M^{}_{\rm L}$ is present in Eq. (26), one will get
the type-II seesaw relation $M^{}_\nu = M^{}_{\rm L}$. More detailed
discussions about various seesaw mechanisms and their
phenomenological consequences will be presented in sections 6, 7 and
8.

\section{Diagnosis of CP Violation}

\subsection{C, P and T transformations}

We begin with a brief summary of the transformation properties of
quantum fields under the discrete space-time symmetries of parity
(P), charge conjugation (C) and time reversal (T). The parity
transformation changes the space coordinates $\vec{x}$ into
$-\vec{x}$. The charge conjugation flips the signs of internal
charges of a particle, such as the electric charge and the lepton
(baryon) number. The time reversal reflects the time coordinate $t$
into $-t$.

A free Dirac spinor $\psi(t, \vec{x})$ or $\overline{\psi}(t,
\vec{x})$ transforms under C, P and T as
\footnote{For simplicity, here we have omitted a phase factor
associated with each transformation. Because one is always
interested in the spinor bilinears, the relevant phase factor
usually plays no physical role.}
\begin{eqnarray}
\psi(t, \vec{x}) \stackrel{\rm C}{\longrightarrow} {\cal C}
\overline{\psi}^T(t, \vec{x}) \; , ~~~~~\;\;
\nonumber \\
\overline{\psi}(t, \vec{x}) \stackrel{\rm C}{\longrightarrow}
-\psi^T(t, \vec{x}) {\cal C}^{-1} \; , \;
\nonumber \\
\psi(t, \vec{x}) \stackrel{\rm P}{\longrightarrow} {\cal P} \psi(t,
-\vec{x}) \; , ~~~~\;\;
\nonumber \\
\overline{\psi}(t, \vec{x}) \stackrel{\rm P}{\longrightarrow}
\overline{\psi}(t, -\vec{x}) {\cal P}^\dagger \; , ~~~~
\nonumber \\
\psi(t, \vec{x}) \stackrel{\rm T}{\longrightarrow} {\cal T} \psi(-t,
\vec{x}) \; , ~~~~\;\;
\nonumber \\
\overline{\psi}(t, \vec{x}) \stackrel{\rm T}{\longrightarrow}
\overline{\psi}(-t, \vec{x}) {\cal T}^\dagger \; , ~~~~
\end{eqnarray}
where ${\cal C} = i \gamma^{}_2 \gamma^{}_0$, ${\cal P} =
\gamma^{}_0$ and ${\cal T} = \gamma^{}_1 \gamma^{}_3$ in the
Dirac-Pauli representation. These transformation properties can
simply be deduced from the requirement that the Dirac equation $i
\gamma^{}_\mu
\partial^\mu \psi(t, \vec{x}) = m \psi(t, \vec{x})$ be invariant
under C, P or T operation. Note that all the classical numbers (or
c-numbers), such as the coupling constants and $\gamma$-matrix
elements, must be complex-conjugated under T. Note also that the
charge-conjugation matrix $\cal C$ satisfies the conditions given in
Eq. (4). It is very important to figure out how the Dirac spinor
bilinears transform under C, P and T, because both leptons and
quarks are described by spinor fields and they always appear in the
bilinear forms in a Lorentz-invariant Lagrangian. Let us consider
the following scalar-, pseudoscalar-, vector-, pseudovector- and
tensor-like spinor bilinears: $\overline{\psi^{}_1} \psi^{}_2$, $i
\overline{\psi^{}_1} \gamma^{}_5 \psi^{}_2$, $\overline{\psi^{}_1}
\gamma^{}_\mu \psi^{}_2$, $\overline{\psi^{}_1} \gamma^{}_\mu
\gamma^{}_5 \psi^{}_2$ and $\overline{\psi^{}_1} \sigma^{}_{\mu\nu}
\psi^{}_2$, where $\sigma^{}_{\mu\nu} \equiv i [\gamma^{}_\mu,
\gamma^{}_\nu]/2$ is defined. One may easily verify that all these
bilinears are Hermitian. Under C, P and T, for example,
\begin{eqnarray}
\overline{\psi^{}_1} \gamma^{}_\mu \psi^{}_2 \stackrel{\rm
C}{\longrightarrow} - \psi^T_1 {\cal C}^{-1} \gamma^{}_\mu {\cal C}
\overline{\psi^{}_2}^T = \psi^T_1 \gamma^T_\mu
\overline{\psi^{}_2}^T = - \left[ \overline{\psi^{}_2} \gamma^{}_\mu
\psi^{}_1 \right]^T = - \overline{\psi^{}_2} \gamma^{}_\mu \psi^{}_1
\; ,
\nonumber \\
\overline{\psi^{}_1} \gamma^{}_\mu \psi^{}_2 \stackrel{\rm
P}{\longrightarrow} \overline{\psi^{}_1} \gamma^{}_0 \gamma^{}_\mu
\gamma^{}_0 \psi^{}_2 = \overline{\psi^{}_1} \gamma^\mu \psi^{}_2 \;
, ~~~~~~~~~~~~~~~~~~~~~~~~~~~~~~~
\nonumber \\
\overline{\psi^{}_1} \gamma^{}_\mu \psi^{}_2 \stackrel{\rm
T}{\longrightarrow} \overline{\psi^{}_1} \left(\gamma^{}_1
\gamma^{}_3\right)^\dagger \gamma^*_\mu \left(\gamma^{}_1
\gamma^{}_3\right) \psi^{}_2 = \overline{\psi^{}_1} \gamma^\mu
\psi^{}_2 \; ; ~~~~~~~~~~~~~~~~~~~~~~
\end{eqnarray}
and thus
\begin{eqnarray}
\overline{\psi^{}_1} \gamma^{}_\mu \psi^{}_2 \stackrel{\rm
CP}{\longrightarrow} - \overline{\psi^{}_2} \gamma^\mu \psi^{}_1 \;
,
\nonumber \\
\overline{\psi^{}_1} \gamma^{}_\mu \psi^{}_2 \stackrel{\rm
CPT}{\longrightarrow} - \overline{\psi^{}_2} \gamma^{}_\mu \psi^{}_1
\; ,
\end{eqnarray}
with $\vec{x} \rightarrow -\vec{x}$ under P and $t \rightarrow -t$
under T for $\psi^{}_1$ and $\psi^{}_2$. The transformation
properties of five spinor bilinears under C, P, T, CP and CPT are
summarized in Table 1, where one should keep in mind that all the
c-numbers are complex-conjugated under T and CPT.
\begin{table*}[t]
\begin{center}
\caption{Transformation properties of the scalar-, pseudoscalar-,
vector-, pseudovector- and tensor-like spinor bilinears under C, P
and T. Here $\vec{x} \rightarrow -\vec{x}$ under P, CP and CPT,
together with $t\rightarrow -t$ under T and CPT, is hidden and
self-explaining for $\psi^{}_1$ and $\psi^{}_2$.} \vspace{0.3cm}
\begin{tabular}{|c|c|c|c|c|c|}
  \hline
  \hline
  & $\overline{\psi^{}_1} \psi^{}_2$
  & $i \overline{\psi^{}_1} \gamma^{}_5 \psi^{}_2$
  & $\overline{\psi^{}_1} \gamma^{}_\mu \psi^{}_2$
  & $\overline{\psi^{}_1} \gamma^{}_\mu \gamma^{}_5 \psi^{}_2$
  & $\overline{\psi^{}_1} \sigma^{}_{\mu\nu} \psi^{}_2$ \\
  \hline
  C & $\overline{\psi^{}_2} \psi^{}_1$
  & $i \overline{\psi^{}_2} \gamma^{}_5 \psi^{}_1$
  & $-\overline{\psi^{}_2} \gamma^{}_\mu \psi^{}_1$
  & $\overline{\psi^{}_2} \gamma^{}_\mu \gamma^{}_5 \psi^{}_1$
  & $-\overline{\psi^{}_2} \sigma^{}_{\mu\nu} \psi^{}_1$ \\
  [2mm]
  P & $\overline{\psi^{}_1} \psi^{}_2$
  & $-i \overline{\psi^{}_1} \gamma^{}_5 \psi^{}_2$
  & $\overline{\psi^{}_1} \gamma^\mu \psi^{}_2$
  & $-\overline{\psi^{}_1} \gamma^\mu \gamma^{}_5 \psi^{}_2$
  & $\overline{\psi^{}_1} \sigma^{\mu\nu} \psi^{}_2$ \\
  [2mm]
  T & $\overline{\psi^{}_1} \psi^{}_2$
  & $-i \overline{\psi^{}_1} \gamma^{}_5 \psi^{}_2$
  & $\overline{\psi^{}_1} \gamma^\mu \psi^{}_2$
  & $\overline{\psi^{}_1} \gamma^\mu \gamma^{}_5 \psi^{}_2$
  & $-\overline{\psi^{}_1} \sigma^{\mu\nu} \psi^{}_2$ \\
  [2mm]
  CP & $\overline{\psi^{}_2} \psi^{}_1$
  & $-i \overline{\psi^{}_2} \gamma^{}_5 \psi^{}_1$
  & $-\overline{\psi^{}_2} \gamma^\mu \psi^{}_1$
  & $-\overline{\psi^{}_2} \gamma^\mu \gamma^{}_5 \psi^{}_1$
  & $-\overline{\psi^{}_2} \sigma^{\mu\nu} \psi^{}_1$ \\
  [2mm]
  CPT & $\overline{\psi^{}_2} \psi^{}_1$
  & $i \overline{\psi^{}_2} \gamma^{}_5 \psi^{}_1$
  & $-\overline{\psi^{}_2} \gamma^{}_\mu \psi^{}_1$
  & $-\overline{\psi^{}_2} \gamma^{}_\mu \gamma^{}_5 \psi^{}_1$
  & $\overline{\psi^{}_2} \sigma^{}_{\mu\nu} \psi^{}_1$ \\
  \hline
\end{tabular}
\end{center}
\end{table*}

It is well known that CPT is a good symmetry in a local quantum
field theory which is Lorentz-invariant and possesses a Hermitian
Lagrangian. The latter is necessary in order to have a unitary
transition operator (i.e., the $S$-matrix). The CPT invariance of a
theory implies that CP and T must be simultaneously conserving or
broken, as already examined in the quark sector of the SM via the
$K^0$-$\bar{K}^0$ mixing system. After a slight modification of the
SM by introducing the Dirac or Majorana mass term for three
neutrinos, one may also look at possible sources of CP or T
violation in the lepton sector.

\subsection{The source of CP violation}

The SM of electroweak interactions is based on the $SU(2)^{}_{\rm L}
\times U(1)^{}_{\rm Y}$ gauge symmetry and the Higgs mechanism. The
latter triggers the spontaneous symmetry breaking $SU(2)^{}_{\rm L}
\times U(1)^{}_{\rm Y} \rightarrow U(1)^{}_{\rm Q}$, such that three
gauge bosons, three charged leptons and six quarks can all acquire
masses. But this mechanism itself does not spontaneously break CP,
and thus one may examine the source of CP violation in the SM either
before or after spontaneous symmetry breaking.

The Lagrangian of the SM ${\cal L} = {\cal L}^{}_{\rm G} + {\cal
L}^{}_{\rm H} + {\cal L}^{}_{\rm F} + {\cal L}^{}_{\rm Y}$ is
composed of four parts: the kinetic term of the gauge fields and
their self-interactions (${\cal L}^{}_{\rm G}$), the kinetic term of
the Higgs doublet and its potential and interactions with the gauge
fields (${\cal L}^{}_{\rm H}$), the kinetic term of the fermion
fields and their interactions with the gauge fields (${\cal
L}^{}_{\rm F}$), and the Yukawa interactions of the fermion fields
with the Higgs doublet (${\cal L}^{}_{\rm Y}$):
\begin{eqnarray}
&& {\cal L}^{}_{\rm G} = -\frac{1}{4} \left( W^{i {\mu \nu}}
W^i_{\mu \nu} + B^{\mu \nu} B^{}_{\mu \nu} \right) \; ,
\nonumber \\
&& {\cal L}^{}_{\rm H} = \left(D^\mu H\right)^\dagger \left(D^{}_\mu
H\right) - \mu^2 H^\dagger H  - \lambda \left(H^\dagger H\right)^2
\; ,
\nonumber \\
&& {\cal L}^{}_{\rm F} = \overline{Q^{}_{\rm L}} i\slashed{D}
Q^{}_{\rm L} + \overline{\ell^{}_{\rm L}} i\slashed{D} \ell^{}_{\rm
L} + \overline{U^{}_{\rm R}} i\slashed{\partial}^\prime U^{}_{\rm R}
+ \overline{D^{}_{\rm R}} i\slashed{\partial}^\prime D^{}_{\rm R} +
\overline{E^{}_{\rm R}} i\slashed{\partial}^\prime E^{}_{\rm R} \; ,
\nonumber \\
&& {\cal L}^{}_{\rm Y} = - \overline{Q^{}_{\rm L}} Y^{}_{\rm u}
\tilde{H} U^{}_{\rm R} - \overline{Q^{}_{\rm L}} Y^{}_{\rm d} H
D^{}_{\rm R} - \overline{\ell^{}_{\rm L}} Y^{}_l H E^{}_{\rm R} +
{\rm h.c.} \; ,
\end{eqnarray}
whose notations are self-explanatory. To accommodate massive
neutrinos, the simplest way is to slightly modify the ${\cal
L}^{}_{\rm F}$ and ${\cal L}^{}_{\rm Y}$ parts (e.g., by introducing
three right-handed neutrinos into the SM and allowing for the Yukawa
interactions between neutrinos and the Higgs doublet). CP violation
is due to the coexistence of ${\cal L}^{}_{\rm F}$ and ${\cal
L}^{}_{\rm Y}$.

We first show that ${\cal L}^{}_{\rm G}$ is always invariant under
CP. The transformation properties of gauge fields $B^{}_\mu$ and
$W^i_\mu$ under C and P are
\begin{eqnarray}
&& \left[ B^{}_\mu, \; W^1_\mu, \; W^2_\mu, \; W^3_\mu \right]
\stackrel{\rm C}{\longrightarrow} \left[ -B^{}_\mu, \; -W^1_\mu, \;
+W^2_\mu, \; -W^3_\mu \right] \; , ~~~~~
\nonumber \\
&& \left[ B^{}_\mu, \; W^1_\mu, \; W^2_\mu, \; W^3_\mu \right]
\stackrel{\rm P}{\longrightarrow} \left[ B^\mu, \; W^{1 \mu}, \;
W^{2 \mu}, \; W^{3 \mu} \right] \; , ~~~~~~~~~
\nonumber \\
&& \left[ B^{}_\mu, \; W^1_\mu, \; W^2_\mu, \; W^3_\mu \right]
\stackrel{\rm CP}{\longrightarrow} \left[ -B^\mu, \; -W^{1 \mu}, \;
+W^{2 \mu}, \; -W^{3 \mu} \right] \; ~
\end{eqnarray}
with $\vec{x} \rightarrow -\vec{x}$ under P and CP for relevant
fields. Then the gauge field tensors $B^{}_{\mu\nu}$ and
$W^i_{\mu\nu}$ transform under CP as follows:
\begin{eqnarray}
\left[ B^{}_{\mu\nu}, \; W^1_{\mu\nu}, \; W^2_{\mu\nu}, \;
W^3_{\mu\nu} \right] \stackrel{\rm CP}{\longrightarrow} \left[
-B^{\mu\nu}, \; -W^{1 \mu\nu}, \; +W^{2 \mu\nu}, \; -W^{3 \mu\nu}
\right] \; .
\end{eqnarray}
Hence ${\cal L}^{}_{\rm G}$ is formally invariant under CP.

We proceed to show that ${\cal L}^{}_{\rm H}$ is also invariant
under CP. The Higgs doublet $H$ contains two scalar components
$\phi^+$ and $\phi^0$; i.e.,
\begin{equation}
H = \left( \begin{matrix} \phi^+ \cr \phi^0 \cr \end{matrix} \right)
\; , ~~~~ H^\dagger = \left( \begin{matrix} \phi^- & {\phi^0}^* \cr
\end{matrix} \right) \; .
\end{equation}
Therefore,
\begin{equation}
H(t, \vec{x}) \stackrel{\rm CP}{\longrightarrow} H^*(t, -\vec{x}) =
\left( \begin{matrix} \phi^- \cr {\phi^0}^* \cr \end{matrix} \right)
\; .
\end{equation}
It is very trivial to prove that the $H^\dagger H$ and $(H^\dagger
H)^2$ terms of ${\cal L}^{}_{\rm H}$ are CP-invariant. To examine
how the $(D^\mu H)^\dagger (D^{}_\mu H)$ term of ${\cal L}^{}_{\rm
H}$ transforms under CP, we explicitly write out
\begin{eqnarray}
D^{}_\mu H = \left( \partial^{}_\mu - i g \tau^k W^k_\mu - i
g^\prime Y B^{}_\mu \right) H = \left( \begin{matrix}
\partial^{}_\mu \phi^+ - i X^+_\mu \phi^0 - i Y^+_\mu \phi^+ \cr
\partial^{}_\mu \phi^0 - i X^-_\mu \phi^+ + i Y^-_\mu \phi^0 \cr
\end{matrix} \right) \; ~~~~~
\end{eqnarray}
with $X^\pm_\mu \equiv g W^\pm_\mu/\sqrt{2} = g (W^1_\mu \mp i
W^2_\mu)/2$, $Y^\pm \equiv \pm g^\prime Y B^{}_\mu + g W^3_\mu/2$,
and $k=1,2,3$. Note that
\begin{equation}
X^\pm_\mu \stackrel{\rm CP}{\longrightarrow} -X^{\mp\mu} \; , ~~~~
Y^\pm_\mu \stackrel{\rm CP}{\longrightarrow} -Y^{\pm\mu} \; ,
\end{equation}
together with $\partial^{}_\mu \rightarrow \partial^\mu$, $\phi^\pm
\rightarrow \phi^\mp$ and $\phi^0 \rightarrow {\phi^0}^*$ under CP.
So it is easy to check that $(D^\mu H)^\dagger (D^{}_\mu H)$ is also
CP-invariant. Therefore, ${\cal L}^{}_{\rm H}$ is formally invariant
under CP.

The next step is to examine the CP invariance of ${\cal L}^{}_{\rm
F}$. To be more specific, we divide ${\cal L}^{}_{\rm F}$ into the
quark sector and the lepton sector; i.e., ${\cal L}^{}_{\rm F} =
{\cal L}^{}_q + {\cal L}^{}_l$. We only analyze the CP property of
${\cal L}^{}_q$ in the following, because that of ${\cal L}^{}_l$
can be analyzed in the same way. The explicit form of ${\cal
L}^{}_q$ reads
\begin{eqnarray}
{\cal L}^{}_q = \overline{Q^{}_{\rm L}} i\slashed{D} Q^{}_{\rm L} +
\overline{U^{}_{\rm R}} i\slashed{\partial}^\prime U^{}_{\rm R} +
\overline{D^{}_{\rm R}} i\slashed{\partial}^\prime D^{}_{\rm R} =
\sum^3_{j=1} \left\{ \frac{g}{2} \left[ \overline{q^\prime_j}
\gamma^\mu P^{}_{\rm L} W^1_\mu q^{}_j + \overline{q^{}_j}
\gamma^\mu P^{}_{\rm L} W^1_\mu q^\prime_j \right] \right .
\nonumber \\
+ \frac{g}{2} \left[ i \overline{q^\prime_j} \gamma^\mu P^{}_{\rm L}
W^2_\mu q^{}_j - i \overline{q^{}_j} \gamma^\mu P^{}_{\rm L} W^2_\mu
q^\prime_j \right] ~~~~~~
\nonumber \\
+ \frac{g}{2} \left[ \overline{q^{}_j} \gamma^\mu P^{}_{\rm L}
W^3_\mu q^{}_j - \overline{q^\prime_j} \gamma^\mu P^{}_{\rm L}
W^3_\mu q^\prime_j \right] ~~~~~~~~~
\nonumber \\
+ i \left[ \overline{q^{}_j} \gamma^\mu P^{}_{\rm L} \left(
\partial^{}_\mu - i \frac{g^\prime}{6} B^{}_\mu \right) q^{}_j
\right] ~~~~~~~~~~~~~~~
\nonumber \\
+ i \left[ \overline{q^\prime_j} \gamma^\mu P^{}_{\rm L} \left(
\partial^{}_\mu - i \frac{g^\prime}{6} B^{}_\mu \right) q^\prime_j
\right] ~~~~~~~~~~~~~~~
\nonumber \\
+ i \left[ \overline{q^{}_j} \gamma^\mu P^{}_{\rm R} \left(
\partial^{}_\mu - i \frac{2g^\prime}{3} B^{}_\mu \right) q^{}_j
\right] ~~~~~~~~~~~~~
\nonumber \\
\left . + i \left[ \overline{q^\prime_j} \gamma^\mu P^{}_{\rm R}
\left( \partial^{}_\mu + i \frac{g^\prime}{3} B^{}_\mu \right)
q^\prime_j \right] \right\} \; , ~~~~~~~~~
\end{eqnarray}
where $q^{}_j$ and $q^\prime_j$ (for $j=1,2,3$) run over $(u, c, t)$
and $(d, s, b)$, respectively. The transformation properties of
gauge fields $B^{}_\mu$ and $W^i_\mu$ under C and P have been given
in Eq. (43). With the help of Table 1, one can see that the relevant
spinor bilinears transform under C and P as follows:
\begin{eqnarray}
\overline{\psi^{}_1} \gamma^{}_\mu \left( 1 \pm \gamma^{}_5 \right)
\psi^{}_2 \stackrel{\rm C}{\longrightarrow} -\overline{\psi^{}_2}
\gamma^{}_\mu \left( 1 \mp \gamma^{}_5 \right) \psi^{}_1 \; ,
\nonumber \\
\overline{\psi^{}_1} \gamma^{}_\mu \left( 1 \pm \gamma^{}_5 \right)
\psi^{}_2 \stackrel{\rm P}{\longrightarrow} +\overline{\psi^{}_1}
\gamma^\mu \left( 1 \mp \gamma^{}_5 \right) \psi^{}_2 \; ,
\nonumber \\
\overline{\psi^{}_1} \gamma^{}_\mu \left( 1 \pm \gamma^{}_5 \right)
\psi^{}_2 \stackrel{\rm CP}{\longrightarrow} -\overline{\psi^{}_2}
\gamma^\mu \left( 1 \pm \gamma^{}_5 \right) \psi^{}_1 \; ,
\end{eqnarray}
with $\vec{x} \rightarrow -\vec{x}$ under P and CP for $\psi^{}_1$
and $\psi^{}_2$. Furthermore,
\begin{eqnarray}
\overline{\psi^{}_1} \gamma^{}_\mu \left( 1 \pm \gamma^{}_5 \right)
\partial^\mu \psi^{}_2 \stackrel{\rm C}{\longrightarrow}
\overline{\psi^{}_2} \gamma^{}_\mu \left( 1 \mp \gamma^{}_5 \right)
\partial^\mu \psi^{}_1 \; ,
\nonumber \\
\overline{\psi^{}_1} \gamma^{}_\mu \left( 1 \pm \gamma^{}_5 \right)
\partial^\mu \psi^{}_2 \stackrel{\rm P}{\longrightarrow}
\overline{\psi^{}_1} \gamma^\mu \left( 1 \mp \gamma^{}_5 \right)
\partial^{}_\mu \psi^{}_2 \; ,
\nonumber \\
\overline{\psi^{}_1} \gamma^{}_\mu \left( 1 \pm \gamma^{}_5 \right)
\partial^\mu \psi^{}_2 \stackrel{\rm CP}{\longrightarrow}
\overline{\psi^{}_2} \gamma^\mu \left( 1 \pm \gamma^{}_5 \right)
\partial^{}_\mu \psi^{}_1 \; ,
\end{eqnarray}
with $\vec{x} \rightarrow -\vec{x}$ under P and CP for $\psi^{}_1$
and $\psi^{}_2$. It is straightforward to check that ${\cal L}^{}_q$
in Eq. (49) is formally invariant under CP. Following the same
procedure and using Eqs. (49), (50) and (51), one can easily show
that ${\cal L}^{}_l = \overline{\ell^{}_{\rm L}} i\slashed{D}
\ell^{}_{\rm L} + \overline{E^{}_{\rm R}} i\slashed{\partial}^\prime
E^{}_{\rm R}$ is also CP-invariant. Thus we conclude that ${\cal
L}^{}_{\rm F}$ is invariant under CP.

The last step is to examine whether ${\cal L}^{}_{\rm Y}$ is
CP-conserving or not. Explicitly,
\begin{eqnarray}
-{\cal L}^{}_{\rm Y} = \overline{Q^{}_{\rm L}} Y^{}_{\rm u}
\tilde{H} U^{}_{\rm R} + \overline{Q^{}_{\rm L}} Y^{}_{\rm d} H
D^{}_{\rm R} + \overline{\ell^{}_{\rm L}} Y^{}_l H E^{}_{\rm R} +
{\rm h.c.}
\nonumber \\
= \sum^3_{j,k=1} \left\{ (Y^{}_{\rm u})^{}_{jk} \left[
\overline{q^{}_j} P^{}_{\rm R} q^{}_k {\phi^0}^* -
\overline{q^\prime_j} P^{}_{\rm R} q^{}_k \phi^- \right] \right .
~~~~~~~
\nonumber \\
+ (Y^{}_{\rm u})^*_{jk} \left[ \overline{q^{}_k} P^{}_{\rm L} q^{}_j
\phi^0 - \overline{q^{}_k} P^{}_{\rm L} q^\prime_j \phi^+ \right]
~~~~~~~~~~~~~~~
\nonumber \\
+ (Y^{}_{\rm d})^{}_{jk} \left[ \overline{q^{}_j} P^{}_{\rm R}
q^\prime_k \phi^+ + \overline{q^\prime_j} P^{}_{\rm R} q^\prime_k
\phi^0 \right] ~~~~~~~~~~~~~~
\nonumber \\
+ (Y^{}_{\rm d})^*_{jk} \left[ \overline{q^\prime_k} P^{}_{\rm L}
q^{}_j \phi^- + \overline{q^\prime_k} P^{}_{\rm L} q^\prime_j
{\phi^0}^* \right] ~~~~~~~~~~~~~
\nonumber \\
+ (Y^{}_l)^{}_{jk} \left[ \overline{\nu^{}_j} P^{}_{\rm R} l^{}_k
\phi^+ + \overline{l^{}_j} P^{}_{\rm R} l^{}_k \phi^0 \right]
~~~~~~~~~~~~~~~~
\nonumber \\
\left . + (Y^{}_l)^*_{jk} \left[ \overline{l^{}_k} P^{}_{\rm L}
\nu^{}_j \phi^- + \overline{l^{}_k} P^{}_{\rm L} l^{}_j {\phi^0}^*
\right] \right\} \; , ~~~~~~~~~
\end{eqnarray}
where $q^{}_j$ and $q^\prime_j$ (for $j=1,2,3$) run over $(u, c, t)$
and $(d, s, b)$, respectively; while $\nu^{}_j$ and $l^{}_j$ (for
$j=1,2,3$) run over $(\nu^{}_e, \nu^{}_\mu, \nu^{}_\tau)$ and $(e,
\mu, \tau)$, respectively. Because of $\phi^\pm \rightarrow
\phi^\mp$, $\phi^0 \rightarrow {\phi^0}^*$ and $\overline{\psi^{}_1}
(1 \pm \gamma^{}_5) \psi^{}_2 \rightarrow \overline{\psi^{}_2} (1
\mp \gamma^{}_5) \psi^{}_1$ under CP, we immediately arrive at
\begin{eqnarray}
-{\cal L}^{}_{\rm Y} \stackrel{\rm CP}{\longrightarrow}
\sum^3_{j,k=1} \left\{ (Y^{}_{\rm u})^{}_{jk} \left[
\overline{q^{}_k} P^{}_{\rm L} q^{}_j \phi^0 \right . -
\overline{q^{}_k} P^{}_{\rm L} q^\prime_j \phi^+ \right]
\nonumber \\
+ (Y^{}_{\rm u})^*_{jk} \left[ \overline{q^{}_j} P^{}_{\rm R} q^{}_k
{\phi^0}^* - \overline{q^\prime_j} P^{}_{\rm R} q^{}_k \phi^-
\right] \
\nonumber \\
+ (Y^{}_{\rm d})^{}_{jk} \left[ \overline{q^\prime_k} P^{}_{\rm L}
q^{}_j \phi^- + \overline{q^\prime_k} P^{}_{\rm L} q^\prime_j
{\phi^0}^* \right] \ \
\nonumber \\
+ (Y^{}_{\rm d})^*_{jk} \left[ \overline{q^{}_j} P^{}_{\rm R}
q^\prime_k \phi^+ + \overline{q^\prime_j} P^{}_{\rm R} q^\prime_k
\phi^0 \right] \; ~ \
\nonumber \\
+ (Y^{}_l)^{}_{jk} \left[ \overline{l^{}_k} P^{}_{\rm L} \nu^{}_j
\phi^- + \overline{l^{}_k} P^{}_{\rm L} l^{}_j {\phi^0}^* \right]
~~~~
\nonumber \\
\left . + (Y^{}_l)^*_{jk} \left[ \overline{\nu^{}_j} P^{}_{\rm R}
l^{}_k \phi^+ + \overline{l^{}_j} P^{}_{\rm R} l^{}_k \phi^0 \right]
\right\} \; ,
\end{eqnarray}
with $\vec{x} \rightarrow -\vec{x}$ for both scalar and spinor
fields under consideration. Comparing between Eqs. (52) and (53), we
see that ${\cal L}^{}_{\rm Y}$ will be formally invariant under CP
if the conditions
\begin{eqnarray}
(Y^{}_{\rm u})^{}_{jk} = (Y^{}_{\rm u})^*_{jk} \; , ~~~~ (Y^{}_{\rm
d})^{}_{jk} = (Y^{}_{\rm d})^*_{jk} \; , ~~~~ (Y^{}_l)^{}_{jk} =
(Y^{}_l)^*_{jk}
\end{eqnarray}
are satisfied. In other words, the Yukawa coupling matrices
$Y^{}_{\rm u}$, $Y^{}_{\rm d}$ and $Y^{}_l$ must be real to
guarantee the CP invartiance of ${\cal L}^{}_{\rm Y}$. Given three
massless neutrinos in the SM, it is always possible to make $Y^{}_l$
real by redefining the phases of charged-lepton fields. But it is in
general impossible to make both $Y^{}_{\rm u}$ and $Y^{}_{\rm d}$
real for three families of quarks, and thus CP violation can only
appear in the quark sector.

Given massive neutrinos beyond the SM, ${\cal L}^{}_{\rm Y}$ must be
modified. The simplest way is to introduce three right-handed
neutrinos and incorporate the Dirac neutrino mass term in Eq. (6)
into ${\cal L}^{}_{\rm Y}$. In this case one should also add the
kinetic term of three right-handed neutrinos into ${\cal L}^{}_{\rm
F}$. It is straightforward to show that the conditions of CP
invariance in the lepton sector turn out to be
\begin{equation}
Y^{}_\nu = Y^*_\nu \; , ~~~~~~ Y^{}_l = Y^*_l \; ,
\end{equation}
exactly in parallel with the quark sector. If an effective Majorana
mass term is introduced into ${\cal L}^{}_{\rm Y}$, as shown in Eq.
(14), then the conditions of CP invariance in the lepton sector
become
\begin{equation}
M^{}_{\rm L} = M^*_{\rm L} \; , ~~~~ Y^{}_l = Y^*_l \; ,
\end{equation}
where $M^{}_{\rm L}$ is the effective Majorana neutrino mass matrix.
One may diagonalize both $Y^{}_\nu$ (or $M^{}_{\rm L}$) and $Y^{}_l$
to make them real and positive, but such a treatment will transfer
CP violation from the Yukawa interactions to the weak
charged-current interactions. Then lepton flavor mixing and CP
violation are described by the $3\times 3$ unitary matrix $V$ given
in Eq. (12), analogous to the $3\times 3$ unitary matrix of quark
flavor mixing and CP violation. In other words, the source of CP
violation is the irremovable complex phase(s) in the flavor mixing
matrix of quarks or leptons. That is why we claim that CP violation
stems from the coexistence of ${\cal L}^{}_{\rm F}$ and ${\cal
L}^{}_{\rm Y}$ within the SM and, in most cases, beyond the SM.

It is worth reiterating that the process of spontaneous gauge
symmetry breaking in the SM does not spontaneously violate CP. After
the Higgs doublet $H$ acquires its vacuum expectation value (i.e.,
$\phi^+ \rightarrow 0$ and $\phi^0 \rightarrow v/\sqrt{2}$ with $v$
being real), we obtain three massive gauge bosons $W^\pm_\mu$ and
$Z^{}_\mu$ as well as one massless gauge boson $A^{}_\mu$. According
to their relations with $W^i_\mu$ and $B^{}_\mu$, it is easy to find
out the transformation properties of these physical fields under CP:
\begin{eqnarray}
W^\pm_\mu \stackrel{\rm CP}{\longrightarrow} -W^{\mp \mu} \; , ~~~~
Z^{}_\mu \stackrel{\rm CP}{\longrightarrow} -Z^\mu \; , ~~~~
A^{}_\mu \stackrel{\rm CP}{\longrightarrow} -A^\mu \; ,
\end{eqnarray}
with $\vec{x} \rightarrow -\vec{x}$ under P and CP for each field.
In contrast, the neutral Higgs boson $h$ is a CP-even particle.
After spontaneous electroweak symmetry breaking, we are left with
the quark mass matrices $M^{}_{\rm u} = v Y^{}_{\rm u}/\sqrt{2}$ and
$M^{}_{\rm d} = v Y^{}_{\rm d}/\sqrt{2}$ or the lepton mass matrices
$M^{}_{\rm D} = v Y^{}_\nu/\sqrt{2}$ and $M^{}_l = v
Y^{}_l/\sqrt{2}~$. The conditions of CP invariance given above can
therefore be replaced with the corresponding mass matrices.

\section{Electromagnetic Properties}

\subsection{Electromagnetic form factors}

Although a neutrino does not possess any electric charge, it can
have electromagnetic interactions via quantum loops. One may
summarize such interactions by means of the following effective
interaction term:
\begin{equation}
{\cal L}^{}_{\rm EM} = \overline{\psi}\Gamma^{}_\mu \psi A^\mu
\equiv J^{}_\mu (x) A^\mu(x) \; ,
\end{equation}
where the form of the electromagnetic current $J^{}_\mu(x)$ is our
present concern. Dirac and Majorana neutrinos couple to the photon
in different ways, which are described by their respective
electromagnetic form factors.

For an arbitrary Dirac particle (e.g., a Dirac neutrino), let us
write down the matrix element of $J^{}_\mu(x)$ between two
one-particle states:
\begin{eqnarray}
\langle \psi(p^\prime)|J^{}_\mu (x)|\psi(p)\rangle = e^{-iqx}
\langle \psi(p^\prime)|J^{}_\mu (0)|\psi(p)\rangle = e^{-iqx}
\overline{u}(\vec{p}^\prime) \Gamma^{}_\mu (p, p^\prime) u(\vec{p})
\;
\end{eqnarray}
with $q = p - p^\prime$. Because $J^{}_\mu(x)$ is a Lorentz vector,
the electromagnetic vertex function $\Gamma^{}_\mu (p, p^\prime)$
must be a Lorentz vector too. The electromagnetic current
conservation (or $U(1)^{}_{\rm Q}$ gauge symmetry) requires
$\partial^\mu J^{}_\mu (x) =0$, leading to
\begin{eqnarray}
\langle \psi(p^\prime)|\partial^\mu J^{}_\mu (x)|\psi(p)\rangle =
\left( -i q^\mu \right) e^{-iqx} \overline{u}(\vec{p}^\prime)
\Gamma^{}_\mu (p, p^\prime) u(\vec{p}) = 0 \; .
\end{eqnarray}
Thus
\begin{equation}
q^\mu \overline{u}(\vec{p}^\prime) \Gamma^{}_\mu (p, p^\prime)
u(\vec{p}) = 0
\end{equation}
holds as one of the model-independent constraints on the form of
$\Gamma^{}_\mu (p, p^\prime)$. In addition, the Hermiticity of
$J^{}_\mu (x)$ or its matrix element implies
\begin{eqnarray}
&& e^{-iqx} \overline{u}(\vec{p}^\prime) \Gamma^{}_\mu (p, p^\prime)
u(\vec{p})  = e^{+iqx} \left[ \overline{u}(\vec{p}^\prime)
\Gamma^{}_\mu (p, p^\prime) u(\vec{p}) \right]^\dagger
\nonumber \\
&& = e^{+iqx} \overline{u}(\vec{p}) \left[ \gamma^{}_0
\Gamma^\dagger_\mu (p,p^\prime) \gamma^{}_0 \right]
u(\vec{p}^\prime) = e^{-iqx} \overline{u}(\vec{p}^\prime) \left[
\gamma^{}_0 \Gamma^\dagger_\mu (p^\prime,p) \gamma^{}_0 \right]
u(\vec{p}) \; ,
\end{eqnarray}
from which we immediately arrive at the second constraint on
$\Gamma^{}_\mu (p, p^\prime)$:
\begin{equation}
\Gamma^{}_\mu (p, p^\prime) = \gamma^{}_0 \Gamma^\dagger_\mu
(p^\prime,p) \gamma^{}_0 \; .
\end{equation}
Because of $p^2 = {p^\prime}^2 = m^2$ with $m$ being the fermion
mass, we have $(p + p^\prime)^2 = 4m^2 -q^2$. Hence $\Gamma^{}_\mu
(p, p^\prime)$ depends only on the Lorentz-invariant quantity $q^2$.

A careful analysis of the Lorentz structure of
$\overline{u}(\vec{p}^\prime) \Gamma^{}_\mu (p, p^\prime)
u(\vec{p})$, with the help of the Gordon-like identities and the
constraints given above, shows that $\Gamma^{}_\mu (p, p^\prime)$
may in general consist of four independent terms:
\begin{eqnarray}
\Gamma^{}_\mu(p, p^\prime) = f^{}_{\rm Q} (q^2) \gamma^{}_\mu +
f^{}_{\rm M} (q^2) i \sigma^{}_{\mu\nu} q^\nu + f^{}_{\rm E} (q^2)
\sigma^{}_{\mu\nu} q^\nu \gamma^{}_5 + f^{}_{\rm A} (q^2) \left( q^2
\gamma^{}_\mu - q^{}_\mu \slashed{q} \right) \gamma^{}_5 \; ,
\end{eqnarray}
where $f^{}_{\rm Q} (q^2)$, $f^{}_{\rm M} (q^2)$, $f^{}_{\rm E}
(q^2)$ and $f^{}_{\rm A} (q^2)$ are usually referred to as the
charge, magnetic dipole, electric dipole and anapole form factors,
respectively. In the non-relativistic limit of ${\cal L}^{}_{\rm
EM}$, it is easy to find that $f^{}_{\rm Q}(0) = Q$ represents the
electric charge of the particle, $f^{}_{\rm M}(0) \equiv \mu$
denotes the magnetic dipole moment of the particle (i.e., ${\cal
L}^{}_{\rm EM}(f^{}_{\rm M}) = - \mu \vec{\sigma}\cdot\vec{B}$ with
$\vec{B}$ being the static magnetic field), $f^{}_{\rm E}(0) \equiv
\epsilon$ stands for the electric dipole moment of the particle
(i.e., ${\cal L}^{}_{\rm EM}(f^{}_{\rm E}) = - \epsilon
\vec{\sigma}\cdot\vec{E}$ with $\vec{E}$ being the static electric
field), and $f^{}_{\rm A}(0)$ corresponds to the Zeldovich anapole
moment of the particle (i.e., ${\cal L}^{}_{\rm EM}(f^{}_{\rm A})
\propto f^{}_{\rm A}(0) \vec{\sigma}\cdot [ \nabla \times \vec{B} -
\dot{\vec E}]$). One can observe that these form factors are not
only Lorentz-invariant but also real (i.e., ${\rm Im}f^{}_{\rm Q} =
{\rm Im}f^{}_{\rm M} = {\rm Im}f^{}_{\rm E} = {\rm Im}f^{}_{\rm A}
=0$). The latter is actually guaranteed by the Hermiticity condition
in Eq. (62).

Given the form of $\Gamma^{}_\mu$ in Eq. (64), it is straightforward
to check the CP properties of ${\cal L}^{}_{\rm EM}$ in Eq. (58).
Note that the photon field transforms as $A^\mu \to -A^{}_\mu$ under
CP, and
\footnote{Taking account of ${\cal C}^{-1} \sigma^{}_{\mu\nu} {\cal
C} = -\sigma^T_{\mu\nu}$ and ${\cal C}^{-1} \gamma^{}_5 {\cal C} =
\gamma^T_5$, one may easily prove that $\overline{\psi}
\sigma^{}_{\mu\nu} \gamma^{}_5 \psi$ is odd under both C and P. Thus
$\overline{\psi} \sigma^{}_{\mu\nu} \gamma^{}_5 \psi$ is CP-even.}
\begin{eqnarray}
\overline{\psi} \gamma^{}_\mu \psi \stackrel{\rm
CP}{\longrightarrow} - \overline{\psi} \gamma^\mu \psi \; ,
~~~~~~~\;\;
\nonumber \\
\overline{\psi} \gamma^{}_\mu \gamma^{}_5 \psi \stackrel{\rm
CP}{\longrightarrow} - \overline{\psi} \gamma^\mu \gamma^{}_5 \psi
\; , ~~~
\nonumber \\
\overline{\psi} \sigma^{}_{\mu\nu} \psi \stackrel{\rm
CP}{\longrightarrow} - \overline{\psi} \sigma^{\mu\nu} \psi \; ,
~~~~~~
\nonumber \\
\overline{\psi} \sigma^{}_{\mu\nu} \gamma^{}_5 \psi \stackrel{\rm
CP}{\longrightarrow} + \overline{\psi} \sigma^{\mu\nu} \gamma^{}_5
\psi \; . \;
\end{eqnarray}
Hence only the term proportional to $f^{}_{\rm E}$ in ${\cal
L}^{}_{\rm EM}$ is CP-violating. If CP were conserved, then this
term would vanish (i.e., $f^{}_{\rm E} = 0$ would hold). Although
there is no experimental hint at CP violation in the lepton sector,
we expect that it should exist as in the quark sector. In any case,
all four form factors are finite for a Dirac neutrino.

If neutrinos are massive Majorana particles, their electromagnetic
properties will be rather different. The reason is simply that
Majorana particles are their own antiparticles and thus can be
described by using a smaller number of degrees of freedom. A free
Majorana neutrino field $\psi$ is by definition equal to its
charge-conjugate field $\psi^c = C\overline{\psi}^T$ up to a global
phase. Then
\begin{eqnarray}
\overline{\psi}\Gamma^{}_\mu \psi = \overline{\psi^c}\Gamma^{}_\mu
\psi^c = \psi^T {\cal C} \Gamma^{}_\mu {\cal C} \overline{\psi}^T =
\left( \psi^T {\cal C} \Gamma^{}_\mu {\cal C} \overline{\psi}^T
\right)^T = - \overline{\psi} {\cal C}^T \Gamma^T_\mu {\cal C}^T
\psi \; ,
\end{eqnarray}
from which one arrives at
\begin{equation}
\Gamma^{}_\mu = - {\cal C}^T \Gamma^T_\mu {\cal C}^T = {\cal C}
\Gamma^T_\mu {\cal C}^{-1} \; .
\end{equation}
Substituting Eq. (64) into the right-hand side of Eq. (67) and
taking account of ${\cal C} \gamma^T_\mu {\cal C}^{-1} =
-\gamma^{}_\mu$, ${\cal C} (\gamma^{}_\mu \gamma^{}_5)^T {\cal
C}^{-1} = +\gamma^{}_\mu \gamma^{}_5$, ${\cal C} \sigma^T_{\mu\nu}
{\cal C}^{-1} = -\sigma^{}_{\mu\nu}$ and ${\cal C}
(\sigma^{}_{\mu\nu} \gamma^{}_5)^T {\cal C}^{-1} =
-\sigma^{}_{\mu\nu} \gamma^{}_5$, we obtain
\begin{eqnarray}
\Gamma^{}_\mu(p, p^\prime) = -f^{}_{\rm Q} (q^2) \gamma^{}_\mu -
f^{}_{\rm M} (q^2) i \sigma^{}_{\mu\nu} q^\nu - f^{}_{\rm E} (q^2)
\sigma^{}_{\mu\nu} q^\nu \gamma^{}_5 + f^{}_{\rm A} (q^2) \left( q^2
\gamma^{}_\mu - q^{}_\mu \slashed{q} \right) \gamma^{}_5 \; .
\end{eqnarray}
A comparison between Eqs. (64) and (68) yields
\begin{equation}
f^{}_{\rm Q}(q^2) = f^{}_{\rm M}(q^2) = f^{}_{\rm E}(q^2) = 0 \; .
\end{equation}
This result means that a Majorana neutrino only has the anapole form
factor $f^{}_{\rm A}(q^2)$.
\begin{figure*}[t]
\centering
\includegraphics[bb = 235 655 380 673,scale=1]{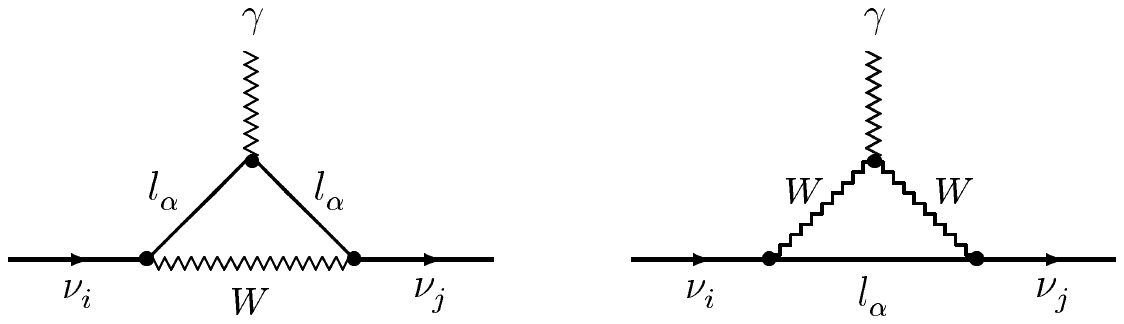}
\vspace{2.8cm} \caption{One-loop Feynman diagrams contributing to
the magnetic and electric dipole moments of massive Dirac neutrinos,
where $\alpha = e, \mu, \tau$ and $i, j = 1, 2, 3$.}
\end{figure*}

More generally, one may write out the matrix elements of the
electromagnetic current $J^{}_\mu(x)$ between two different states
(i.e., the incoming and outgoing particles are different):
\begin{eqnarray}
\langle \psi^{}_j(p^\prime)|J^{}_\mu (x)|\psi^{}_i(p)\rangle =
e^{-iqx} \overline{u^{}_j}(\vec{p}^\prime) \Gamma^{ij}_\mu (p,
p^\prime) u^{}_i(\vec{p}) \; ,
\end{eqnarray}
where $q = p - p^\prime$ together with $p^2 = m^2_i$ and
${p^\prime}^2 = m^2_j$ (for $i\neq j$). Here the electromagnetic
vertex matrix $\Gamma^{}_\mu(p, p^\prime)$ can be decomposed into
the following Lorentz-invariant form in terms of four form factors:
\begin{eqnarray}
\Gamma^{}_\mu(p, p^\prime) = F^{}_{\rm Q} (q^2) \left( q^2
\gamma^{}_\mu - q^{}_\mu \slashed{q} \right)  + F^{}_{\rm M} (q^2) i
\sigma^{}_{\mu\nu} q^\nu + F^{}_{\rm E} (q^2) \sigma^{}_{\mu\nu}
q^\nu \gamma^{}_5 + F^{}_{\rm A} (q^2) \left( q^2 \gamma^{}_\mu -
q^{}_\mu \slashed{q} \right) \gamma^{}_5 \; ,
\end{eqnarray}
where $F^{}_{\rm Q}$, $F^{}_{\rm M}$, $F^{}_{\rm E}$ and $F^{}_{\rm
A}$ are all the $2\times 2$ matrices in the space of neutrino mass
eigenstates. The diagonal case (i.e., $i=j$) has been discussed
above, from Eq. (59) to Eq. (69). In the off-diagonal case (i.e.,
$i\neq j$), the Hermiticity of $J^{}_\mu(x)$ is no more a constraint
on $\Gamma^{}_\mu(p, p^\prime)$ for Dirac neutrinos because Eq. (62)
only holds for $i=j$. It is now possible for Majorana neutrinos to
have finite {\it transition} dipole moments, simply because Eqs.
(66)---(69) do not hold when $\psi^{}_i$ and $\psi^{}_j$ represent
different flavors.

We conclude that Dirac neutrinos may have both electric and magnetic
dipole moments, while Majorana neutrinos have neither electric nor
magnetic dipole moments. But massive Majorana neutrinos can have
{\it transition} dipole moments which involve two different neutrino
flavors in the initial and final states, so can massive Dirac
neutrinos.

\subsection{Magnetic and electric dipole moments}

The magnetic and electric dipole moments of massive neutrinos,
denoted as $\mu \equiv F^{}_{\rm M}(0)$ and $\epsilon \equiv
F^{}_{\rm E}(0)$, are interesting in both theories and experiments
because they are closely related to the dynamics of neutrino mass
generation and to the characteristic of new physics.

Let us consider a minimal extension of the SM in which three
right-handed neutrinos are introduced and lepton number conservation
is required. In this case massive neutrinos are Dirac particles and
their magnetic and electric dipole moments can be evaluated by
calculating the Feynman diagrams in Fig. 1. Taking account of the
smallness of both $m^2_\alpha/M^2_W$ and $m^2_i/M^2_W$, where
$m^{}_\alpha$ (for $\alpha = e, \mu, \tau$) and $m^{}_i$ (for $i=1,
2, 3$) stand respectively for the charged-lepton and neutrino
masses, one obtains
\begin{eqnarray}
&& \mu^{\rm D}_{ij} = \frac{3e G^{}_{\rm F} m^{}_i}{32\sqrt{2}
\pi^2} \left( 1 + \frac{m^{}_j}{m^{}_i} \right) \times \sum_\alpha
\left( 2 - \frac{m^2_\alpha}{M^2_W} \right) V^{}_{\alpha i}
V^*_{\alpha j} \; ,
\nonumber \\
&& \epsilon^{\rm D}_{ij} = \frac{3e G^{}_{\rm F} m^{}_i}{32\sqrt{2}
\pi^2} \left( 1 - \frac{m^{}_j}{m^{}_i} \right) \times \sum_\alpha
\left( 2 - \frac{m^2_\alpha}{M^2_W} \right) V^{}_{\alpha i}
V^*_{\alpha j} \; ,
\end{eqnarray}
to an excellent degree of accuracy. Here $V^{}_{\alpha i}$ and
$V^{}_{\alpha j}$ are the elements of the unitary lepton flavor
mixing matrix $V$. Some discussions are in order.

(1) In the diagonal case (i.e., $i=j$), we are left with vanishing
electric dipole moments (i.e., $\epsilon^{\rm D}_{ii} =0$). The
magnetic dipole moments $\mu^{\rm D}_{ii}$ are finite and
proportional to the neutrino masses $m^{}_i$ (for $i=1,2,3$):
\begin{equation}
\mu^{\rm D}_{ii} = \frac{3e G^{}_{\rm F} m^{}_i}{8\sqrt{2} \pi^2}
\left( 1 - \frac{1}{2} \sum_\alpha \frac{m^2_\alpha}{M^2_W}
|V^{}_{\alpha i}|^2 \right) \; .
\end{equation}
Hence a massless Dirac neutrino in the SM has no magnetic dipole
moment. In the leading-order approximation, $\mu^{\rm D}_{ii}$ are
independent of the strength of lepton flavor mixing and have tiny
values
\begin{equation}
\mu^{\rm D}_{ii} \approx \frac{3e G^{}_{\rm F} m^{}_i}{8\sqrt{2}
\pi^2} \approx 3 \times 10^{-19} \left( \frac{m^{}_i}{1 ~ {\rm eV}}
\right) \mu^{}_{\rm B} \; ,
\end{equation}
where $\mu^{}_{\rm B} = e\hbar/(2 m^{}_e)$ is the Bohr magneton.
Given $m^{}_i \leq 1$ eV, the magnitude of $\mu^{\rm D}_{ii}$ is far
below its present experimental upper bound ($< {\rm a ~ few} \times
10^{-11} \mu^{}_{\rm B}$).

(2) In the off-diagonal case (i.e., $i\neq j$), the unitarity of $V$
allows us to simplify Eq. (72) to
\begin{eqnarray}
&& \mu^{\rm D}_{ij} = -\frac{3e G^{}_{\rm F} m^{}_i}{32\sqrt{2}
\pi^2} \left( 1 + \frac{m^{}_j}{m^{}_i} \right) \sum_\alpha
\frac{m^2_\alpha}{M^2_W} V^{}_{\alpha i} V^*_{\alpha j} \;
, \nonumber \\
&& \epsilon^{\rm D}_{ij} = -\frac{3e G^{}_{\rm F} m^{}_i}{32\sqrt{2}
\pi^2} \left( 1 - \frac{m^{}_j}{m^{}_i} \right) \sum_\alpha
\frac{m^2_\alpha}{M^2_W} V^{}_{\alpha i} V^*_{\alpha j} \; .
\end{eqnarray}
We see that the magnitudes of $\mu^{\rm D}_{ij}$ and $\epsilon^{\rm
D}_{ij}$ (for $i\neq j$), compared with that of $\mu^{\rm D}_{ii}$,
are further suppressed due to the smallness of $m^2_\alpha/M^2_W$.
Similar to the expression given in Eq. (74),
\begin{eqnarray}
&& \mu^{\rm D}_{ij} \approx -4 \times 10^{-23} \left( \frac{m^{}_i +
m^{}_j}{1 ~ {\rm eV}} \right) \times \left( \sum_\alpha
\frac{m^2_\alpha}{m^2_\tau} V^{}_{\alpha i} V^*_{\alpha j} \right)
\mu^{}_{\rm B} \; ,
\nonumber \\
&& \epsilon^{\rm D}_{ij} \approx -4 \times 10^{-23} \left(
\frac{m^{}_i - m^{}_j}{1 ~ {\rm eV}} \right) \times \left(
\sum_\alpha \frac{m^2_\alpha}{m^2_\tau} V^{}_{\alpha i} V^*_{\alpha
j} \right) \mu^{}_{\rm B} \; ,
\end{eqnarray}
which can illustrate how small $\mu^{\rm D}_{ij}$ and $\epsilon^{\rm
D}_{ij}$ are.

(3) Although Majorana neutrinos do not have intrinsic ($i=j$)
magnetic and electric dipole moments, they may have finite
transition ($i\neq j$) dipole moments. Because of the fact that
Majorana neutrinos are their own antiparticles, their magnetic and
electric dipole moments can also get contributions from two
additional one-loop Feynman diagrams involving the charge-conjugate
fields of $\nu^{}_i$, $\nu^{}_j$, $l^{}_\alpha$, $W^\pm$ and
$\gamma$ shown in Fig. 1
\footnote{Here we confine ourselves to a simple extension of the SM
with three known neutrinos to be massive Majorana particles.}.
In this case one obtains
\begin{eqnarray}
&& \mu^{\rm M}_{ij} = -\frac{3e G^{}_{\rm F} i}{16\sqrt{2} \pi^2}
\left( m^{}_i + m^{}_j \right) \times \sum_\alpha
\frac{m^2_\alpha}{M^2_W} {\rm Im} \left( V^{}_{\alpha i} V^*_{\alpha
j} \right) \; ,
\nonumber \\
&& \epsilon^{\rm M}_{ij} = -\frac{3e G^{}_{\rm F}}{16\sqrt{2} \pi^2}
\left( m^{}_i - m^{}_j \right) \times \sum_\alpha
\frac{m^2_\alpha}{M^2_W} {\rm Re} \left( V^{}_{\alpha i} V^*_{\alpha
j} \right) \; ,
\end{eqnarray}
where $m^{}_i \neq m^{}_j$ must hold. Comparing between Eqs. (75)
and (77), we observe that the magnitudes of $\mu^{\rm M}_{ij}$ and
$\epsilon^{\rm M}_{ij}$ are the same order as those of $\mu^{\rm
D}_{ij}$ and $\epsilon^{\rm D}_{ij}$ in most cases, although the
CP-violating phases hidden in $V^{}_{\alpha i} V^*_{\alpha j}$ are
possible to give rise to significant cancellations in some cases.

(4) The fact that $\mu^{}_{ij}$ and $\epsilon^{}_{ij}$ are
proportional to $m^{}_i$ or $m^{}_j$ can be understood in the
following way. Note that both tensor- and pseudotensor-like spinor
bilinears are chirality-changing operators, which link the
left-handed state to the right-handed one
\footnote{That is why both magnetic and electric dipole moments must
vanish for a Weyl neutrino, because it is massless and does not
possess the right-handed component.}:
\begin{eqnarray}
\overline{\psi} \sigma^{}_{\mu\nu} \psi = \overline{\psi^{}_{\rm L}}
\sigma^{}_{\mu\nu} \psi^{}_{\rm R} + {\rm h.c.} \; , ~~~~~
\nonumber \\
\overline{\psi} \sigma^{}_{\mu\nu} \gamma^{}_5 \psi =
\overline{\psi^{}_{\rm L}} \sigma^{}_{\mu\nu} \gamma^{}_5
\psi^{}_{\rm R} - {\rm h.c.} \; .
\end{eqnarray}
Note also that the same relations hold when $\psi$ is replaced by
its charge-conjugate field $\psi^c$ for Majorana neutrinos. Because
$(\nu^{}_i)^{}_{\rm R}$ and $(\nu^{}_j)^{}_{\rm R}$ do not have any
interactions with $W^\pm$ in Fig. 1, it seems that only
$(\nu^{}_i)^{}_{\rm L}$ and $(\nu^{}_j)^{}_{\rm L}$ are flowing
along the external fermion lines. To obtain a chirality-changing
contribution from the effective (one-loop) electromagnetic vertex,
one has to put a mass insertion on one of the external legs in the
Feynman diagrams. As a result, the magnetic and electric dipole
moments must involve $m^{}_i$ and $m^{}_j$, the masses of $\nu^{}_i$
and $\nu^{}_j$ neutrinos.

(5) Is the magnetic or electric dipole moment of a neutrino always
proportional to its mass? The answer is negative if new physics
beyond the $SU(2)^{}_{\rm L} \times U(1)^{}_{\rm Y}$ gauge theory is
involved. For instance, a new term proportional to the
charged-lepton mass can contribute to the magnetic dipole moment of
a massive Dirac neutrino in the $SU(2)^{}_{\rm L} \times
SU(2)^{}_{\rm R} \times U(1)^{}_{\rm Y}$ model with broken
left-right symmetry. Depending on the details of this model, such a
term might cancel or exceed the term proportional to the neutrino
mass in the expression of the magnetic dipole moment.

Finite magnetic and electric dipole moments of massive neutrinos may
produce a variety of new processes beyond the SM. For example, (a)
radiative neutrino decays $\nu^{}_i \to \nu^{}_j + \gamma$ can
happen, so can the Cherenkov radiation of neutrinos in an external
electromagnetic field; (b) the elastic neutrino-electron or
neutrino-nucleon scattering can be mediated by the magnetic and
electric dipole moments; (c) the phenomenon of precession of the
neutrino spin can occur in an external magnetic field; (d) the
photon (or plasmon) can decay into a neutrino-antineutrino pair in a
plasma (i.e., $\gamma^* \to \nu \overline{\nu}$). Of course,
non-vanishing electromagnetic dipole moments contribute to neutrino
masses too.

\subsection{Radiative neutrino decays}

If the electromagnetic moments of a massive neutrino $\nu^{}_i$ are
finite, it can decay into a lighter neutrino $\nu^{}_j$ and a photon
$\gamma$. The Lorentz-invariant vertex matrix of this $\nu^{}_i \to
\nu^{}_j + \gamma$ process is in general described by $\Gamma^{}_\mu
(p, p^\prime)$ in Eq. (71). Because $q^2 =0$ and $q^{}_\mu
\varepsilon^\mu =0$ hold for a real photon $\gamma$, where
$\varepsilon^\mu$ represents the photon polarization, the form of
$\Gamma^{}_\mu (p, p^\prime)$ can be simplified to
\begin{equation}
\Gamma^{}_\mu(p, p^\prime) = \left[ i F^{}_{\rm M} (0) + F^{}_{\rm
E} (0) \gamma^{}_5 \right] \sigma^{}_{\mu\nu} q^\nu \; .
\end{equation}
By definition, $F^{ij}_{\rm M}(0) \equiv \mu^{}_{ij}$ and
$F^{ij}_{\rm E}(0) \equiv \epsilon^{}_{ij}$ are just the magnetic
and electric transition dipole moments between $\nu^{}_i$ and
$\nu^{}_j$ neutrinos. Given the transition matrix element
$\overline{u^{}_j} (\vec{p}^\prime) \Gamma^{ij}_\mu (p, p^\prime)
u^{}_i (\vec{p})$, it is straightforward to calculate the decay
rate. In the rest frame of the decaying neutrino $\nu^{}_i$,
\begin{equation}
\Gamma^{}_{\nu^{}_i \to \nu^{}_j + \gamma} = \frac{\left( m^2_i -
m^2_j\right)^3}{8\pi m^3_i} \left( \left|\mu^{}_{ij}\right|^2 +
\left|\epsilon^{}_{ij}\right|^2 \right) \; .
\end{equation}
This result is valid for both Dirac and Majorana neutrinos.

In the $SU(2)^{}_{\rm L} \times U(1)^{}_{\rm Y}$ gauge theory with
three massive Dirac (or Majorana) neutrinos, the radiative decay
$\nu^{}_i \to \nu^{}_j + \gamma$ is mediated by the one-loop Feynman
diagrams (and their charge-conjugate diagrams) shown in Fig. 1. The
explicit expressions of $\mu^{}_{ij}$ and $\epsilon^{}_{ij}$ have
been given in Eq. (75) for Dirac neutrinos and in Eq. (77) for
Majorana neutrinos. Hence
\begin{eqnarray}
\Gamma^{\rm (D)}_{\nu^{}_i \to \nu^{}_j + \gamma} = \frac{\left(
m^2_i - m^2_j\right)^3}{8\pi m^3_i} \left( \left|\mu^{\rm
D}_{ij}\right|^2 + \left|\epsilon^{\rm D}_{ij}\right|^2 \right) =
\frac{9 \alpha G^2_{\rm F} m^5_i}{2^{11} \pi^4} \left( 1 -
\frac{m^2_j}{m^2_i} \right)^3 \left( 1 + \frac{m^2_j}{m^2_i} \right)
\nonumber \\
\times \left| \sum_\alpha \frac{m^2_\alpha}{M^2_W} V^{}_{\alpha i}
V^*_{\alpha j} \right|^2 \; , ~~~~~~~~~~~~~~~~~~~~~
\end{eqnarray}
for Dirac neutrinos; or
\begin{eqnarray}
\Gamma^{\rm (M)}_{\nu^{}_i \to \nu^{}_j + \gamma} = \frac{\left(
m^2_i - m^2_j\right)^3}{8\pi m^3_i} \left( \left|\mu^{\rm
M}_{ij}\right|^2 + \left|\epsilon^{\rm M}_{ij}\right|^2 \right) =
\frac{9 \alpha G^2_{\rm F} m^5_i}{2^{10} \pi^4} \left( 1 -
\frac{m^2_j}{m^2_i} \right)^3 \left\{ \left( 1 +
\frac{m^{}_j}{m^{}_i} \right)^2 \right .
\nonumber \\
\left . \times \left[ \sum_\alpha \frac{m^2_\alpha}{M^2_W} {\rm Im}
\left( V^{}_{\alpha i} V^*_{\alpha j} \right) \right]^2 + \left( 1 -
\frac{m^{}_j}{m^{}_i} \right)^2 \left[ \sum_\alpha
\frac{m^2_\alpha}{M^2_W} {\rm Re} \left( V^{}_{\alpha i} V^*_{\alpha
j} \right) \right]^2 \right \} \; ,
\end{eqnarray}
for Majorana neutrinos, where $\alpha = e^2/(4\pi)$ denotes the
electromagnetic fine-structure constant.

To compare $\Gamma^{}_{\nu^{}_i \to \nu^{}_j + \gamma}$ with the
experimental data in a simpler way, one may define an effective
magnetic dipole moment
\begin{equation}
\mu^{}_{\rm eff} \equiv \sqrt{\left|\mu^{}_{ij} \right|^2 +
\left|\epsilon^{}_{ij} \right|^2} \;\; .
\end{equation}
Eq. (80) can then be expressed as
\begin{eqnarray}
\Gamma^{}_{\nu^{}_i \to \nu^{}_j + \gamma} = 5.3 \times \left( 1 -
\frac{m^2_j}{m^2_i} \right)^3 \left( \frac{m^{}_i}{1 ~ {\rm eV}}
\right)^3 \times \left( \frac{\mu^{}_{\rm eff}}{\mu^{}_{\rm
B}}\right)^2 {\rm s}^{-1} \; .
\end{eqnarray}
Although $\mu^{}_{\rm eff}$ is extremely small in some simple
extensions of the SM, it could be sufficiently large in some more
complicated or exotic scenarios beyond the SM, such as a class of
extra-dimension models. Experimentally, radiative decays of massive
neutrinos can be constrained by seeing no emission of the photons
from solar $\nu^{}_e$ and reactor $\overline{\nu}^{}_e$ fluxes. Much
stronger constraints on $\mu^{}_{\rm eff}$ can be obtained from the
Supernova 1987A limit on the neutrino decay and from the
astrophysical limit on distortions of the cosmic microwave
background (CMB) radiation. A brief summary of these limits is
$$
\frac{\mu^{}_{\rm eff}}{\mu^{}_{\rm B}} < \left\{
\begin{array}{lll} 0.9 \times 10^{-1} \left( \displaystyle\frac{{\rm
eV}}{m^{}_\nu} \right)^2 && {\rm Reactor} \\
0.5 \times 10^{-5} \left( \displaystyle\frac{{\rm eV}}{m^{}_\nu}
\right)^2 && {\rm Sun} \\
1.5 \times 10^{-8} \left( \displaystyle\frac{{\rm eV}}{m^{}_\nu}
\right)^2 && {\rm SN ~ 1987A} \\
1.0 \times 10^{-11} \left( \displaystyle\frac{{\rm eV}}{m^{}_\nu}
\right)^{9/4} && {\rm CMB}
\end{array}
\right .
$$
where $m^{}_\nu$ denotes the effective mass of the decaying neutrino
(i.e., $m^{}_\nu = m^{}_i$).

\subsection{Electromagnetic $\nu^{}_e$-$e$ scattering}

In practice, the most sensitive way of probing the electromagnetic
dipole moments of a massive neutrino is to measure the cross section
of elastic neutrino-electron (or antineutrino-electron) scattering,
which can be expressed as a sum of the contribution from the SM
($\sigma^{}_0$) and that from the electromagnetic dipole moments of
massive neutrinos ($\sigma^{}_\mu$):
\begin{equation}
\frac{{\rm d} \sigma}{{\rm d} T} = \frac{{\rm d} \sigma^{}_0}{{\rm
d} T} + \frac{{\rm d} \sigma^{}_\mu}{{\rm d} T} \; ,
\end{equation}
where $T = E^{}_e - m^{}_e$ denotes the kinetic energy of the recoil
electron in this process. We have
\begin{eqnarray}
\frac{{\rm d} \sigma^{}_0}{{\rm d} T} = \frac{G^2_{\rm F}
m^{}_e}{2\pi} \left[ g^2_+ + g^2_- \left( 1 - \frac{T}{E^{}_\nu}
\right)^2 - g^{}_+ g^{}_- \frac{m^{}_e T}{E^2_\nu} \right]
\end{eqnarray}
for neutrino-electron scattering, where $g^{}_+ =
2\sin^2\theta^{}_{\rm w} + 1$ for $\nu^{}_e$, $g^{}_+ =
2\sin^2\theta^{}_{\rm w} -1$ for $\nu^{}_\mu$ and $\nu^{}_\tau$, and
$g^{}_- = 2\sin^2\theta^{}_{\rm w}$ for all flavors. Note that Eq.
(86) is also valid for antineutrino-electron scattering if one
simply exchanges the positions of $g^{}_+$ and $g^{}_-$. On the
other hand,
\begin{equation}
\frac{{\rm d} \sigma^{}_\mu}{{\rm d} T} = \frac{\alpha^2 \pi}{m^2_e}
\left( \frac{1}{T} - \frac{1}{E^{}_\nu} \right) \left(
\frac{\mu^{}_\nu}{\mu^{}_{\rm B}} \right)^2 \;
\end{equation}
with $\mu^2_\nu \equiv |\mu^{\rm D}_{ii}|^2 + |\epsilon^{\rm
D}_{ii}|^2$ (for $i=1$, $2$ or $3$), which holds for both neutrinos
and antineutrinos. In obtaining Eqs. (86) and (87) one has assumed
the scattered neutrino to be a Dirac particle and omitted the
effects of finite neutrino masses and flavor mixing (i.e., $\nu^{}_e
= \nu^{}_1$, $\nu^{}_\mu = \nu^{}_2$ and $\nu^{}_\tau = \nu^{}_3$
have been taken). Hence there is no interference between the
contributions coming from the SM and electromagnetic dipole moments
--- the latter leads to a helicity flip of the neutrino but the
former is always helicity-conserving. While an interference term
will appear if one takes account of neutrino masses and flavor
mixing, its magnitude linearly depends on the neutrino masses and
thus is strongly suppressed in comparison with the pure weak and
electromagnetic terms. So the incoherent sum of ${\rm
d}\sigma^{}_0/{\rm d}T$ and ${\rm d}\sigma^{}_\mu/{\rm d}T$ in Eq.
(85) is actually an excellent approximation of ${\rm d}\sigma/{\rm
d}T$.

It is obvious that the two terms of ${\rm d}\sigma/{\rm d}T$ depend
on the kinetic energy of the recoil electron in quite different
ways. In particular, ${\rm d}\sigma^{}_\mu/{\rm d}T$ grows rapidly
with decreasing values of $T$. Hence a measurement of smaller $T$
can probe smaller $\mu^{}_\nu$ in this kind of experiments. The
magnitude of ${\rm d}\sigma^{}_\mu/{\rm d}T$ becomes larger than
that of ${\rm d}\sigma^{}_0/{\rm d}T$ if the condition
\begin{equation}
T \leq \frac{\alpha^2 \pi^2}{G^2_{\rm F} m^3_e} \left(
\frac{\mu^{}_\nu}{\mu^{}_{\rm B}} \right)^2 \approx 3 \times 10^{22}
\left( \frac{\mu^{}_\nu}{\mu^{}_{\rm B}} \right)^2 {\rm keV} \;
\end{equation}
is roughly satisfied, as one can easily see from Eqs. (86) and (87).
No distortion of the recoil electron energy spectrum of
$\nu^{}_\alpha e^-$ or $\overline{\nu}^{}_\alpha e^-$ scattering
(for $\alpha = e, \mu, \tau$) has so far been observed in any direct
laboratory experiments, and thus only the upper bounds on
$\mu^{}_\nu$ can be derived. For instance, an analysis of the
$T$-spectrum in the Super-Kamiokande experiment yields $\mu^{}_\nu <
1.1 \times 10^{-10} \mu^{}_{\rm B}$. More stringent bounds on
$\mu^{}_\nu$ can hopefully be achieved in the future.

In view of current experimental data on neutrino oscillations, we
know that neutrinos are actually massive. Hence the effects of
finite neutrino masses and flavor mixing should be taken into
account in calculating the cross section of elastic
neutrino-electron or antineutrino-electron scattering. Here let us
illustrate how the neutrino oscillation may affect the weak and
electromagnetic terms of elastic $\overline{\nu}^{}_e e^-$
scattering in a reactor experiment, where the antineutrinos are
produced from the beta decay of fission products and detected by
their elastic scattering with electrons in a detector. The
antineutrino state created in this beta decay (via $W^- \to e^- +
\overline{\nu}^{}_e$) at the reactor is a superposition of three
antineutrino mass eigenstates:
\begin{equation}
|\overline{\nu}^{}_e (0)\rangle = \sum^3_{j=1} V^{}_{e j}
|\overline{\nu}^{}_j \rangle \; .
\end{equation}
Such a $\overline{\nu}^{}_e$ beam propagates over the distance $L$
to the detector,
\begin{equation}
|\overline{\nu}^{}_e (L)\rangle = \sum^3_{j=1} e^{i q^{}_j L}
V^{}_{e j} |\overline{\nu}^{}_j \rangle \; ,
\end{equation}
in which $q^{}_j = \sqrt{E^2_\nu - m^2_j}~$ is the momentum of
$\nu^{}_j$ with $E^{}_\nu$ being the beam energy and $m^{}_j$ being
the mass of $\nu^{}_j$. After taking account of the effect of
neutrino oscillations, one obtains the differential cross section of
elastic antineutrino-electron scattering as follows:
\begin{equation}
\frac{{\rm d} \sigma^\prime}{{\rm d} T} = \frac{{\rm d}
\sigma^\prime_0}{{\rm d} T} + \frac{{\rm d} \sigma^\prime_\mu}{{\rm
d} T} \; ,
\end{equation}
where
\begin{eqnarray}
\frac{{\rm d} \sigma^\prime_0}{{\rm d} T} = \frac{G^2_{\rm F}
m^{}_e}{2\pi} \left\{ g^2_- + \left (g^{}_- - 1 \right)^2 \left( 1 -
\frac{T}{E^{}_\nu} \right)^2 - g^{}_- \left( g^{}_- - 1 \right)
\frac{m^{}_e T}{E^2_\nu} \right .
\nonumber \\
\left . + 2 g^{}_- \left| \sum^3_{j=1} e^{iq^{}_j L} |V^{}_{e j}|^2
\right|^2 \left[ 2 \left( 1 - \frac{T}{E^{}_\nu} \right)^2 -
\frac{m^{}_e T}{E^2_\nu} \right] \right\} ~
\end{eqnarray}
with $g^{}_- = 2 \sin^2\theta^{}_{\rm w}$ for $\overline{\nu}^{}_e$,
and
\begin{eqnarray}
\frac{{\rm d} \sigma^\prime_\mu}{{\rm d} T} = \frac{\alpha^2
\pi}{m^2_e} \sum^3_{k=1} \left| \sum^3_{j=1} e^{iq^{}_j L} V^{}_{e
j} \frac{\epsilon^{}_{jk} + i \mu^{}_{jk}}{\mu^{}_{\rm B}} \right|^2
\times \left( \frac{1}{T} - \frac{1}{E^{}_\nu} \right) \;
\end{eqnarray}
with $\mu^{}_{jk}$ and $\epsilon^{}_{jk}$ being the magnetic and
electric transition dipole moments between $\nu^{}_j$ and $\nu^{}_k$
neutrinos as defined in Eq. (79). Because different neutrino mass
eigenstates are in principle distinguishable in the electromagnetic
$\overline{\nu}^{}_e e^-$ scattering, their contributions to the
total cross section are incoherent. Eq. (93) shows that it is in
general difficult to determine or constrain the magnitudes of
$\mu^{}_{jk}$ and $\epsilon^{}_{jk}$ (for $j, k = 1, 2, 3$) from a
single measurement.

\section{Lepton Flavor Mixing and CP Violation}

Regardless of the dynamical origin of tiny neutrino masses
\footnote{For simplicity, here we do not consider possible
non-unitarity of the $3\times 3$ neutrino mixing matrix because its
effects are either absent or very small.},
we may discuss lepton flavor mixing by taking account of the
effective mass terms of charged leptons and Majorana neutrinos at
low energies
\footnote{As for Dirac neutrinos, the corresponding mass term is the
same as that given in Eq. (7). In this case the neutrino mass matrix
$M^{}_\nu$ is in general not symmetric and can be diagonalized by
means of the transformation $V^\dagger_\nu M^{}_\nu U^{}_\nu = {\rm
Diag}\{m^{}_1, m^{}_2, m^{}_3 \}$, where both $V^{}_\nu$ and
$U^{}_\nu$ are unitary.},
\begin{eqnarray}
-{\cal L}^\prime_{\rm lepton} = \overline{\left(e~ \mu~
\tau\right)^{}_{\rm L}} ~ M^{}_l \left( \begin{matrix} e \cr \mu \cr
\tau \end{matrix} \right)^{}_{\rm R} + \frac{1}{2}
\overline{\left(\nu^{}_e ~ \nu^{}_\mu ~ \nu^{}_\tau \right)^{}_{\rm
L}} ~ M^{}_\nu \left( \begin{matrix} \nu^c_e \cr \nu^c_\mu \cr
\nu^c_\tau \end{matrix} \right)^{}_{\rm R} + {\rm h.c.} \; .
\end{eqnarray}
The phenomenon of lepton flavor mixing arises from a mismatch
between the diagonalizations of $M^{}_l$ and $M^{}_\nu$ in an
arbitrary flavor basis: $V^\dagger_l M^{}_l U^{}_l = {\rm
Diag}\{m^{}_e, \mu^{}_\mu, m^{}_\tau \}$ and $V^\dagger_\nu M^{}_\nu
V^*_\nu = {\rm Diag}\{ m^{}_1, m^{}_2, m^{}_3 \}$, where $V^{}_l$,
$U^{}_l$ and $V^{}_\nu$ are the $3\times 3$ unitary matrices. In the
basis of mass eigenstates, it is the unitary matrix $V = V^\dagger_l
V^{}_\nu$ that will appear in the weak charged-current interactions
in Eq. (12). Although the basis of $M^{}_l = {\rm Diag}\{ m^{}_e,
m^{}_\mu, m^{}_\tau \}$ with $V^{}_l = {\bf 1}$ and $V = V^{}_\nu$
is often chosen in neutrino phenomenology, one should keep in mind
that both the charged-lepton and neutrino sectors may in general
contribute to lepton flavor mixing. In other words, both $V^{}_l$
and $V^{}_\nu$ are not fully physical, and only their product $V =
V^\dagger_l V^{}_\nu$ is a physical description of lepton flavor
mixing and CP violation at low energies.

\subsection{Parametrizations of $V$}

Flavor mixing among $n$ different lepton families can be described
by an $n\times n$ unitary matrix $V$, whose number of independent
parameters relies on the nature of neutrinos. If neutrinos are Dirac
particles, one may make use of $n (n-1)/2$ rotation angles and
$(n-1)(n-2)/2$ phase angles to parametrize $V$. If neutrinos are
Majorana particles, however, a full parametrization of $V$ needs $n
(n-1)/2$ rotation angles and the same number of phase angles
\footnote{No matter whether neutrinos are Dirac or Majorana
particles, the $n\times n$ unitary flavor mixing matrix has
$(n-1)^2(n-2)^2/4$ Jarlskog invariants of CP violation defined as
${\cal J}^{ij}_{\alpha\beta} \equiv {\rm Im} \left(V^{}_{\alpha i}
V^{}_{\beta j} V^*_{\alpha j} V^*_{\beta i} \right)$.}.
The flavor mixing between charged leptons and Dirac neutrinos is
completely analogous to that of quarks, for which a number of
different parametrizations have been proposed and classified in the
literature. Here we classify all possible parametrizations for the
flavor mixing between charged leptons and Majorana neutrinos with
$n=3$. Regardless of the freedom of phase reassignments, we find
that there are nine structurally different parametrizations for the
$3\times 3$ lepton flavor mixing matrix $V$.

The $3\times 3$ lepton flavor mixing matrix $V$, which is often
called the Pontecorvo-Maki-Nakagawa-Sakata (PMNS) matrix, can be
expressed as a product of three unitary matrices $O^{}_1$, $O^{}_2$
and $O^{}_3$. They correspond to simple rotations in the complex
(1,2), (2,3) and (3,1) planes:
\begin{eqnarray}
O^{}_1 = \left ( \begin{matrix} c^{}_1 e^{i\alpha^{}_1} & s^{}_1
e^{-i\beta^{}_1} & 0 \cr -s^{}_1 e^{i\beta^{}_1} & c^{}_1
e^{-i\alpha^{}_1} & 0 \cr 0 & 0 & e^{i\gamma^{}_1} \cr \end{matrix}
\right ) \; ,
\nonumber \\
O^{}_2 = \left ( \begin{matrix} e^{i\gamma^{}_2} & 0 & 0 \cr 0 &
c^{}_2 e^{i\alpha^{}_2} & s^{}_2 e^{-i\beta^{}_2} \cr 0 & -s^{}_2
e^{i\beta^{}_2} & c^{}_2 e^{-i\alpha^{}_2} \cr \end{matrix} \right )
\; ,
\nonumber \\
O^{}_3 = \left ( \begin{matrix} c^{}_3 e^{i\alpha^{}_3} & 0 & s^{}_3
e^{-i\beta^{}_3} \cr 0 & e^{i\gamma^{}_3} & 0 \cr -s^{}_3
e^{i\beta^{}_3} & 0 & c^{}_3 e^{-i\alpha^{}_3} \cr \end{matrix}
\right ) \; ,
\end{eqnarray}
where $s^{}_i \equiv \sin\theta^{}_i$ and $c^{}_i \equiv
\cos\theta^{}_i$ (for $i = 1, 2, 3$). Obviously $O^{}_i O^\dagger_i
= O^\dagger_i O^{}_i = {\bf 1}$ holds, and any two rotation matrices
do not commute with each other. We find twelve different ways to
arrange the product of $O^{}_1$, $O^{}_2$ and $O^{}_3$, which can
cover the whole $3\times 3$ space and provide a full description of
$V$. Explicitly, six of the twelve different combinations of
$O^{}_i$ belong to the type
\begin{eqnarray}
V = O^{}_i(\theta^{}_i, \alpha^{}_i, \beta^{}_i, \gamma^{}_i)
\otimes O^{}_j(\theta^{}_j, \alpha^{}_j, \beta^{}_j, \gamma^{}_j)
\otimes O^{}_i(\theta^\prime_i, \alpha^\prime_i, \beta^\prime_i,
\gamma^\prime_i)
\end{eqnarray}
with $i\neq j$, where the complex rotation matrix $O^{}_i$ occurs
twice; and the other six belong to the type
\begin{eqnarray}
V  =  O^{}_i(\theta^{}_i, \alpha^{}_i, \beta^{}_i, \gamma^{}_i)
\otimes O^{}_j(\theta^{}_j, \alpha^{}_j, \beta^{}_j, \gamma^{}_j)
\otimes O^{}_k(\theta^{}_k, \alpha^{}_k, \beta^{}_k, \gamma^{}_k)
\end{eqnarray}
with $i\neq j\neq k$, in which the rotations take place in three
different complex planes. The products $O^{}_i O^{}_j O^{}_i$ and
$O^{}_i O^{}_k O^{}_i$ (for $i\neq k$) in Eq. (97) are correlated
with each other, if the relevant phase parameters are switched off.
Hence only nine of the twelve parametrizations, three from Eq. (96)
and six from Eq. (97), are structurally different.

In each parametrization of $V$, there apparently exist nine phase
parameters. Some of them or their combinations can be absorbed by
redefining the relevant phases of charged-lepton and neutrino
fields. If neutrinos are Dirac particles, $V$ contains only a single
irremovable CP-violating phase $\delta$. If neutrinos are Majorana
particles, however, there is no freedom to rearrange the relative
phases of three Majorana neutrino fields. Hence $V$ may in general
contain three irremovable CP-violating phases in the Majorana case
($\delta$ and two Majorana phases). Both CP- and T-violating effects
in neutrino oscillations depend only upon the Dirac-like phase
$\delta$.

Different parametrizations of $V$ are mathematically equivalent, so
adopting any of them does not directly point to physical
significance. But it is very likely that one particular
parametrization is more useful and transparent than the others in
studying the neutrino phenomenology and (or) exploring the
underlying dynamics responsible for lepton mass generation and CP
violation. Here we highlight two particular parametrizations of the
PMNS matrix $V$. The first one is the so-called ``standard"
parametrization advocated by the Particle Data Group:
\begin{eqnarray}
V = \left ( \begin{matrix} 1 & 0 & 0 \cr 0 & c^{}_{23} & s^{}_{23}
\cr 0 & -s^{}_{23} & c^{}_{23} \cr \end{matrix} \right ) \left (
\begin{matrix} c^{}_{13} & 0 & s^{}_{13} e^{-i\delta} \cr 0 & 1 & 0
\cr -s^{}_{13} e^{i\delta} & 0 & c^{}_{13} \cr
\end{matrix} \right ) \left ( \begin{matrix} c^{}_{12} & s^{}_{12} & 0 \cr
-s^{}_{12} & c^{}_{12} & 0 \cr 0 & 0 & 1 \cr \end{matrix} \right )
P^\prime \; ,
\end{eqnarray}
where $c^{}_{ij} \equiv \cos\theta^{}_{ij}$ and $s^{}_{ij} \equiv
\sin\theta^{}_{ij}$ (for $ij = 12, 13, 23$) together with the
Majorana phase matrix $P^\prime = {\rm Diag} \{e^{i\rho},
e^{i\sigma}, 1\}$. Without loss of generality, the three mixing
angles ($\theta^{}_{12}, \theta^{}_{13}, \theta^{}_{23}$) can all be
arranged to lie in the first quadrant. Arbitrary values between $0$
and $2\pi$ are allowed for three CP-violating phases ($\delta, \rho,
\sigma$). A remarkable merit of this parametrization is that its
three mixing angles are approximately equivalent to the mixing
angles of solar ($\theta^{}_{12}$), atmospheric ($\theta^{}_{23}$)
and CHOOZ reactor ($\theta^{}_{13}$) neutrino oscillation
experiments. Another useful parametrization is the Fritzsch-Xing
(FX) parametrization proposed originally for quark mixing and later
for lepton mixing:
\begin{eqnarray}
V = \left ( \begin{matrix} c^{}_l & s^{}_l   & 0 \cr -s^{}_l &
c^{}_l & 0 \cr 0   & 0 & 1 \cr \end{matrix} \right ) \left (
\begin{matrix} e^{-i\phi}  & 0 & 0 \cr 0   & c & s \cr 0   & -s & c
\cr \end{matrix} \right ) \left (
\begin{matrix} c^{}_{\nu} & -s^{}_{\nu} & 0 \cr s^{}_{\nu} &
c^{}_{\nu} & 0 \cr 0   & 0 & 1 \cr \end{matrix} \right ) P^\prime \;
,
\end{eqnarray}
where $c^{}_{l,\nu} \equiv \cos\theta^{}_{l,\nu}$, $s^{~}_{l,\nu}
\equiv \sin\theta_{l,\nu}$, $c \equiv \cos\theta$, $s \equiv
\sin\theta$, and $P^\prime$ is a diagonal phase matrix containing
two nontrivial CP-violating phases. Although the form of $V$ in Eq.
(99) is apparently different from that in Eq. (98), their
corresponding flavor mixing angles ($\theta^{}_l, \theta^{}_\nu,
\theta$) and ($\theta^{}_{12}, \theta^{}_{13}, \theta^{}_{23}$) have
quite similar meanings in interpreting the experimental data on
neutrino oscillations. In the limit $\theta^{}_l = \theta^{}_{13} =
0$, one easily arrives at $\theta^{}_\nu = \theta^{}_{12}$ and
$\theta = \theta^{}_{23}$. As a natural consequence of very small
$\theta^{}_l$, three mixing angles of the FX parametrization can
also be related to those of solar ($\theta^{}_\nu$), atmospheric
($\theta$) and CHOOZ reactor ($\theta^{}_l\sin\theta$) neutrino
oscillation experiments in the leading-order approximation. A
striking merit of this parametrization is that its six parameters
have very simple renormalization-group equations when they run from
a superhigh-energy scale to the electroweak scale or vice versa.

\subsection{Democratic or tri-bimaximal mixing?}

Current neutrino oscillation data indicate the essential feature of
lepton flavor mixing: two mixing angles are quite large
($\theta^{}_{12} \sim 34^\circ$ and $\theta^{}_{23} \sim 45^\circ$)
while the third one is very small ($\theta^{}_{13} < 10^\circ$).
Such a flavor mixing pattern is far beyond the original imagination
of most people because it is rather different from the well-known
quark mixing pattern ($\vartheta^{}_{12} \approx 14.5^\circ$,
$\vartheta^{}_{23} \approx 2.6^\circ$, $\vartheta^{}_{13} \approx
0.23^\circ$ and $\delta =76.5^\circ$) described by the same
parametrization of the Cabibbo-Kobayashi-Maskawa (CKM) matrix. To
understand this difference, a number of constant lepton mixing
patterns have been proposed as the starting point of model building.
Possible flavor symmetries and their spontaneous or explicit
breaking mechanisms hidden in those constant patterns might finally
help us pin down the dynamics responsible for lepton mass generation
and flavor mixing. To illustrate, let us first comment on the
``democratic" neutrino mixing pattern and then pay more attention to
the ``tri-bimaximal" neutrino mixing pattern.

The ``democratic" lepton flavor mixing pattern
\begin{equation}
U^{}_0 = \left( \begin{matrix} \frac{1}{\sqrt{2}} &
\frac{1}{\sqrt{2}} & 0 \cr \frac{-1}{\sqrt{6}} & \frac{1}{\sqrt{6}}
& \frac{\sqrt{2}}{\sqrt{3}} \cr \frac{1}{\sqrt{3}} &
~\frac{-1}{\sqrt{3}} ~ & \frac{1}{\sqrt{3}} \cr \end{matrix} \right)
\end{equation}
was originally obtained by Fritzsch and Xing as the leading term of
the $3\times 3$ lepton mixing matrix from the breaking of flavor
democracy or $S(3)^{}_{\rm L} \times S(3)^{}_{\rm R}$ symmetry of
the charged-lepton mass matrix in the basis where the Majorana
neutrino mass matrix is diagonal and possesses the $S(3)$ symmetry.
Its naive predictions $\theta^{}_{12} = 45^\circ$ and
$\theta^{}_{23} \approx 54.7^\circ$ are no more favored today, but
they may receive proper corrections from the symmetry-breaking
perturbations so as to fit current neutrino oscillation data.

Today's most popular constant pattern of neutrino mixing is the
``tri-bimaximal" mixing matrix:
\begin{equation}
V^{}_0 = \left( \begin{matrix} \frac{\sqrt{2}}{\sqrt{3}} &
\frac{1}{\sqrt{3}} & 0 \cr \frac{-1}{\sqrt{6}} & \frac{1}{\sqrt{3}}
& \frac{1}{\sqrt{2}} \cr \frac{1}{\sqrt{6}} & ~\frac{-1}{\sqrt{3}} ~
& \frac{1}{\sqrt{2}} \cr \end{matrix} \right) \;
\end{equation}
which looks like a twisted form of the democratic mixing pattern
with the same entries. Its strange name comes from the fact that
this flavor mixing pattern is actually a product of the
``tri-maximal" mixing matrix and a ``bi-maximal" mixing matrix:
\begin{eqnarray}
V^\prime_0 = \left( \begin{matrix} \frac{1}{\sqrt{3}} &
\frac{1}{\sqrt{3}} & \frac{1}{\sqrt{3}} \cr \frac{1}{\sqrt{3}} &
\frac{\omega}{\sqrt{3}} & \frac{\omega^2}{\sqrt{3}} \cr
\frac{1}{\sqrt{3}} & ~\frac{\omega^2}{\sqrt{3}} ~ &
\frac{\omega}{\sqrt{3}} \cr \end{matrix} \right) \left(
\begin{matrix} \frac{1}{\sqrt{2}} & ~ 0 ~ & \frac{-1}{\sqrt{2}} \cr 0
& 1 & 0 \cr \frac{1}{\sqrt{2}} & 0 & \frac{1}{\sqrt{2}} \cr
\end{matrix} \right) = P V^{}_0 P^\prime \; ,
\end{eqnarray}
where $\omega = e^{i2\pi/3}$ denotes the complex cube-root of unity
(i.e., $\omega^3 =1$), and $P = {\rm Diag} \{1, \omega, \omega^2 \}$
and $P^\prime = {\rm Diag} \{1, 1, i\}$ are two diagonal phase
matrices. $V^{}_0$ or $V^\prime_0$ predicts $\theta^{}_{12} =
\arctan(1/\sqrt{2}) \approx 35.3^\circ$, $\theta^{}_{13} = 0^\circ$
and $\theta^{}_{23} = 45^\circ$, consistent quite well with current
neutrino oscillation data. Because the entries of $U^{}_0$ or
$V^{}_0$ are all formed from small integers (0, 1, 2 and 3) and
their square roots, it is often suggestive of certain discrete
flavor symmetries in the language of group theories. That is why the
democratic or tri-bimaximal neutrino mixing pattern can serve as a
good starting point of model building based on a variety of flavor
symmetries, such as $Z^{}_2$, $Z^{}_3$, $S^{}_3$, $S^{}_4$,
$A^{}_4$, $D^{}_4$, $D^{}_5$, $Q^{}_4$, $Q^{}_6$, $\Delta (27)$ and
$\Sigma (81)$. In particular, a lot of interest has been paid to the
derivation of $V^{}_0$ with the help of the non-Abelian discrete
$A^{}_4$ symmetry.

Note that the democratic mixing matrix $U^{}_0$ and the
tri-bimaximal mixing matrix $V^{}_0$ are related with each other via
the following transformation:
\begin{eqnarray}
V^{}_0 = \left( \begin{matrix} 1 & 0 & 0 \cr 0 & ~\cos\theta^{}_0 ~
& -\sin\theta^{}_0 \cr 0 & \sin\theta^{}_0 & \cos\theta^{}_0 \cr
\end{matrix} \right) U^{}_0
\left( \begin{matrix} \cos\theta^{}_0 & ~ -\sin\theta^{}_0 ~ & 0 \cr
\sin\theta^{}_0 & \cos\theta^{}_0 & 0 \cr 0 & 0 & 1 \cr
\end{matrix} \right) \; ,
\end{eqnarray}
where $\theta^{}_0 = \arctan (\sqrt{2} -1)^2 \approx 9.7^\circ$.
This angle is actually a measure of the difference between the
mixing angles of $U^{}_0$ and $V^{}_0$ (namely, $45^\circ -
35.3^\circ = 54.7^\circ - 45^\circ = 9.7^\circ$). In this sense, we
argue that it is worthwhile to explore possible flavor symmetries
behind both $V^{}_0$ and $U^{}_0$ so as to build realistic models
for neutrino mass generation and lepton flavor mixing.

Let us remark that a specific constant mixing pattern should be
regarded as the leading-order approximation of the ``true'' lepton
flavor mixing matrix, whose mixing angles should in general depend
on both the ratios of charged-lepton masses and those of neutrino
masses. We may at least make the following naive speculation about
how to phenomenologically understand the observed pattern of lepton
flavor mixing:
\begin{itemize}
\item     Large values of $\theta^{}_{12}$ and $\theta^{}_{23}$
could arise from a weak hierarchy or a near degeneracy of the
neutrino mass spectrum, because the strong hierarchy of
charged-lepton masses implies that $m^{}_e/m^{}_\mu$ and
$m^{}_\mu/m^{}_\tau$ at the electroweak scale are unlikely to
contribute to $\theta^{}_{12}$ and $\theta^{}_{23}$ in a dominant
way.

\item     Special values of $\theta^{}_{12}$ and $\theta^{}_{23}$
might stem from an underlying flavor symmetry of the charged-lepton
mass matrix or the neutrino mass matrix. Then the contributions of
lepton mass ratios to flavor mixing angles, due to flavor symmetry
breaking, are expected to serve as perturbative corrections to
$U^{}_0$ or $V^{}_0$, or another constant mixing pattern.

\item     Vanishing or small $\theta^{}_{13}$ could be a natural
consequence of the explicit textures of lepton mass matrices. It
might also be related to the flavor symmetry which gives rise to
sizable $\theta^{}_{12}$ and $\theta^{}_{23}$ (e.g., in $U^{}_0$ or
$V^{}_0$).

\item    Small corrections to a constant flavor mixing
pattern may also result from the renormalization-group running
effects of leptons and quarks, e.g., from a superhigh-energy scale
to low energies or vice versa.
\end{itemize}
There are too many possibilities of linking the observed pattern of
lepton flavor mixing to a certain flavor symmetry, and none of them
is unique from the theoretical point of view. In this sense, flavor
symmetries should not be regarded as a perfect guiding principle of
model building.
\begin{figure*}[t]
\centering
\includegraphics[width=0.95\linewidth]{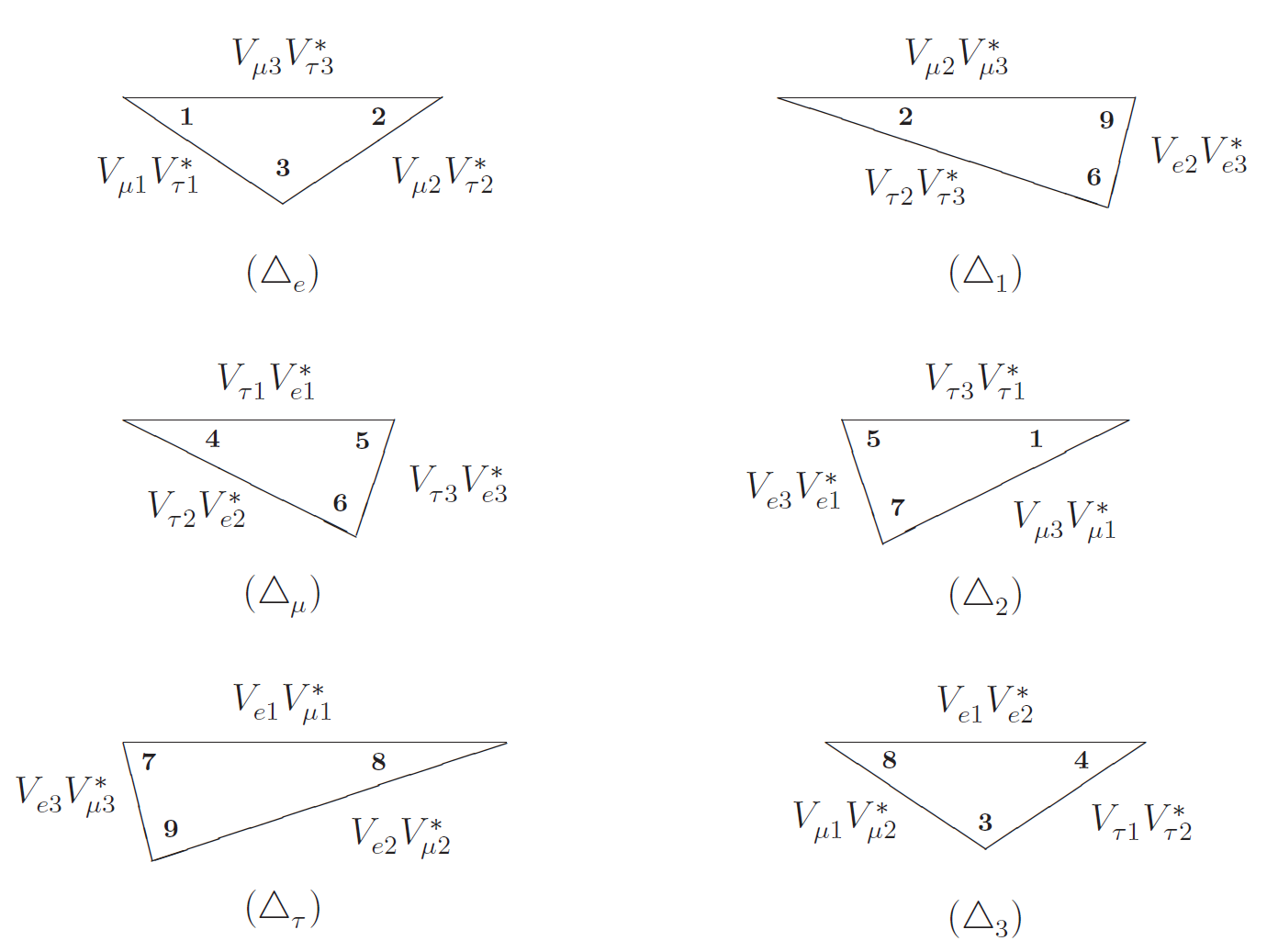}
\caption{Unitarity triangles of the $3\times 3$ PMNS
matrix in the complex plane. Each triangle is named by the index
that does not manifest in its three sides.}
\end{figure*}

\subsection{Leptonic unitarity triangles}

In the basis where the flavor eigenstates of charged leptons are
identified with their mass eigenstates, the PMNS matrix $V$ relates
the neutrino mass eigenstates $(\nu^{}_1, \nu^{}_2, \nu^{}_3)$ to
the neutrino flavor eigenstates $(\nu^{}_e, \nu^{}_\mu,
\nu^{}_\tau)$:
\begin{equation}
\left ( \begin{matrix} \nu^{}_e \cr \nu^{}_\mu \cr \nu^{}_\tau \cr
\end{matrix} \right ) = \left ( \begin{matrix} V^{}_{e1}  & V^{}_{e2} & V^{}_{e3}
\cr V^{}_{\mu 1} & V^{}_{\mu 2} & V^{}_{\mu 3} \cr V^{}_{\tau 1} &
V^{}_{\tau 2} & V^{}_{\tau 3} \cr \end{matrix} \right ) \left (
\begin{matrix} \nu^{}_1 \cr \nu^{}_2 \cr \nu^{}_3 \cr \end{matrix}
\right ) \; .
\end{equation}
The unitarity of $V$ represents two sets of normalization and
orthogonality conditions:
\begin{equation}
\sum_i \left ( V^{}_{\alpha i} V^*_{\beta i} \right ) =  ~
\delta^{}_{\alpha\beta} \; , ~~~~ \sum_\alpha \left ( V^{}_{\alpha
i} V^*_{\alpha j} \right ) = \delta^{}_{ij} \; ,
\end{equation}
where Greek and Latin subscripts run over $(e, \mu, \tau)$ and $(1,
2, 3)$, respectively. In the complex plane the six orthogonality
relations in Eq. (105) define six triangles $(\triangle^{}_e,
\triangle^{}_\mu, \triangle^{}_\tau)$ and $(\triangle^{}_1,
\triangle^{}_2, \triangle^{}_3)$ shown in Fig. 2, the so-called
unitarity triangles. These six triangles have eighteen different
sides and nine different inner (or outer) angles. But the unitarity
of $V$ requires that all six triangles have the same area amounting
to ${\cal J}/2$, where $\cal J$ is the Jarlskog invariant of CP
violation defined through
\begin{equation}
{\rm Im} \left ( V^{}_{\alpha i} V^{}_{\beta j} V^*_{\alpha j}
V^*_{\beta i} \right ) = {\cal J} \sum_{\gamma}
\epsilon^{}_{\alpha\beta\gamma} \sum_{k} \epsilon^{}_{ijk} \; .
\end{equation}
One has ${\cal J} = c^{}_{12} s^{}_{12} c^2_{13} s^{}_{13} c^{}_{23}
s^2_{23} \sin\delta$ in the standard parametrization of $V$ as well
as ${\cal J} = c^{}_l s^{}_l c^{}_\nu s^{}_\nu c s^2 \sin\phi$ in
the FX parametrization of $V$. No matter whether neutrinos are Dirac
or Majorana particles, the strength of CP or T violation in neutrino
oscillations depends only upon $\cal J$.

To show why the areas of six unitarity triangles are identical with
one another, let us take triangles $\triangle^{}_\tau$ and
$\triangle^{}_3$ for example. They correspond to the orthogonality
relations
\begin{eqnarray}
V^{}_{e1} V^{*}_{\mu 1} + V^{}_{e2} V^{*}_{\mu 2} + V^{}_{e3}
V^{*}_{\mu
3} \; & = & \; 0 \; , \nonumber \\
V^{}_{e1} V^{*}_{e2} + V^{}_{\mu 1} V^{*}_{\mu 2} + V^{}_{\tau 1}
V^{*}_{\tau 2} \; & = & \; 0 \; .
\end{eqnarray}
Multiplying these two equations by $V^{}_{\mu 2}V^*_{e2}$ and
$V^{}_{\mu 2} V^*_{\mu 1}$ respectively, we arrive at two rescaled
triangles which share the side
\begin{eqnarray}
V^{}_{e1} V^{}_{\mu 2} V^*_{e2} V^*_{\mu 1} = -|V^{}_{e2} V^{}_{\mu
2}|^2 - V^{}_{e3} V^{}_{\mu 2} V^*_{e2} V^*_{\mu 3} = -|V^{}_{\mu 1}
V^{}_{\mu 2}|^2 - V^{}_{\mu 2} V^{}_{\tau 1} V^*_{\mu 1} V^*_{\tau
2} \; .
\end{eqnarray}
This result is consistent with the definition of ${\cal J}$ in Eq.
(106); i.e., ${\rm Im}(V^{}_{e1} V^{}_{\mu 2} V^*_{e2} V^*_{\mu 1})
= {\cal J}$ and ${\rm Im} (V^{}_{e3} V^{}_{\mu 2} V^*_{e2} V^*_{\mu
3}) = {\rm Im}(V^{}_{\mu 2} V^{}_{\tau 1} V^*_{\mu 1} V^*_{\tau 2})
= -{\cal J}$. The latter simultaneously implies that the areas of
$\triangle^{}_\tau$ and $\triangle^{}_3$ are equal to ${\cal J}/2$.
One may analogously prove that all the six unitarity triangles have
the same area ${\cal J}/2$. If CP or T were an exact symmetry,
${\cal J} =0$ would hold and those unitarity triangles would
collapse into lines in the complex plane. Note that the shape and
area of each unitarity triangle are irrelevant to the nature of
neutrinos; i.e., they are the same for Dirac and Majorana neutrinos.

Because of $V^*_{e1} V^{}_{\mu 1} + V^*_{e2} V^{}_{\mu 2} =
-V^*_{e3} V^{}_{\mu 3}$ or equivalently $|V^{}_{e1} V^*_{\mu 1} +
V^{}_{e2} V^*_{\mu 2}|^2 = |V^{}_{e3} V^*_{\mu 3}|^2$, it is easy to
obtain
\begin{eqnarray}
2{\rm Re}\left( V^{}_{e1} V^{}_{\mu 2} V^*_{e2} V^*_{\mu 1} \right)
= |V^{}_{e3}|^2|V^{}_{\mu 3}|^2 - |V^{}_{e1}|^2 |V^{}_{\mu 1}|^2 -
|V^{}_{e2}|^2 |V^{}_{\mu 2}|^2 \; .
\end{eqnarray}
Combining $V^{}_{e1} V^{}_{\mu 2} V^*_{e2} V^*_{\mu 1} = {\rm Re}
(V^{}_{e1} V^{}_{\mu 2} V^*_{e2} V^*_{\mu 1}) + i{\cal J}$ with Eq.
(109) leads us to the result
\begin{eqnarray}
{\cal J}^2 = |V^{}_{e1}|^2 |V^{}_{\mu 2}|^2 |V^{}_{e2}|^2 |V^{}_{\mu
1}|^2 - \frac{1}{4} \left( |V^{}_{e3}|^2|V^{}_{\mu 3}|^2 -
|V^{}_{e1}|^2 |V^{}_{\mu 1}|^2 - |V^{}_{e2}|^2 |V^{}_{\mu 2}|^2
\right)^2
\nonumber \\
= |V^{}_{e1}|^2 |V^{}_{\mu 2}|^2 |V^{}_{e2}|^2 |V^{}_{\mu 1} |^2 -
\frac{1}{4} \left ( 1 + |V^{}_{e1}|^2 |V^{}_{\mu 2}|^2 +
|V^{}_{e2}|^2 |V^{}_{\mu 1}|^2 \right . ~~~~~~~~~~~~~~~~~~~ \ \
\nonumber \\
\left . - |V^{}_{e1}|^2 - |V^{}_{\mu 2}|^2 - |V^{}_{e2}|^2 -
|V^{}_{\mu 1}|^2 \right )^2 \; . ~~~~~~
\end{eqnarray}
As a straightforward generalization of Eq. (110), ${\cal J}^2$ can
be expressed in terms of the moduli of any four independent matrix
elements of $V$:
\begin{eqnarray}
{\cal J}^2 = |V^{}_{\alpha i}|^2 |V^{}_{\beta j}|^2 |V^{}_{\alpha
j}|^2 |V^{}_{\beta i} |^2 - \frac{1}{4} \left ( 1 + |V^{}_{\alpha
i}|^2 |V^{}_{\beta j}|^2 + |V^{}_{\alpha j}|^2 |V^{}_{\beta i}|^2
\right . ~~~~~~~~~~~~~~~
\nonumber \\
\left . - |V^{}_{\alpha i}|^2 - |V^{}_{\beta j}|^2 - |V^{}_{\alpha
j}|^2 - |V^{}_{\beta i}|^2 \right )^2 \; ,
\end{eqnarray}
in which $\alpha \neq \beta$ running over $(e, \mu, \tau)$ and $i
\neq j$ running over $(1, 2, 3)$. The implication of this result is
very obvious: the information about leptonic CP violation can in
principle be extracted from the measured moduli of the neutrino
mixing matrix elements.

As a consequence of the unitarity of $V$, two interesting relations
can be derived from the normalization conditions in Eq. (105):
\begin{eqnarray}
&& |V^{}_{e 2}|^2 - |V^{}_{\mu 1}|^2 = |V^{}_{\mu 3}|^2 -
|V^{}_{\tau 2}|^2 = |V^{}_{\tau 1}|^2 - |V^{}_{e 3}|^2 \equiv
\Delta^{}_{\rm L} \; ,
\nonumber \\
&& |V^{}_{e 2}|^2 - |V^{}_{\mu 3}|^2 = |V^{}_{\mu 1}|^2 -
|V^{}_{\tau 2}|^2 = |V^{}_{\tau 3}|^2 - |V^{}_{e 1}|^2 \equiv
\Delta^{}_{\rm R} \; .
\end{eqnarray}
The off-diagonal asymmetries $\Delta^{}_{\rm L}$ and $\Delta^{}_{\rm
R}$ characterize the geometrical structure of $V$ about its
$V^{}_{e1}$-$V^{}_{\mu 2}$-$V^{}_{\tau 3}$ and $V^{}_{e
3}$-$V^{}_{\mu 2}$-$V^{}_{\tau 1}$ axes, respectively. For instance,
$\Delta^{}_{\rm L} = 1/6$ and $\Delta^{}_{\rm R} = -1/6$ hold for
the tri-bimaximal neutrino mixing pattern $V^{}_0$. If
$\Delta^{}_{\rm L} = 0$ (or $\Delta^{}_{\rm R} = 0$) held, $V$ would
be symmetric about the $V^{}_{e1}$-$V^{}_{\mu 2}$-$V^{}_{\tau 3}$
(or $V^{}_{e3}$-$V^{}_{\mu 2}$-$V^{}_{\tau 1}$) axis. Geometrically
this would correspond to the congruence between two unitarity
triangles; i.e.,
\begin{eqnarray}
\Delta^{}_{\rm L} = 0 : \; \triangle^{}_e \cong \triangle^{}_1 \; ,
\triangle^{}_\mu \cong \triangle^{}_2 \; , \triangle^{}_\tau \cong
\triangle^{}_3 \; ; \;\;
\nonumber \\
\Delta^{}_{\rm R} = 0 : \; \triangle^{}_e \cong \triangle^{}_3 \; ,
\triangle^{}_\mu \cong \triangle^{}_2 \; , \triangle^{}_\tau \cong
\triangle^{}_1 \; . ~~
\end{eqnarray}
Indeed the counterpart of $\Delta^{}_{\rm L}$ in the quark sector is
only of ${\cal O}(10^{-5})$; i.e., the CKM matrix is almost
symmetric about its $V^{}_{ud}$-$V^{}_{cs}$-$V^{}_{tb}$ axis. An
exactly symmetric flavor mixing matrix might hint at an underlying
flavor symmetry, from which some deeper understanding of the fermion
mass texture could be achieved.

\subsection{Flavor problems in particle physics}

In the subatomic world the fundamental building blocks of matter
have twelve flavors: six quarks and six leptons (and their
antiparticles). Table 2 is a brief list of some important
discoveries in flavor physics, which can partly give people a
ball-park feeling of a century of developments in particle physics.
The SM of electromagnetic and weak interactions contain thirteen
free parameters in its lepton and quark sectors: three
charged-lepton masses, six quark masses, three quark flavor mixing
angles and one CP-violating phase. If three known neutrinos are
massive Majorana particles, one has to introduce nine free
parameters to describe their flavor properties: three neutrino
masses, three lepton flavor mixing angles and three CP-violating
phases. Thus an effective theory of electroweak interactions at low
energies totally consists of twenty-two flavor parameters which can
only be determined from experiments. Why is the number of degrees of
freedom so big in the flavor sector? What is the fundamental physics
behind these parameters? Such puzzles constitute the flavor problems
in particle physics.
\begin{table*}[t]
\begin{center}
\caption{Some important discoveries in the developments of flavor
physics.} \vspace{0.3cm}
\begin{tabular}{c|cl}
  \hline
  \hline
  && Discoveries of lepton flavors, quark flavors and CP violation \\
\hline
  1897 && electron (Thomson, 1897)  \\
  1919 && proton (up and down quarks) (Rutherford, 1919) \\
  1932 && neutron (up and down quarks) (Chadwick, 1932) \\
  1933 && positron (Anderson, 1933) \\
  1936 && muon (Neddermeyer and Anderson, 1937) \\
  1947 && Kaon (strange quark) (Rochester and Butler, 1947) \\
  1956 && electron antineutrino (Cowan {\it et al.}, 1956) \\
  1962 && muon neutrino (Danby {\it et al.}, 1962) \\
  1964 && CP violation in $s$-quark decays (Christenson {\it et al.}, 1964) \\
  1974 && charm quark (Aubert {\it et al.}, 1974; Abrams {\it et al.}, 1974) \\
  1975 && tau (Perl {\it et al.}, 1975) \\
  1977 && bottom quark (Herb {\it et al.}, 1977) \\
  1995 && top quark (Abe {\it et al.}, 1995; Abachi {\it et al.}, 1995) \\
  2000 && tau neutrino (Kodama {\it et al.}, 2000) \\
  2001 && CP violation in $b$-quark decays (Aubert {\it et al.}, 2001; Abe {\it et al.}, 2001) \\
  \hline
\end{tabular}
\end{center}
\end{table*}

Current experimental data on neutrino oscillations can only tell us
$m^{}_1 < m^{}_2$. It remains unknown whether $m^{}_3$ is larger
than $m^{}_2$ (normal hierarchy) or smaller than $m^{}_1$ (inverted
hierarchy). The possibility $m^{}_1 \approx m^{}_2 \approx m^{}_3$
(near degeneracy) cannot be excluded at present. In contrast, three
families of charged fermions have very strong mass hierarchies:
\begin{eqnarray}
\frac{m^{}_e}{m^{}_\mu} \sim \frac{m^{}_u}{m^{}_c} \sim
\frac{m^{}_c}{m^{}_t} \sim \lambda^4 \; , \nonumber
\\
\frac{m^{}_\mu}{m^{}_\tau} \sim \frac{m^{}_d}{m^{}_s} \sim
\frac{m^{}_s}{m^{}_b} \sim \lambda^2 \; ,
\end{eqnarray}
where $\lambda \equiv \sin\theta^{}_{\rm C} \approx 0.22$ with
$\theta^{}_{\rm C}$ being the Cabibbo angle of quark flavor mixing.
In the standard parametrization of the CKM matrix, three quark
mixing angles exhibit an impressive hierarchy:
\begin{equation}
\vartheta^{}_{12} \sim \lambda \; , ~~~~ \vartheta^{}_{23} \sim
\lambda^2 \; , ~~~~ \vartheta^{}_{13} \sim \lambda^4 \; .
\end{equation}
These two kinds of hierarchies might intrinsically be related to
each other, because the flavor mixing angles actually measure a
mismatch between the mass and flavor eigenstates of up- and
down-type quarks. For example, the relations $\vartheta^{}_{12}
\approx \sqrt{m^{}_d/m^{}_s}~$, $\vartheta^{}_{23} \approx
\sqrt{m^{}_d/m^{}_b}$ and $\vartheta^{}_{13} \approx
\sqrt{m^{}_u/m^{}_t}$ are compatible with Eqs. (114) and (115). They
can be derived from a specific pattern of up- and down-type quark
mass matrices with five texture zeros. On the other hand, it seems
quite difficult to find a simple way of linking two large lepton
flavor mixing angles $\theta^{}_{12} \sim \pi/6$ and $\theta^{}_{23}
\sim \pi/4$ to small $m^{}_e/m^{}_\mu$ and $m^{}_\mu/m^{}_\tau$. One
might ascribe the largeness of $\theta^{}_{12}$ and $\theta^{}_{23}$
to a very weak hierarchy of three neutrino masses and the smallness
of $\theta^{}_{13}$ to the strong mass hierarchy in the
charged-lepton sector. There are of course many possibilities of
model building to understand the observed lepton flavor mixing
pattern, but none of them has experimentally and theoretically been
justified.

Among a number of concrete flavor puzzles that are currently facing
us, the following three are particularly intriguing.
\begin{itemize}
\item     The pole masses of three charged leptons satisfy the
equality
\begin{equation}
\frac{m^{}_e + m^{}_\mu + m^{}_\tau}{\left( \sqrt{m^{}_e} +
\sqrt{m^{}_\mu} + \sqrt{m^{}_\tau} \right)^2} = \frac{2}{3} \;
\end{equation}
to an amazingly good degree of accuracy --- its error bar is only of
${\cal O}(10^{-5})$.

\item     There are two quark-lepton ``complementarity" relations
in flavor mixing:
\begin{equation}
\theta^{}_{12} + \vartheta^{}_{12} \approx \theta^{}_{23} +
\vartheta^{}_{23} \approx \frac{\pi}{4} \; ,
\end{equation}
which are compatible with the present experimental data.

\item     Two unitarity triangles of the CKM matrix, defined by
the orthogonality conditions $V^{}_{ud} V^*_{ub} + V^{}_{cd}
V^*_{cb} + V^{}_{td} V^*_{tb} = 0$ and $V^{}_{tb} V^*_{ub} +
V^{}_{ts} V^*_{us} + V^{}_{td} V^*_{ud} = 0$, are almost the right
triangles. Namely, the common inner angle of these two triangles
satisfies
\begin{equation}
\alpha \equiv \arg \left(
-\frac{V^{}_{ud}V^*_{ub}}{V^{}_{td}V^*_{tb}} \right) \approx
\frac{\pi}{2} \; ,
\end{equation}
indicated by current experimental data on quark mixing and CP
violation.
\end{itemize}
Such special numerical relations might just be accidental. One or
two of them might also be possible to result from a certain
(underlying) flavor symmetry.

\section{Running of Neutrino Mass Parameters}

\subsection{One-loop RGEs}

The spirit of seesaw mechanisms is to attribute the small masses of
three known neutrinos to the existence of some heavy degrees of
freedom, such as the $SU(2)^{}_{\rm L}$ gauge-singlet fermions, the
$SU(2)^{}_{\rm L}$ gauge-triplet scalars or the $SU(2)^{}_{\rm L}$
gauge-triplet fermions. All of them point to the unique dimension-5
Weinberg operator in an effective theory after the corresponding
heavy particles are integrated out:
\begin{equation}
\frac{{\cal L}^{}_{\rm d =5}}{\Lambda} = \frac{1}{2}
\kappa^{}_{\alpha \beta} \overline{\ell^{}_{\alpha \rm L}} \tilde{H}
\tilde{H}^T \ell^c_{\beta \rm L} + {\rm h.c.} \; ,
\end{equation}
where $\Lambda$ is the cutoff scale, $\ell^{}_{\rm L}$ denotes the
left-handed lepton doublet, $\tilde{H} \equiv i\sigma^{}_2 H^*$ with
$H$ being the SM Higgs doublet, and $\kappa$ stands for the
effective neutrino coupling matrix. After spontaneous gauge symmetry
breaking, $\tilde{H}$ gains its vacuum expectation value $\langle
\tilde{H} \rangle = v/\sqrt{2}$ with $v \approx 246$ GeV. We are
then left with the effective Majorana mass matrix $M^{}_\nu = \kappa
v^2/2$ for three light neutrinos from Eq. (119). If the dimension-5
Weinberg operator is obtained in the framework of the minimal
supersymmetric standard model (MSSM), one will be left with
$M^{}_\nu = \kappa (v \sin\beta)^2/2$, where $\tan\beta$ denotes the
ratio of the vacuum expectation values of two MSSM Higgs doublets.

Eq. (119) or its supersymmetric counterpart can provide a simple but
generic way of generating tiny neutrino masses. There are a number
of interesting possibilities of building renormalizable gauge models
to realize the effective Weinberg mass operator, either radiatively
or at the tree level. The latter case is just associated with the
well-known seesaw mechanisms to be discussed in section 6. Here we
assume that ${\cal L}^{}_{\rm d =5}/\Lambda$ arises from an
underlying seesaw model, whose lightest heavy particle has a mass of
${\cal O}(\Lambda)$. In other words, $\Lambda$ characterizes the
seesaw scale. Above $\Lambda$ there may exist one or more energy
thresholds corresponding to the masses of heavier seesaw particles.
Below $\Lambda$ the energy dependence of the effective neutrino
coupling matrix $\kappa$ is described by its renormalization-group
equation (RGE). The evolution of $\kappa$ from $\Lambda$ down to the
electroweak scale is formally independent of any details of the
relevant seesaw model from which $\kappa$ is derived.

At the one-loop level $\kappa$ obeys the RGE
\begin{equation}
16\pi^2 \frac{{\rm d}\kappa}{{\rm d}t} = \alpha^{}_\kappa \kappa +
C^{}_\kappa \left[ (Y^{}_lY^\dagger_l) \kappa + \kappa (Y^{}_l
Y^\dagger_l)^T \right]
\end{equation}
where $t\equiv \ln (\mu/\Lambda)$ with $\mu$ being an arbitrary
renormalization scale between the electroweak scale and the seesaw
scale, and $Y^{}_l$ is the charged-lepton Yukawa coupling matrix.
The RGE of $Y^{}_l$ and those of $Y^{}_{\rm u}$ (up-type quarks) and
$Y^{}_{\rm d}$ (down-type quarks) are given by
\begin{eqnarray}
&& 16\pi^2 \frac{{\rm d}Y^{}_l}{{\rm d}t} = \left[ \alpha^{}_l +
C^l_l (Y^{}_lY^\dagger_l) \right] Y^{}_l \; ,
\nonumber \\
&& 16\pi^2 \frac{{\rm d}Y^{}_{\rm u}}{{\rm d}t} = \left[
\alpha^{}_{\rm u} + C^{\rm u}_{\rm u} (Y^{}_{\rm u} Y^\dagger_{\rm
u}) + C^{\rm d}_{\rm u} (Y^{}_{\rm d} Y^\dagger_{\rm d})\right]
Y^{}_{\rm u} \; ,
\nonumber \\
&& 16\pi^2 \frac{{\rm d}Y^{}_{\rm d}}{{\rm d}t} = \left[
\alpha^{}_{\rm d} + C^{\rm u}_{\rm d} (Y^{}_{\rm u} Y^\dagger_{\rm
u}) + C^{\rm d}_{\rm d} (Y^{}_{\rm d} Y^\dagger_{\rm d})\right]
Y^{}_{\rm d} \; .
\end{eqnarray}
In the framework of the SM we have
\begin{eqnarray}
C^{}_\kappa = C^{\rm d}_{\rm u} = C^{\rm u}_{\rm d} = -\frac{3}{2}
\; ,
\nonumber \\
C^l_l = C^{\rm u}_{\rm u} = C^{\rm d}_{\rm d} = + \frac{3}{2} \; ,
\;
\end{eqnarray}
and
\begin{eqnarray}
&& \alpha^{}_\kappa = -3 g^2_2 + \lambda + 2 {\rm Tr} \left[ 3
(Y^{}_{\rm u} Y^\dagger_{\rm u}) + 3 (Y^{}_{\rm d} Y^\dagger_{\rm
d}) + (Y^{}_l Y^\dagger_l) \right] \; ,
\nonumber \\
&& \alpha^{}_l = -\frac{9}{4} g^2_1 -\frac{9}{4} g^2_2 + {\rm Tr}
\left[ 3 (Y^{}_{\rm u} Y^\dagger_{\rm u}) + 3 (Y^{}_{\rm d}
Y^\dagger_{\rm d}) + (Y^{}_l Y^\dagger_l) \right] \; ,
\nonumber \\
&& \alpha^{}_{\rm u} = - \frac{17}{20} g^2_1 - \frac{9}{4} g^2_2 - 8
g^2_3 + {\rm Tr} \left[ 3 (Y^{}_{\rm u} Y^\dagger_{\rm u}) + 3
(Y^{}_{\rm d} Y^\dagger_{\rm d}) + (Y^{}_l Y^\dagger_l) \right] \; ,
\nonumber \\
&& \alpha^{}_{\rm d} = -\frac{1}{4} g^2_1 - \frac{9}{4} g^2_2 - 8
g^2_3 + {\rm Tr} \left[ 3 (Y^{}_{\rm u} Y^\dagger_{\rm u}) + 3
(Y^{}_{\rm d} Y^\dagger_{\rm d}) + (Y^{}_l Y^\dagger_l) \right] \; ;
\end{eqnarray}
and in the framework of the MSSM we have
\begin{eqnarray}
C^{}_\kappa = C^{\rm d}_{\rm u} = C^{\rm u}_{\rm d} = +1 \; ,
\nonumber \\
C^l_l = C^{\rm u}_{\rm u} = C^{\rm d}_{\rm d} = +3 \; , \;
\end{eqnarray}
and
\begin{eqnarray}
&& \alpha^{}_\kappa = -\frac{6}{5} g^2_1 -6 g^2_2 + 6 {\rm Tr}
(Y^{}_{\rm u} Y^\dagger_{\rm u}) \; ,
\nonumber \\
&& \alpha^{}_l = -\frac{9}{5} g^2_1 -3 g^2_2 + {\rm Tr} \left[ 3
(Y^{}_{\rm d} Y^\dagger_{\rm d}) + (Y^{}_l Y^\dagger_l) \right] \; ,
\nonumber \\
&& \alpha^{}_{\rm u} = - \frac{13}{15} g^2_1 - 3 g^2_2 -
\frac{16}{3} g^2_3 + 3 {\rm Tr} (Y^{}_{\rm u} Y^\dagger_{\rm u}) \;
,
\nonumber \\
&& \alpha^{}_{\rm d} = - \frac{7}{15} g^2_1 - 3 g^2_2 - \frac{16}{3}
g^2_3 + {\rm Tr} \left[ 3 (Y^{}_{\rm d} Y^\dagger_{\rm d}) + (Y^{}_l
Y^\dagger_l) \right] \; .
\end{eqnarray}
Here $g^{}_1$, $g^{}_2$ and $g^{}_3$ are the gauge couplings and
satisfy their RGEs
\begin{equation}
16\pi^2 \frac{{\rm d} g^{}_i}{{\rm d} t} = b^{}_i g^3_i \; ,
\end{equation}
where $(b^{}_1, b^{}_2, b^{}_3) = (41/10, -19/6, -7)$ in the SM or
$(33/5, 1, -3)$ in the MSSM. In addition, $\lambda$ is the Higgs
self-coupling parameter of the SM and obeys the RGE
\begin{eqnarray}
16\pi^2 \frac{{\rm d} \lambda}{{\rm d} t} = 6 \lambda^2 - 3 \lambda
\left( \frac{3}{5} g^2_1 + 3 g^2_2 \right) + \frac{3}{2} \left(
\frac{3}{5} g^2_1 + g^2_2 \right)^2 + 3 g^4_2
\nonumber \\
+ 4 \lambda {\rm Tr} \left[ 3 (Y^{}_{\rm u} Y^\dagger_{\rm u}) + 3
(Y^{}_{\rm d} Y^\dagger_{\rm d}) + (Y^{}_l Y^\dagger_l) \right]
~~~~~
\nonumber \\
- 8 {\rm Tr} \left[ 3 (Y^{}_{\rm u} Y^\dagger_{\rm u})^2 + 3
(Y^{}_{\rm d} Y^\dagger_{\rm d})^2 + (Y^{}_l Y^\dagger_l)^2 \right]
\; .
\end{eqnarray}
The relation between $\lambda$ and the Higgs mass $M^{}_h$ is given
by $\lambda = M^2_h/(2v^2)$, where $v \approx 246$ GeV is the vacuum
expectation value of the Higgs field.

The above RGEs allow us to evaluate the running behavior of $\kappa$
together with those of $Y^{}_l$, $Y^{}_{\rm u}$ and $Y^{}_{\rm d}$,
from the seesaw scale to the electroweak scale or vice versa. We
shall examine the evolution of neutrino masses, lepton flavor mixing
angles and CP-violating phases in the following.

\subsection{Running neutrino mass parameters}

Without loss of any generality, we choose the flavor basis where
$Y^{}_l$ is diagonal: $Y^{}_l = D^{}_l \equiv {\rm Diag}\{y^{}_e,
y^{}_\mu, y^{}_\tau \}$ with $y^{}_\alpha$ being the eigenvalues of
$Y^{}_l$. In this case the effective Majorana neutrino coupling
matrix $\kappa$ can be diagonalized by the PMNS matrix $V$; i.e.,
$V^\dagger \kappa V^* = \widehat{\kappa} \equiv {\rm Diag}\{
\kappa^{}_1, \kappa^{}_2, \kappa^{}_3 \}$ with $\kappa^{}_i$ being
the eigenvalues of $\kappa$. Then
\begin{eqnarray}
\frac{{\rm d} \kappa}{{\rm d} t} = \dot{V} \widehat{\kappa} V^T + V
\dot{\widehat{\kappa}} V^T + V \widehat{\kappa} \dot{V}^T =
\frac{1}{16\pi^2} \left[ \alpha^{}_\kappa V \widehat{\kappa} V^T +
C^{}_\kappa \left( D^2_l V \widehat{\kappa} V^T + V \widehat{\kappa}
V^T D^2_l \right) \right] \; ,
\end{eqnarray}
with the help of Eq. (120). After a definition of the Hermitian
matrix $S \equiv V^\dagger D^2_l V$ and the anti-Hermitian matrix $T
\equiv V^\dagger \dot{V}$, Eq. (128) leads to
\begin{equation}
\dot{\widehat{\kappa}} = \frac{1}{16\pi^2} \left[ \alpha^{}_\kappa
\widehat{\kappa} + C^{}_\kappa ( S \widehat{\kappa} +
\widehat{\kappa} S^* ) \right] - T \widehat{\kappa} +
\widehat{\kappa} T^* .
\end{equation}
Because $\widehat{\kappa}$ is by definition diagonal and real, the
left- and right-hand sides of Eq. (129) must be diagonal and real.
We can therefore arrive at
\begin{equation}
\dot{\kappa}^{}_i = \frac{1}{16\pi^2} \left( \alpha^{}_\kappa + 2
C^{}_\kappa {\rm Re} S^{}_{ii} \right ) \kappa^{}_i \; ,
\end{equation}
together with ${\rm Im}T^{}_{ii} = {\rm Re}T^{}_{ii} ={\rm
Im}S^{}_{ii} = 0$ (for $i= 1, 2, 3$). As the off-diagonal parts of
Eq. (129) are vanishing, we have
\begin{equation}
T^{}_{ij} \kappa^{}_j - \kappa^{}_i T^*_{ij} =
\frac{C^{}_\kappa}{16\pi^2} \left( S^{}_{ij} \kappa^{}_j +
\kappa^{}_i S^*_{ij} \right) \;
\end{equation}
with $i\neq j$. Therefore,
\begin{eqnarray}
{\rm Re}T^{}_{ij} = - \frac{C^{}_\kappa}{16\pi^2} \frac{\kappa^{}_i
+ \kappa^{}_j}{\kappa^{}_i - \kappa^{}_j} {\rm Re} S^{}_{ij} \; ,
\nonumber \\
{\rm Im}T^{}_{ij} = - \frac{C^{}_\kappa}{16\pi^2} \frac{\kappa^{}_i
- \kappa^{}_j}{\kappa^{}_i + \kappa^{}_j} {\rm Im} S^{}_{ij} \; .
\end{eqnarray}
Due to $\dot{V} = VT$, Eq. (132) actually governs the evolution of
$V$ with energies.

We proceed to define $V \equiv P U P^\prime$, in which $P \equiv
{\rm Diag} \{ e^{i\phi^{}_e}, e^{i\phi^{}_\mu}, e^{i\phi^{}_\tau}
\}$, $P^\prime \equiv {\rm Diag} \{ e^{i\rho}, e^{i\sigma}, 1 \}$,
and $U$ is the CKM-like matrix containing three neutrino mixing
angles and one CP-violating phase. Although $P$ does not have any
physical meaning, its phases have their own RGEs. In contrast,
$P^\prime$ serves for the Majorana phase matrix. We find
\begin{eqnarray}
T^\prime \equiv P^\prime T {P^\prime}^\dagger = P^\prime V^\dagger
\dot{V} {P^\prime}^\dagger = \dot{P}^\prime {P^\prime}^\dagger +
U^\dagger \dot{U} + U^\dagger P^\dagger \dot{P} U \; ,
\end{eqnarray}
from which we can obtain six independent constraint equations:
\begin{eqnarray}
&& T^\prime_{11} = i \dot{\rho} + \sum_\alpha \left[ U^*_{\alpha 1}
\dot{U}^{}_{\alpha 1} + i U^{}_{\alpha 1} \dot{\phi}^{}_\alpha
\right] \; ,
\nonumber \\
&& T^\prime_{22} = i \dot{\sigma} + \sum_\alpha \left[ U^*_{\alpha
2} \dot{U}^{}_{\alpha 2} + i U^{}_{\alpha 2} \dot{\phi}^{}_\alpha
\right] \; ,
\nonumber \\
&& T^\prime_{33} = \sum_\alpha \left[ U^*_{\alpha 3}
\dot{U}^{}_{\alpha 3} + i U^{}_{\alpha 3} \dot{\phi}^{}_\alpha
\right] \; ;
\nonumber \\
&& T^\prime_{12} = \sum_\alpha \left[ U^*_{\alpha 1}
\dot{U}^{}_{\alpha 2} + i U^{}_{\alpha 2} \dot{\phi}^{}_\alpha
\right] \; ,
\nonumber \\
&& T^\prime_{13} = \sum_\alpha \left[ U^*_{\alpha 1}
\dot{U}^{}_{\alpha 3} + i U^{}_{\alpha 3} \dot{\phi}^{}_\alpha
\right] \; ,
\nonumber \\
&& T^\prime_{23} = \sum_\alpha \left[ U^*_{\alpha 2}
\dot{U}^{}_{\alpha 3} + i U^{}_{\alpha 3} \dot{\phi}^{}_\alpha
\right] \; ,
\end{eqnarray}
where $\alpha$ runs over $e$, $\mu$ and $\tau$. Note that $T^{}_{ii}
= 0$ holds and $T^{}_{ij}$ is given by Eq. (132). In view of $y^{}_e
\ll y^{}_\mu \ll y^{}_\tau$, we take $D^2_l \approx {\rm Diag} \{ 0,
0, y^2_\tau \}$ as an excellent approximation. Then $S^{}_{ij}$,
$T^{}_{ij}$ and $T^\prime_{ij}$ can all be expressed in terms of
$y^2_\tau$ and the parameters of $U$ and $P^\prime$. After a
straightforward calculation, we obtain the explicit expressions of
Eqs. (130) and (134) as follows:
\begin{equation}
\dot{\kappa}^{}_i = \frac{\kappa^{}_i}{16\pi^2} \left(
\alpha^{}_\kappa + 2 C^{}_\kappa y^2_\tau |U^{}_{\tau i}|^2 \right)
\; , \;
\end{equation}
and
\begin{eqnarray}
&& \sum_\alpha \left[ U^*_{\alpha 1} \left( i \dot{U}_{\alpha 1} -
U^{}_{\alpha 1} \dot{\phi}^{}_\alpha \right) \right] = \dot{\rho} \;
,
\nonumber \\
&& \sum_\alpha \left[ U^*_{\alpha 2} \left( i \dot{U}_{\alpha 2} -
U^{}_{\alpha 2} \dot{\phi}^{}_\alpha \right) \right] = \dot{\sigma}
\; ,
\nonumber \\
&& \sum_\alpha \left[ U^*_{\alpha 3} \left( i \dot{U}_{\alpha 3} -
U^{}_{\alpha 3} \dot{\phi}^{}_\alpha \right) \right] = 0 \; ,
\nonumber \\
&& \sum_\alpha \left[ U^*_{\alpha 1} \left( \dot{U}_{\alpha 2} + i
U^{}_{\alpha 2} \dot{\phi}^{}_\alpha \right) \right] = -
\frac{C^{}_\kappa y^2_\tau}{16\pi^2} e^{i \left( \rho - \sigma
\right)} \left[ \zeta^{-1}_{12} {\rm Re} \left( U^*_{\tau 1}
U^{}_{\tau 2} e^{i \left( \sigma - \rho \right)} \right) + i
\zeta^{}_{12} {\rm Im} \left( U^*_{\tau 1} U^{}_{\tau 2} e^{i \left(
\sigma - \rho \right)} \right) \right] \; ,
\nonumber \\
&& \sum_\alpha \left[ U^*_{\alpha 1} \left( \dot{U}_{\alpha 3} + i
U^{}_{\alpha 3} \dot{\phi}^{}_\alpha \right) \right] = -
\frac{C^{}_\kappa y^2_\tau}{16\pi^2} e^{i \rho} \left[
\zeta^{-1}_{13} {\rm Re} \left( U^*_{\tau 1} U^{}_{\tau 3} e^{-i
\rho} \right) + i \zeta^{}_{13} {\rm Im} \left( U^*_{\tau 1}
U^{}_{\tau 3} e^{-i \rho} \right) \right] \; ,
\nonumber \\
&& \sum_\alpha \left[ U^*_{\alpha 2} \left( \dot{U}_{\alpha 3} + i
U^{}_{\alpha 3} \dot{\phi}^{}_\alpha \right) \right] = -
\frac{C^{}_\kappa y^2_\tau}{16\pi^2} e^{i \sigma} \left[
\zeta^{-1}_{23} {\rm Re} \left( U^*_{\tau 2} U^{}_{\tau 3} e^{-i
\sigma} \right) + i \zeta^{}_{23} {\rm Im} \left( U^*_{\tau 2}
U^{}_{\tau 3} e^{-i \sigma} \right) \right] \; ,
\end{eqnarray}
where $\zeta^{}_{ij} \equiv (\kappa^{}_i - \kappa^{}_j)/(\kappa^{}_i
+ \kappa^{}_j)$ with $i\neq j$. One can see that those
$y^2_\tau$-associated terms only consist of the matrix elements
$U^{}_{\tau i}$ (for $i=1, 2, 3$). If a parametrization of $U$
assures $U^{}_{\tau i}$ to be as simple as possible, the resultant
RGEs of neutrino mixing angles and CP-violating phases will be very
concise. We find that the FX parametrization advocated in Eq. (99)
with
$$
U = \left ( \begin{matrix} s^{}_l s^{}_{\nu} c + c^{}_l c^{}_{\nu}
e^{-i\phi} & s^{}_l c^{}_{\nu} c - c^{}_l s^{}_{\nu} e^{-i\phi} &
s^{}_l s \cr c^{}_l s^{}_{\nu} c - s^{}_l c^{}_{\nu} e^{-i\phi} &
c^{}_l c^{}_{\nu} c + s^{}_l s^{}_{\nu} e^{-i\phi}  & c^{}_l s \cr -
s^{}_{\nu} s   & - c^{}_{\nu} s   & c \cr \end{matrix} \right )
$$
accords with the above observation, while the ``standard"
parametrization in Eq. (98) does not. That is why the RGEs of
neutrino mixing angles and CP-violating phases in the standard
parametrization are rather complicated.

Here we take the FX form of $U$ to derive the RGEs of neutrino mass
and mixing parameters. Combining Eqs. (135), (136) and the FX form
of $U$, we arrive at
\begin{eqnarray}
\dot{\kappa}^{}_1 = \frac{\kappa^{}_1}{16\pi^2} \left (
\alpha^{}_\kappa + 2 C^{}_\kappa y^2_\tau s^2_\nu s^2 \right ) \; ,
\nonumber \\
\dot{\kappa}^{}_2 = \frac{\kappa^{}_2}{16\pi^2} \left (
\alpha^{}_\kappa + 2 C^{}_\kappa y^2_\tau c^2_\nu s^2 \right ) \; ,
\nonumber \\
\dot{\kappa}^{}_3 = \frac{\kappa^{}_3}{16\pi^2} \left (
\alpha^{}_\kappa + 2 C^{}_\kappa y^2_\tau c^2 \right ) \; , ~~~
\end{eqnarray}
where $\alpha^{}_\kappa \approx -3g^2_2 + 6y^2_t + \lambda$ (SM) or
$\alpha^{}_\kappa \approx -1.2g^2_1 - 6g^2_2 + 6 y^2_t$ (MSSM); and
\begin{eqnarray}
&& \dot{\theta}^{}_l = \frac{C^{}_\kappa y^2_\tau}{16\pi^2} ~
c^{}_\nu s^{}_\nu c \left [ \zeta^{-1}_{13} c^{}_\rho c^{}_{(\rho
-\phi)} + \zeta^{}_{13} s^{}_\rho s^{}_{(\rho - \phi)} -
\zeta^{-1}_{23} c^{}_\sigma c^{}_{(\sigma -\phi)} - \zeta^{}_{23}
s^{}_\sigma s^{}_{(\sigma - \phi)} \right ] \; ,
\nonumber \\
&& \dot{\theta}^{}_\nu = \frac{C^{}_\kappa y^2_\tau}{16\pi^2} ~
c^{}_\nu s^{}_\nu \left [ s^2 \left ( \zeta^{-1}_{12} c^2_{(\sigma
-\rho)} + \zeta^{}_{12} s^2_{(\sigma -\rho)} \right ) + c^2 \left (
\zeta^{-1}_{13} c^2_\rho + \zeta^{}_{13} s^2_\rho \right ) - c^2
\left ( \zeta^{-1}_{23} c^2_\sigma + \zeta^{}_{23} s^2_\sigma \right
) \right ] \; ,
\nonumber \\
&& \dot{\theta} = \frac{C^{}_\kappa y^2_\tau}{16\pi^2} ~ c s \left [
s^2_\nu \left ( \zeta^{-1}_{13} c^2_\rho + \zeta^{}_{13} s^2_\rho
\right ) + c^2_\nu \left ( \zeta^{-1}_{23} c^2_\sigma +
\zeta^{}_{23} s^2_\sigma \right ) \right ] \; ;
\end{eqnarray}
as well as
\begin{eqnarray}
&& \dot{\rho} = \frac{C^{}_\kappa y^2_\tau}{16\pi^2} \left [
\widehat{\zeta}^{}_{12} c^2_\nu s^2 c^{}_{(\sigma -\rho)}
s^{}_{(\sigma -\rho)} + \widehat{\zeta}^{}_{13} \left (s^2_\nu s^2 -
c^2 \right ) c^{}_\rho s^{}_\rho + \widehat{\zeta}^{}_{23} c^2_\nu
s^2 c^{}_\sigma s^{}_\sigma \right ] \; ,
\nonumber \\
&& \dot{\sigma} = \frac{C^{}_\kappa y^2_\tau}{16\pi^2} \left [
\widehat{\zeta}^{}_{12} s^2_\nu s^2 c^{}_{(\sigma -\rho)}
s^{}_{(\sigma -\rho)} + \widehat{\zeta}^{}_{13} s^2_\nu s^2
c^{}_\rho s^{}_\rho + \widehat{\zeta}^{}_{23} \left ( c^2_\nu s^2 -
c^2 \right ) c^{}_\sigma s^{}_\sigma \right ] \; ,
\nonumber \\
&& \dot{\phi} = \frac{C^{}_\kappa y^2_\tau}{16\pi^2} \left [ \left (
c^2_l -s^2_l \right ) c^{-1}_l s^{-1}_l c^{}_\nu s^{}_\nu c \left(
\zeta^{-1}_{13} c^{}_\rho s^{}_{(\rho -\phi)} - \zeta^{}_{13}
s^{}_\rho c^{}_{(\rho -\phi)} - \zeta^{-1}_{23} c^{}_\sigma
s^{}_{(\sigma -\phi)} + \zeta^{}_{23} s^{}_\sigma c^{}_{(\sigma
-\phi)} \right ) \right .
\nonumber \\
&& \left . ~~~~~~~~~~~~~~~~~ + \widehat{\zeta}^{}_{12} s^2
c^{}_{(\sigma -\rho)} s^{}_{(\sigma -\rho)} +
\widehat{\zeta}^{}_{13} \left (s^2_\nu - c^2_\nu c^2 \right )
c^{}_\rho s^{}_\rho + \widehat{\zeta}^{}_{23} \left (c^2_\nu -
s^2_\nu c^2 \right ) c^{}_\sigma s^{}_\sigma \right ] \; ,
\end{eqnarray}
where $\widehat{\zeta}^{}_{ij} \equiv \zeta^{-1}_{ij} -
\zeta^{}_{ij} = 4 \kappa^{}_i \kappa^{}_j/\left ( \kappa^2_i -
\kappa^2_j \right )$, $c^{}_a \equiv \cos a$ and $s^{}_a \equiv \sin
a$ (for $a = \rho$, $\sigma$, $\sigma -\rho$, $\rho -\phi$ or
$\sigma -\phi$).

Some discussions on the basic features of RGEs of three neutrino
masses, three flavor mixing angles and three CP-violating phases are
in order.

(a) The running behaviors of three neutrino masses $m^{}_i$ (or
equivalently $\kappa^{}_i$) are essentially identical and determined
by $\alpha^{}_\kappa$, unless $\tan\beta$ is large enough in the
MSSM to make the $y^2_\tau$-associated term competitive with the
$\alpha^{}_\kappa$ term. In our phase convention,
$\dot{\kappa}^{}_i$ or $\dot{m}^{}_i$ (for $i=1, 2, 3$) are
independent of the CP-violating phase $\phi$.

(b) Among three neutrino mixing angles, only the derivative of
$\theta^{}_\nu$ contains a term proportional to $\zeta^{-1}_{12}$.
Note that $\zeta^{-1}_{ij} = ( m^{}_i + m^{}_j )^2/\Delta m^2_{ij}$
with $\Delta m^2_{ij} \equiv m^2_i - m^2_j$ holds. Current solar and
atmospheric neutrino oscillation data yield $\Delta m^2_{21} \approx
7.7 \times 10^{-5} ~ {\rm eV}^2$ and $\left | \Delta m^2_{32} \right
| \approx \left | \Delta m^2_{31} \right | \approx 2.4 \times
10^{-3} ~ {\rm eV}^2$. So $\theta^{}_\nu$ is in general more
sensitive to radiative corrections than $\theta^{}_l$ and $\theta$.
The evolution of $\theta^{}_\nu$ can be suppressed through the
fine-tuning of $(\sigma -\rho)$. The smallest neutrino mixing angle
$\theta^{}_l$ may get radiative corrections even if its initial
value is zero, and thus it can be radiatively generated from other
neutrino mixing angles and CP-violating phases.

(c) The running behavior of $\phi$ is quite different from those of
$\rho$ and $\sigma$, because it includes a peculiar term
proportional to $s^{-1}_l$. This term, which dominates $\dot{\phi}$
when $\theta^{}_l$ is sufficiently small, becomes divergent in the
limit $\theta^{}_l \rightarrow 0$. Indeed, $\phi$ is not
well-defined if $\theta^{}_l$ is exactly vanishing. But both
$\theta^{}_l$ and $\phi$ can be radiatively generated. We may
require that $\dot{\phi}$ remain finite when $\theta^{}_l$
approaches zero, implying that the following necessary condition can
be extracted from the expression of $\dot{\phi}$ in Eq. (139):
\begin{eqnarray}
\zeta^{-1}_{13} c^{}_\rho s^{}_{(\rho -\phi)} - \zeta^{}_{13}
s^{}_\rho c^{}_{(\rho -\phi)} - \zeta^{-1}_{23} c^{}_\sigma
s^{}_{(\sigma -\phi)} + \zeta^{}_{23} s^{}_\sigma c^{}_{(\sigma
-\phi)} = 0 \; .
\end{eqnarray}
Note that the initial value of $\theta^{}_l$, if it is exactly zero
or extremely small, may immediately drive $\phi$ to its {\it
quasi-fixed point}. In this case Eq. (140) can be used to understand
the relationship between $\phi$ and two Majorana phases $\rho$ and
$\sigma$ at the quasi-fixed point.

(d) The running behaviors of $\rho$ and $\sigma$ are relatively mild
in comparison with that of $\phi$. A remarkable feature of
$\dot{\rho}$ and $\dot{\sigma}$ is that they will vanish, if both
$\rho$ and $\sigma$ are initially vanishing. This observation
indicates that $\rho$ and $\sigma$ cannot simultaneously be
generated from $\phi$ via the RGEs.

\section{How to Generate Neutrino Masses?}

Neutrinos are assumed or required to be massless in the SM, just
because the structure of the SM itself is too simple to accommodate
massive neutrinos.
\begin{itemize}
\item     Two fundamentals of the SM are the $SU(2)^{}_{\rm L} \times
U(1)^{}_{\rm Y}$ gauge symmetry and the Lorentz invariance. Both of
them are mandatory to guarantee that the SM is a consistent quantum
field theory.

\item     The particle content of the SM is rather economical. There
are no right-handed neutrinos in the SM, so a Dirac neutrino mass
term is not allowed. There is only one Higgs doublet, so a
gauge-invariant Majorana mass term is forbidden.

\item     The SM is a renormalizable quantum field theory. Hence
an effective dimension-5 operator, which may give each neutrino a
Majorana mass, is absent.
\end{itemize}
In other words, the SM accidently possesses the $(B-L)$ symmetry
which assures three known neutrinos to be exactly massless.

But today's experiments have convincingly indicated the existence of
neutrino oscillations. This quantum phenomenon can appear if and
only if neutrinos are massive and lepton flavors are mixed, and thus
it is a kind of new physics beyond the SM. To generate non-zero but
tiny neutrino masses, one or more of the above-mentioned constraints
on the SM must be abandoned or relaxed. It is intolerable to abandon
the gauge symmetry and Lorentz invariance; otherwise, one would be
led astray. Given the framework of the SM as a consistent field
theory, its particle content can be modified and (or) its
renormalizability can be abandoned to accommodate massive neutrinos.
There are several ways to this goal.

\subsection{Relaxing the renormalizability}

In 1979, Weinberg extended the SM by introducing some
higher-dimension operators in terms of the fields of the SM itself:
\begin{equation}
{\cal L}^{}_{\rm eff} = {\cal L}^{}_{\rm SM} + \frac{{\cal
L}^{}_{\rm d=5}}{\Lambda} + \frac{{\cal L}^{}_{\rm d=6}}{\Lambda^2}
+ \cdots \; ,
\end{equation}
where $\Lambda$ denotes the cut-off scale of this effective theory.
Within such a framework, the lowest-dimension operator that violates
the lepton number ($L$) is the unique dimension-5 operator
$HHLL/\Lambda$. After spontaneous gauge symmetry breaking, this
Weinberg operator yields $m^{}_i \sim \langle H\rangle^2/\Lambda$
for neutrino masses, which can be sufficiently small ($\leq 1$ eV)
if $\Lambda$ is not far away from the scale of grand unified
theories ($\Lambda \sim 10^{13}$ GeV for $\langle H\rangle \sim
10^2$ GeV). In this sense we argue that neutrino masses can serve as
a low-energy window onto new physics at superhigh energies.

\subsection{A pure Dirac neutrino mass term?}

Given three right-handed neutrinos, the gauge-invariant and
lepton-number-conserving mass terms of charged leptons and neutrinos
are
\begin{equation}
-{\cal L}^{}_{\rm lepton} = \overline{\ell^{}_{\rm L}} Y^{}_l H
E^{}_{\rm R} + \overline{\ell^{}_{\rm L}} Y^{}_\nu \tilde{H}
N^{}_{\rm R} + {\rm h.c.} \; ,
\end{equation}
where $\tilde{H} \equiv i\sigma^{~}_2 H^*$ is defined and $\ell_{\rm
L}$ denotes the left-handed lepton doublet. After spontaneous gauge
symmetry breaking, we arrive at the charged-lepton mass matrix
$M^{}_l = Y^{}_l v/\sqrt{2}$ and the Dirac neutrino mass matrix
$M^{}_\nu = Y^{}_\nu v/\sqrt{2}$ with $v \simeq 246 ~ {\rm GeV}$. In
this case, the smallness of three neutrino masses $m^{}_i$ (for
$i=1,2,3$) is attributed to the smallness of three eigenvalues of
$Y^{}_\nu$ (denoted as $y^{}_i$ for $i=1,2,3$). Then we encounter a
transparent hierarchy problem: $y^{}_i/y^{}_e = m^{}_i/m^{}_e \leq
0.5 ~{\rm eV}/0.5 ~{\rm MeV} \sim 10^{-6}$. Why is $y^{}_i$ so
small? There is no explanation at all in this Dirac-mass picture.

A speculative way out is to invoke extra dimensions; namely, the
smallness of Dirac neutrino masses is ascribed to the assumption
that three right-handed neutrinos have access to one or more extra
spatial dimensions. The idea is simply to confine the SM particles
onto a brane and to allow $N^{}_{\rm R}$ to travel in the bulk. For
example, the wave-function of $N^{}_{\rm R}$ spreads out over the
extra dimension $y$, giving rise to a suppressed Yukawa interaction
at $y = 0$ (i.e., the location of the brane):
\begin{equation}
\left[ \overline{\ell^{}_{\rm L}} Y^{}_\nu \tilde{H} N^{}_{\rm R}
\right]^{}_{y=0} \sim \frac{1}{\sqrt{L}} \left[
\overline{\ell^{}_{\rm L}} Y^{}_\nu \tilde{H} N^{}_{\rm R}
\right]^{}_{y=L} \; .
\end{equation}
The magnitude of $1/\sqrt{L}$ is measured by
$\Lambda/\Lambda^{}_{\rm Planck}$, and thus it can naturally be
small for an effective theory far below the Planck scale.

\subsection{Seesaw mechanisms}

This approach works at the tree level and reflects the essential
spirit of seesaw mechanisms --- tiny masses of three known neutrinos
are attributed to the existence of heavy degrees of freedom and
lepton number violation.
\begin{itemize}
\item     Type-I seesaw --- three heavy right-handed neutrinos are
added into the SM and the lepton number is violated by their
Majorana mass term:
\begin{eqnarray}
-{\cal L}^{}_{\rm lepton} = \overline{\ell^{}_{\rm L}} Y^{}_l H
E^{}_{\rm R} + \overline{\ell^{}_{\rm L}} Y^{}_\nu \tilde{H}
N^{}_{\rm R} + \frac{1}{2} \overline{N^{\rm c}_{\rm R}} M^{}_{\rm R}
N^{}_{\rm R} + {\rm h.c.} \; ,
\end{eqnarray}
where $M^{}_{\rm R}$ is the Majorana mass matrix.

\item     Type-II seesaw --- one heavy Higgs triplet is added into the
SM and the lepton number is violated by its interactions with both
the lepton doublet and the Higgs doublet:
\begin{eqnarray}
-{\cal L}^{}_{\rm lepton} = \overline{\ell^{}_{\rm L}} Y^{}_l H
E^{}_{\rm R} + \frac{1}{2} \overline{\ell^{}_{\rm L}} Y^{}_\Delta
\Delta i\sigma^{}_2 \ell^{\rm c}_{\rm L} - \lambda^{}_\Delta
M^{}_\Delta H^T i\sigma^{}_2 \Delta H + {\rm h.c.} \; ,
\end{eqnarray}
where
\begin{equation} \Delta \equiv \left(\begin{matrix}
\Delta^- & -\sqrt{2} ~ \Delta^0 \cr \sqrt{2} ~ \Delta^{--} &
-\Delta^- \cr \end{matrix} \right)
\end{equation}
denotes the $SU(2)^{}_{\rm L}$ Higgs triplet.

\item     Type-III seesaw --- three heavy triplet fermions are
added into the SM and the lepton number is violated by their
Majorana mass term:
\begin{eqnarray}
-{\cal L}^{}_{\rm lepton} = \overline{\ell^{}_{\rm L}} Y^{}_l H
E^{}_{\rm R} + \overline{\ell^{}_{\rm L}} \sqrt{2} Y^{}_\Sigma
\Sigma^{\rm c} \tilde{H} + \frac{1}{2} {\rm Tr} \left(
\overline{\Sigma} M^{}_\Sigma \Sigma^{\rm c} \right) + {\rm h.c.} \;
,
\end{eqnarray}
where
\begin{equation}
\Sigma = \left( \begin{matrix} \Sigma^0/\sqrt{2} & \Sigma^+ \cr
\Sigma^- & -\Sigma^0/\sqrt{2} \cr \end{matrix} \right)
\end{equation}
denotes the $SU(2)^{}_{\rm L}$ fermion triplet.
\end{itemize}
Of course, there are a number of variations or combinations of these
three typical seesaw mechanisms in the literature.

For each of the above seesaw pictures, one may arrive at the unique
dimension-5 Weinberg operator of neutrino masses after integrating
out the corresponding heavy degrees of freedom:
$$
\frac{{\cal L}^{}_{\rm d=5}}{\Lambda} = \left\{
\begin{array}{lcl}
\displaystyle \frac{1}{2} \left(Y^{}_\nu M^{-1}_{\rm R}
Y^T_\nu\right)^{}_{\alpha\beta} \overline{\ell^{}_{\alpha \rm L}}
\tilde{H} \tilde{H}^T \ell^{\rm c}_{\beta \rm L} + {\rm h.c.} &&
\vspace{0.15cm} \\
\displaystyle -\frac{\lambda^{}_\Delta}{M^{}_\Delta}
\left(Y^{}_\Delta\right)^{}_{\alpha\beta} \overline{\ell^{}_{\alpha
\rm L}} \tilde{H} \tilde{H}^T \ell^{\rm c}_{\beta \rm L} + {\rm
h.c.} &&
\vspace{0.15cm} \\
\displaystyle \frac{1}{2} \left(Y^{}_\Sigma M^{-1}_{\rm \Sigma}
Y^T_\Sigma\right)^{}_{\alpha\beta} \overline{\ell^{}_{\alpha \rm L}}
\tilde{H} \tilde{H}^T \ell^{\rm c}_{\beta \rm L} + {\rm h.c.} &&
\end{array}
\right .
$$
corresponding to type-I, type-II and type-III seesaws. After
spontaneous gauge symmetry breaking, $\tilde{H}$ achieves its vacuum
expectation value $\langle \tilde{H}\rangle = v/\sqrt{2}$ with $v
\simeq 246$ GeV. Then we are left with the effective Majorana
neutrino mass term for three known neutrinos,
\begin{equation}
-{\cal L}^{}_{\rm mass} = \frac{1}{2} \overline{\nu^{}_{\rm L}}
M^{}_\nu \nu^{\rm c}_{\rm L} + {\rm h.c.} \; ,
\end{equation}
where the Majorana mass matrix $M^{}_\nu$ is given by
\begin{equation}
M^{}_\nu = \left\{
\begin{array}{lcl}
\displaystyle -\frac{1}{2} Y^{}_\nu \frac{v^2}{M^{}_{\rm R}} Y^T_\nu
&& ({\rm Type ~I}) \; , \vspace{0.15cm} \\
\displaystyle \lambda^{}_\Delta Y^{}_\Delta \frac{v^2}{M^{}_\Delta }
&& ({\rm Type
~II}) \; , \vspace{0.15cm} \\
\displaystyle -\frac{1}{2} Y^{}_\Sigma \frac{v^2}{M^{}_\Sigma}
Y^T_\Sigma && ({\rm Type ~III}) \; .
\end{array}
\right .
\end{equation}
It becomes obvious that the smallness of $M^{}_\nu$ can be
attributed to the largeness of $M^{}_{\rm R}$, $M^{}_\Delta$ or
$M^{}_\Sigma$ in the seesaw mechanism.

\subsection{Radiative origin of neutrino masses}

In a seminal paper published in 1972, Weinberg pointed out that ``in
theories with spontaneously broken gauge symmetries, various masses
or mass differences may vanish in zeroth order as a consequence of
the representation content of the fields appearing in the
Lagrangian. These masses or mass differences can then be calculated
as finite higher-order effects." Such a mechanism may allow us to
slightly go beyond the SM and radiatively generate tiny neutrino
masses. A typical example is the well-known Zee model,
\begin{eqnarray}
-{\cal L}^{}_{\rm lepton} = \overline{\ell^{}_{\rm L}} Y^{}_l H
E^{}_{\rm R} + \overline{\ell^{}_{\rm L}} Y^{}_S S^- i\sigma^{}_2
l^{\rm c}_{\rm L} + \tilde{\Phi}^T F S^+ i\sigma^{}_2 \tilde{H} +
{\rm h.c.} \; ,
\end{eqnarray}
where $S^{\pm}$ are charged $SU(2)^{}_{\rm L}$ singlet scalars,
$\Phi$ denotes a new $SU(2)^{}_{\rm L}$ doublet scalar which has the
same quantum number as the SM Higgs doublet $H$, $Y^{}_S$ is an
anti-symmetric matrix, and $F$ represents a mass. Without loss of
generality, we choose the basis of $M^{}_l = Y^{}_l \langle H\rangle
= {\rm Diag}\{m^{}_e, m^{}_\mu, m^{}_\tau \}$. In this model
neutrinos are massless at the tree level, but their masses can
radiatively be generated via the one-loop corrections. Given $M^{}_S
\gg M^{}_H \sim M^{}_\Phi \sim F$ and $\langle \Phi\rangle \sim
\langle H\rangle$, the elements of the effective mass matrix of
three light Majorana neutrinos are
\begin{equation}
\left(M^{}_\nu\right)^{}_{\alpha\beta} \; \sim \;
\frac{M^{}_H}{16\pi^2} \cdot \frac{m^2_\alpha - m^2_\beta}{M^2_S}
\left(Y^{}_S\right)^{}_{\alpha\beta} \; ,
\end{equation}
where $\alpha$ and $\beta$ run over $e$, $\mu$ and $\tau$. The
smallness of $M^{}_\nu$ is therefore ascribed to the smallness of
$Y^{}_S$ and $(m^2_\alpha - m^2_\beta)/M^2_S$. Although the original
version of the Zee model is disfavored by current experimental data
on neutrino oscillations, its extensions or variations at the
one-loop or two-loop level can survive.

\section{On the Scales of Seesaw Mechanisms}

As we have seen, the key point of a seesaw mechanism is to ascribe
the smallness of neutrino masses to the existence of some new
degrees of freedom heavier than the Fermi scale $v \simeq 246$ GeV,
such as heavy Majorana neutrinos or heavy Higgs bosons. The energy
scale where a seesaw mechanism works is crucial, because it is
relevant to whether this mechanism is theoretically natural and
experimentally testable. Between Fermi and Planck scales, there
might exist two other fundamental scales: one is the scale of a
grand unified theory (GUT) at which strong, weak and electromagnetic
forces can be unified, and the other is the TeV scale at which the
unnatural gauge hierarchy problem of the SM can be solved or at
least softened by a kind of new physics.

\subsection{How about a very low seesaw scale?}

In reality, however, there is no direct evidence for a high or
extremely high seesaw scale. Hence eV-, keV-, MeV- and GeV-scale
seesaws are all possible, at least in principle, and they are
technically natural in the sense that their lepton-number-violating
mass terms are naturally small according to 't Hooft's naturalness
criterion
--- ``At any energy scale $\mu$, a set of parameters $\alpha^{}_i
(\mu)$ describing a system can be small, if and only if, in the
limit $\alpha^{}_i (\mu) \rightarrow 0$ for each of these
parameters, the system exhibits an enhanced symmetry." But there are
several potential problems associated with low-scale seesaws: (a) a
low-scale seesaw does not give any obvious connection to a
theoretically well-justified fundamental physical scale (such as the
Fermi scale, the TeV scale, the GUT scale or the Planck scale); (b)
the neutrino Yukawa couplings in a low-scale seesaw model turn out
to be tiny, giving no actual explanation of why the masses of three
known neutrinos are so small; and (c) in general, a very low seesaw
scale does not allow the ``canonical" thermal leptogenesis mechanism
to work.

\subsection{Seesaw-induced hierarchy problem}

Many theorists argue that the conventional seesaw scenarios are
natural because their scales (i.e., the masses of heavy degrees of
freedom) are close to the GUT scale. This argument is reasonable on
the one hand, but it reflects the drawbacks of the conventional
seesaw models on the other hand. In other words, the conventional
seesaw models have no direct experimental testability and involve a
potential hierarchy problem. The latter is usually spoke of when two
largely different energy scales exist in a model, but there is no
symmetry to stabilize the low-scale physics suffering from large
corrections coming from the high-scale physics.

Such a seesaw-induced fine-tuning problem means that the SM Higgs
mass is very sensitive to quantum corrections from the heavy degrees
of freedom in a seesaw mechanism. For example,
$$
\delta M^2_H = \left\{
\begin{array}{lcl}
\displaystyle -\frac{y^2_i}{8\pi^2} \left(\Lambda^2 + M^2_i
\ln\frac{M^2_i}{\Lambda^2} \right) &\hspace{-0.3cm}& ({\rm I})
\vspace{0.2cm} \\
\displaystyle \frac{3}{16\pi^2} \left[ \lambda^{}_3 \left(\Lambda^2
+ M^2_\Delta \ln\frac{M^2_\Delta}{\Lambda^2}\right) + 4
\lambda^2_\Delta M^2_\Delta \ln\frac{M^2_\Delta}{\Lambda^2} \right]
&\hspace{-0.28cm}& ({\rm II})
\vspace{0.2cm} \\
\displaystyle -\frac{3y^2_i}{8\pi^2} \left(\Lambda^2 + M^2_i
\ln\frac{M^2_i}{\Lambda^2} \right) &\hspace{-0.3cm}& ({\rm III})
\end{array}
\right .
$$
in three typical seesaw scenarios, where $\Lambda$ is the regulator
cut-off, $y^{}_i$ and $M^{}_i$ (for $i=1,2,3$) stand respectively
for the eigenvalues of $Y^{}_\nu$ (or $Y^{}_\Sigma$) and $M^{}_{\rm
R}$ (or $M^{}_\Sigma$), and the contributions proportional to $v^2$
and $M^2_H$ have been omitted. The above results show a quadratic
sensitivity to the new scale which is characteristic of the seesaw
model, implying that a high degree of fine-tuning would be necessary
to accommodate the experimental data on $M^{}_H$ if the seesaw scale
is much larger than $v$ (or the Yukawa couplings are not extremely
fine-tuned in type-I and type-III seesaws). Taking the type-I seesaw
scenario for illustration, we assume $\Lambda \sim M^{}_i$ and
require $|\delta M^2_H| \leq 0.1 ~{\rm TeV}^2$. Then the above
equation leads us to the following rough estimate:
\begin{eqnarray}
M^{}_i \sim \left[\frac{(2\pi v)^2 |\delta
M^2_H|}{m^{}_i}\right]^{1/3} \leq 10^7 {\rm GeV} \left[\frac{0.2 ~
{\rm eV}}{m^{}_i}\right]^{1/3} \left[\frac{|\delta M^2_H|}{0.1 ~
{\rm TeV}^2}\right]^{1/3} \; .
\end{eqnarray}
This naive result indicates that a hierarchy problem will arise if
the masses of heavy Majorana neutrinos are larger than about $10^7$
GeV in the type-I seesaw scheme. Because of $m^{}_i \sim y^2_i
v^2/(2M^{}_i)$, the bound $M^{}_i \leq 10^7$ GeV implies $y^{}_i
\sim \sqrt{2m^{}_i M^{}_i}/v \leq 2.6 \times 10^{-4}$ for $m^{}_i
\sim 0.2$ eV. Such a small magnitude of $y^{}_i$ seems to be a bit
unnatural in the sense that the conventional seesaw idea attributes
the smallness of $m^{}_i$ to the largeness of $M^{}_i$ other than
the smallness of $y^{}_i$.

There are two possible ways out of this impasse: one is to appeal
for the supersymmetry, and the other is to lower the seesaw scale.
We shall follow the second way to discuss the TeV seesaw mechanisms
which do not suffer from the above-mentioned hierarchy problem.

\subsection{Why are the TeV seesaws interesting?}

There are several reasons for people to expect some new physics at
the TeV scale. This kind of new physics should be able to stabilize
the Higgs-boson mass and hence the electroweak scale; in other
words, it should be able to solve or soften the unnatural gauge
hierarchy problem. It has also been argued that the
weakly-interacting particle candidates for dark matter should weigh
about one TeV or less. If the TeV scale is really a fundamental
scale, may we argue that the TeV seesaws are natural? Indeed, we are
reasonably motivated to speculate that possible new physics existing
at the TeV scale and responsible for the electroweak symmetry
breaking might also be responsible for the origin of neutrino
masses. It is interesting and meaningful in this sense to
investigate and balance the ``naturalness" and ``testability" of TeV
seesaws at the energy frontier set by the LHC.

As a big bonus of the conventional (type-I) seesaw mechanism, the
thermal leptogenesis mechanism provides us with an elegant dynamic
picture to interpret the cosmological matter-antimatter asymmetry
characterized by the observed ratio of baryon number density to
photon number density, $\eta^{}_B \equiv n^{}_B/n^{}_\gamma = (6.1
\pm 0.2) \times 10^{10}$. When heavy Majorana neutrino masses are
down to the TeV scale, the Yukawa couplings should be reduced by
more than six orders of magnitude so as to generate tiny masses for
three known neutrinos via the type-I seesaw and satisfy the
out-of-equilibrium condition, but the CP-violating asymmetries of
heavy Majorana neutrino decays can still be enhanced by the resonant
effects in order to account for $\eta^{}_B$. This ``resonant
leptogenesis" scenario might work in a specific TeV seesaw model.

Is there a TeV Noah's Ark which can naturally and simultaneously
accommodate the seesaw idea, the leptogenesis picture and the
collider signatures? We are most likely not so lucky and should not
be too ambitious at present. In the following we shall concentrate
on the TeV seesaws themselves and their possible collider signatures
and low-energy consequences.

\section{TeV Seesaws: Natural and Testable?}

The neutrino mass terms in three typical seesaw mechanisms have been
given before. Without loss of generality, we choose the basis in
which the mass eigenstates of three charged leptons are identified
with their flavor eigenstates.

\subsection{Type-I seesaw}

Given $M^{}_{\rm D} = Y^{}_\nu v/\sqrt{2}~$, the approximate type-I
seesaw formula in Eq. (150) can be rewritten as $M^{}_\nu = -
M^{}_{\rm D} M^{-1}_{\rm R} M^T_{\rm D}$. Note that the $3\times 3$
light neutrino mixing matrix $V$ is not exactly unitary in this
seesaw scheme, and its deviation from unitarity is of ${\cal
O}(M^2_{\rm D}/M^2_{\rm R})$. Let us consider two interesting
possibilities. (1) $M^{}_{\rm D} \sim {\cal O}(10^2)$ GeV and
$M^{}_{\rm R} \sim {\cal O}(10^{15})$ GeV to get $M^{}_\nu \sim
{\cal O}(10^{-2})$ eV. In this conventional and {\it natural} case,
$M^{}_{\rm D}/M^{}_{\rm R} \sim {\cal O}(10^{-13})$ holds. Hence the
non-unitarity of $V$ is only at the ${\cal O}(10^{-26})$ level, too
small to be observed. (2) $M^{}_{\rm D} \sim {\cal O}(10^2)$ GeV and
$M^{}_{\rm R} \sim {\cal O}(10^{3})$ GeV to get $M^{}_\nu \sim {\cal
O}(10^{-2})$ eV. In this {\it unnatural} case, a significant
``structural cancellation" has to be imposed on the textures of
$M^{}_{\rm D}$ and $M^{}_{\rm R}$. Because of $M^{}_{\rm
D}/M^{}_{\rm R} \sim {\cal O}(0.1)$, the non-unitarity of $V$ can
reach the percent level and may lead to observable effects.

Now we discuss how to realize the above ``structural cancellation"
for the type-I seesaw mechanism at the TeV scale. For the sake of
simplicity, we take the basis of $M^{}_{\rm R} = {\rm Diag}\{M^{}_1,
M^{}_2, M^{}_3\}$ for three heavy Majorana neutrinos ($N^{}_1,
N^{}_2, N^{}_3$). It is well known that $M^{}_\nu$ vanishes if
\begin{eqnarray}
M^{}_{\rm D} = m \left( \begin{matrix} y^{}_1 & y^{}_2 & y^{}_3 \cr
\alpha y^{}_1 & \alpha y^{}_2 & \alpha y^{}_3 \cr \beta y^{}_1 &
\beta y^{}_2 & \beta y^{}_3 \cr \end{matrix} \right) \; , ~~~~
\sum^3_{i=1} \frac{y^2_i}{M^{}_i} = 0
\end{eqnarray}
simultaneously hold. Tiny neutrino masses can be generated from tiny
corrections to the texture of $M^{}_{\rm D}$ in Eq. (154). For
example, $M^\prime_{\rm D} = M^{}_{\rm D} - \epsilon X^{}_{\rm D}$
with $M^{}_{\rm D}$ given above and $\epsilon$ being a small
dimensionless parameter (i.e., $|\epsilon| \ll 1$) yields
\begin{eqnarray}
M^\prime_\nu = -M^\prime_{\rm D} M^{-1}_{\rm R} M^{\prime T}_{\rm D}
\simeq \epsilon \left( M^{}_{\rm D} M^{-1}_{\rm R} X^T_{\rm D} +
X^{}_{\rm D} M^{-1}_{\rm R} M^T_{\rm D} \right) \; ,
\end{eqnarray}
from which $M^\prime_\nu \sim {\cal O}(10^{-2})$ eV can be obtained
by adjusting the size of $\epsilon$.

A lot of attention has recently been paid to a viable type-I seesaw
model and its collider signatures at the TeV scale. At least the
following lessons can be learnt:
\begin{itemize}
\item     Two necessary conditions must be satisfied in order to
test a type-I seesaw model at the LHC: (a) $M^{}_i$ are of ${\cal
O}(1)$ TeV or smaller; and (b) the strength of light-heavy neutrino
mixing (i.e., $M^{}_{\rm D}/M^{}_{\rm R}$) is large enough.
Otherwise, it would be impossible to produce and detect $N^{}_i$ at
the LHC.

\item     The collider signatures of $N^{}_i$ are essentially
decoupled from the mass and mixing parameters of three light
neutrinos $\nu^{}_i$. For instance, the small parameter $\epsilon$
in Eq. (155) has nothing to do with the ratio $M^{}_{\rm
D}/M^{}_{\rm R}$.

\item     The non-unitarity of $V$ might lead to some
observable effects in neutrino oscillations and other
lepton-flavor-violating or lepton-number-violating processes, if
$M^{}_{\rm D}/M^{}_{\rm R} \leq {\cal O}(0.1)$ holds.

\item     The clean LHC signatures of heavy Majorana neutrinos are
the $\Delta L =2$ like-sign dilepton events, such as $pp \to
W^{*\pm} W^{*\pm} \to \mu^\pm \mu^\pm jj$ and $pp \to W^{*\pm} \to
\mu^\pm N_i \to \mu^\pm \mu^\pm jj$ (a dominant channel due to the
resonant production of $N^{}_i$).
\end{itemize}
Some instructive and comprehensive analyses of possible LHC events
for a single heavy Majorana neutrino have recently been done, but
they only serve for illustration because such a simplified type-I
seesaw scenario is actually unrealistic.

\subsection{Type-II seesaw}

The type-II seesaw formula $M^{}_\nu = Y^{}_\Delta v^{}_\Delta =
\lambda^{}_\Delta Y^{}_\Delta v^2/M^{}_\Delta$ has been given in Eq.
(150). Note that the last term of Eq. (145) violates both $L$ and
$B-L$, and thus the smallness of $\lambda^{}_\Delta$ is naturally
allowed according to 't Hooft's naturalness criterion (i.e., setting
$\lambda^{}_\Delta =0$ will increase the symmetry of ${\cal
L}^{}_{\rm lepton}$). Given $M^{}_\Delta \sim {\cal O}(1)$ TeV, for
example, this seesaw mechanism works to generate $M^{}_\nu \sim
{\cal O}(10^{-2})$ eV provided $\lambda^{}_\Delta Y^{}_\Delta \sim
{\cal O}(10^{-12})$ holds. The neutrino mixing matrix $V$ is exactly
unitary in the type-II seesaw mechanism, simply because the heavy
degrees of freedom do not mix with the light ones.

There are totally seven physical Higgs bosons in the type-II seesaw
scheme: doubly-charged $H^{++}$ and $H^{--}$, singly-charged $H^+$
and $H^-$, neutral $A^0$ (CP-odd), and neutral $h^0$ and $H^0$
(CP-even), where $h^0$ is the SM-like Higgs boson. Except for
$M^2_{h^0}$, we get a quasi-degenerate mass spectrum for other
scalars: $M^2_{H^{\pm \pm}} = M^2_\Delta \approx M^2_{H^0} \approx
M^2_{H^\pm} \approx M^2_{A^0}$. As a consequence, the decay channels
$H^{\pm \pm} \to W^\pm H^\pm$ and $H^{\pm \pm} \to H^\pm H^\pm$ are
kinematically forbidden. The production of $H^{\pm\pm}$ at the LHC
is mainly through $q\bar{q} \to \gamma^*, Z^* \to H^{++}H^{--}$ and
$q\bar{q}^\prime \to W^* \to H^{\pm\pm}H^\mp$ processes, which do
not rely on the small Yukawa couplings.

The typical collider signatures in this seesaw scenario are the
lepton-number-violating $H^{\pm\pm} \to l^\pm_\alpha l^\pm_\beta$
decays as well as $H^+ \to l^+_\alpha \overline{\nu}$ and $H^- \to
l^-_\alpha \nu$ decays. Their branching ratios
\begin{eqnarray}
&& {\cal B}(H^{\pm\pm} \to l^\pm_\alpha l^\pm_\beta) = \frac{
|(M^{}_\nu)^{}_{\alpha\beta}|^2 \left(2 - \delta^{}_{\alpha\beta}
\right)}{\displaystyle \sum_{\rho,\sigma}
|(M^{}_\nu)^{}_{\rho\sigma}|^2} \; ,
\nonumber \\
&& {\cal B}(H^+ \to l^+_\alpha \overline{\nu}) = \frac{\displaystyle
\sum_\beta |(M^{}_\nu)^{}_{\alpha\beta}|^2}{\displaystyle
\sum_{\rho,\sigma} |(M^{}_\nu)^{}_{\rho\sigma}|^2} \;
\end{eqnarray}
are closely related to the masses, flavor mixing angles and
CP-violating phases of three light neutrinos, because $M^{}_\nu = V
\widehat{M}^{}_\nu V^T$ with $\widehat{M}^{}_\nu = {\rm
Diag}\{m^{}_1, m^{}_2, m^{}_3\}$ holds. Some detailed analyses of
such decay modes together with the LHC signatures of $H^{\pm\pm}$
and $H^{\pm}$ bosons have been done in the literature.

It is worth pointing out that the following dimension-6 operator can
easily be derived from the type-II seesaw mechanism,
\begin{eqnarray}
\frac{{\cal L}^{}_{\rm d=6}}{\Lambda^2} =
-\frac{\left(Y^{}_\Delta\right)^{}_{\alpha\beta}
\left(Y^{}_\Delta\right)^\dagger_{\rho\sigma}}{4 M^2_\Delta}
(\overline{\ell^{}_{\alpha \rm L}} \gamma^\mu \ell^{}_{\sigma \rm
L}) (\overline{\ell^{}_{\beta \rm L}} \gamma^{}_\mu \ell^{}_{\rho
\rm L}) \; , \;\;
\end{eqnarray}
which has two immediate low-energy effects: the non-standard
interactions of neutrinos and the lepton-flavor-violating
interactions of charged leptons. An analysis of such effects
provides us with some preliminary information:
\begin{itemize}
\item     The magnitudes of non-standard interactions of
neutrinos and the widths of lepton-flavor-violating tree-level
decays of charged leptons are both dependent on neutrino masses
$m^{}_i$ and flavor-mixing and CP-violating parameters of $V$.

\item     For a long-baseline neutrino oscillation experiment, the
neutrino beam encounters the earth matter and the electron-type
non-standard interaction contributes to the matter potential.

\item     At a neutrino factory, the lepton-flavor-violating
processes $\mu^-\rightarrow e^-\nu^{}_e\overline{\nu}^{}_\mu$ and
$\mu^+\rightarrow e^+\overline{\nu}_e\nu^{}_\mu$ could cause some
wrong-sign muons at a near detector.
\end{itemize}
Current experimental constraints tell us that such low-energy
effects are very small, but they might be experimentally accessible
in the future precision measurements.

\subsection{Type-(I+II) seesaw}

The type-(I+II) seesaw mechanism can be achieved by combining the
neutrino mass terms in Eqs. (144) and (145). After spontaneous gauge
symmetry breaking, we are left with the overall neutrino mass term
\begin{eqnarray}
-{\cal L}^{}_{\rm mass} = \frac{1}{2} ~ \overline{\left( \nu^{}_{\rm
L} N^{\rm c}_{\rm R}\right)} \left( \begin{matrix} M^{}_{\rm L} &
M^{}_{\rm D} \cr M^T_{\rm D} & M^{}_{\rm R} \end{matrix} \right)
\left(
\begin{matrix} \nu^{\rm c}_{\rm L} \cr N^{}_{\rm R} \end{matrix}
\right) + {\rm h.c.} \; ,
\end{eqnarray}
where $M^{}_{\rm D} = Y^{}_\nu v/\sqrt{2}$ and $M^{}_{\rm L} =
Y^{}_\Delta v^{}_\Delta$ with $\langle H \rangle \equiv v/\sqrt{2}$
and $\langle \Delta \rangle \equiv v^{}_\Delta$ corresponding to the
vacuum expectation values of the neutral components of the Higgs
doublet $H$ and the Higgs triplet $\Delta$. The $6\times 6$ mass
matrix in Eq. (158) is symmetric and can be diagonalized by the
unitary transformation done in Eq. (28); i.e.,
\begin{eqnarray}
\left( \begin{matrix} V & R \cr S & U \end{matrix} \right)^\dagger
\left( \begin{matrix} M^{}_{\rm L} & M^{}_{\rm D} \cr M^T_{\rm D} &
M^{}_{\rm R} \end{matrix} \right) \left(\begin{matrix} V & R \cr S &
U \end{matrix} \right)^* = \left( \begin{matrix} \widehat{M}^{}_\nu
& {\bf 0} \cr {\bf 0} & \widehat{M}^{}_N
\end{matrix} \right) \; ,
\end{eqnarray}
where $\widehat{M}^{}_\nu = {\rm Diag}\{m^{}_1, m^{}_2, m^{}_3\}$
and $\widehat{M}^{}_N = {\rm Diag}\{M^{}_1, M^{}_2, M^{}_3\}$.
Needless to say, $V^\dagger V + S^\dagger S = VV^\dagger +
RR^\dagger = {\bf 1}$ holds as a consequence of the unitarity of
this transformation. Hence $V$, the flavor mixing matrix of light
Majorana neutrinos, must be non-unitary if $R$ and $S$ are non-zero.

In the leading-order approximation, the type-(I+II) seesaw formula
reads as
\begin{equation}
M^{}_\nu \approx M^{}_{\rm L} - M^{}_{\rm D} M^{-1}_{\rm R} M^T_{\rm
D} \; .
\end{equation}
Hence type-I and type-II seesaws can be regarded as two extreme
cases of the type-(I+II) seesaw. Note that two mass terms in Eq.
(160) are possibly comparable in magnitude. If both of them are
small, their contributions to $M^{}_\nu$ may have significant
interference effects which make it practically impossible to
distinguish between type-II and type-(I+II) seesaws; but if both of
them are large, their contributions to $M^{}_\nu$ must be
destructive. The latter case unnaturally requires a significant
cancellation between two big quantities in order to obtain a small
quantity, but it is interesting in the sense that it may give rise
to possibly observable collider signatures of heavy Majorana
neutrinos.

Let me briefly describe a particular type-(I+II) seesaw model and
comment on its possible LHC signatures. First, we assume that both
$M^{}_i$ and $M^{}_\Delta$ are of ${\cal O}(1)$ TeV. Then the
production of $H^{\pm\pm}$ and $H^\pm$ bosons at the LHC is
guaranteed, and their lepton-number-violating signatures will probe
the Higgs triplet sector of the type-(I+II) seesaw mechanism. On the
other hand, ${\cal O}(M^{}_{\rm D}/M^{}_{\rm R}) \leq {\cal O}(0.1)$
is possible as a result of ${\cal O}( M^{}_{\rm R}) \sim {\cal
O}(1)$ TeV and ${\cal O}(M^{}_{\rm D}) \leq {\cal O}(v)$, such that
appreciable signatures of $N^{}_i$ can be achieved at the LHC.
Second, the small mass scale of $M^{}_\nu$ implies that the relation
${\cal O}(M^{}_{\rm L}) \sim {\cal O}(M^{}_{\rm D} M^{-1}_{\rm R}
M^T_{\rm D})$ must hold. In other words, it is the significant but
incomplete cancellation between $M^{}_{\rm L}$ and $M^{}_{\rm D}
M^{-1}_{\rm R} M^T_{\rm D}$ terms that results in the non-vanishing
but tiny masses for three light neutrinos. We admit that dangerous
radiative corrections to two mass terms of $M^{}_\nu$ require a
delicate fine-tuning of the cancellation at the loop level. But this
scenario allows us to reconstruct $M^{}_{\rm L}$ via the excellent
approximation $M^{}_{\rm L} = V \widehat{M}^{}_\nu V^T + R
\widehat{M}^{}_N R^T \approx R \widehat{M}^{}_N R^T$, such that the
elements of the Yukawa coupling matrix $Y^{}_\Delta$ read as
follows:
\begin{equation}
\left(Y^{}_\Delta\right)^{}_{\alpha \beta} \; = \;
\frac{\left(M^{}_{\rm L}\right)^{}_{\alpha \beta}}{v^{}_\Delta}
\approx \sum^3_{i=1} \frac{R^{}_{\alpha i} R^{}_{\beta i}
M^{}_i}{v^{}_\Delta} \; ,
\end{equation}
where the subscripts $\alpha$ and $\beta$ run over $e$, $\mu$ and
$\tau$. This result implies that the leptonic decays of $H^{\pm
\pm}$ and $H^\pm$ bosons depend on both $R$ and $M^{}_i$, which
actually determine the production and decays of $N^{}_i$. Thus we
have established an interesting correlation between the singly- or
doubly-charged Higgs bosons and the heavy Majorana neutrinos. To
observe the correlative signatures of $H^\pm$, $H^{\pm\pm}$ and
$N^{}_i$ at the LHC will serve for a direct test of this type-(I+II)
seesaw model.

\subsection{Type-III seesaw}

The lepton mass terms in the type-III seesaw scheme have already
been given in Eq. (147). After spontaneous gauge symmetry breaking,
we are left with
\begin{eqnarray}
&& -{\cal L}^{}_{\rm mass} = \frac{1}{2} ~\overline{\left(
\nu^{}_{\rm L} ~ \Sigma^0 \right)} \left( \begin{matrix} {\bf 0} &
M^{}_{\rm D} \cr M^T_{\rm D} & M^{}_\Sigma \cr \end{matrix} \right)
\left(
\begin{matrix} \nu^{\rm c}_{\rm L} \cr {\Sigma^{0}}^{\rm c} \cr
\end{matrix} \right) + {\rm h.c.} \; ,
\nonumber \\
&& -{\cal L}^\prime_{\rm mass} = \overline{\left( e^{}_{\rm L} ~
\Psi^{}_{\rm L} \right)} \left( \begin{matrix} M^{}_l & \sqrt{2}
M^{}_{\rm D} \cr {\bf 0} & M^{}_\Sigma \cr \end{matrix} \right)
\left( \begin{matrix} E^{}_{\rm R} \cr \Psi^{}_{\rm R} \cr
\end{matrix} \right) + {\rm h.c.} \; , ~~~~~
\end{eqnarray}
respectively, for neutral and charged fermions, where $M^{}_l =
Y^{}_l v/\sqrt{2}~$, $M^{}_{\rm D} = Y^{}_\Sigma v/\sqrt{2}~$, and
$\Psi = \Sigma^- + {\Sigma^+}^{\rm c}$. The symmetric $6\times 6$
neutrino mass matrix can be diagonalized by the following unitary
transformation:
\begin{eqnarray}
\left( \begin{matrix} V & R \cr S & U \cr \end{matrix}
\right)^\dagger \left( \begin{matrix} {\bf 0} & M^{}_{\rm D} \cr
M^T_{\rm D} & M^{}_\Sigma \cr \end{matrix} \right) \left(
\begin{matrix} V & R \cr S & U \cr \end{matrix} \right)^*
= \left( \begin{matrix} \widehat{M}^{}_\nu & {\bf 0} \cr {\bf 0} &
\widehat{M}^{}_\Sigma \cr \end{matrix} \right) \; ,
\end{eqnarray}
where $\widehat{M}^{}_\nu = {\rm Diag}\{m^{}_1, m^{}_2, m^{}_3 \}$
and $\widehat{M}^{}_\Sigma = {\rm Diag}\{M^{}_1, M^{}_2, M^{}_3 \}$.
In the leading-order approximation, this diagonalization yields the
type-III seesaw formula $M^{}_\nu = -M^{}_{\rm D} M^{-1}_\Sigma
M^T_{\rm D}$, which is equivalent to the one derived from the
effective dimension-5 operator in Eq. (150). Let us use one sentence
to comment on the similarities and differences between type-I and
type-III seesaw mechanisms: the non-unitarity of the $3\times 3$
neutrino mixing matrix $V$ has appeared in both cases, although the
modified couplings between the $Z^0$ boson and three light neutrinos
differ and the non-unitary flavor mixing is also present in the
couplings between the $Z^0$ boson and three charged leptons in the
type-III seesaw scenario.

At the LHC, the typical lepton-number-violating signatures of the
type-III seesaw mechanism can be $pp \to \Sigma^+ \Sigma^0 \to
l^+_\alpha l^+_\beta + Z^0W^-(\to 4j)$ and $pp \to \Sigma^- \Sigma^0
\to l^-_\alpha l^-_\beta + Z^0W^+(\to 4j)$ processes. A detailed
analysis of such collider signatures have been done in the
literature. As for the low-energy phenomenology, a consequence of
this seesaw scenario is the non-unitarity of the $3\times 3$ flavor
mixing matrix $N$ ($\approx V$) in both charged- and neutral-current
interactions. Current experimental bounds on the deviation of
$NN^\dagger$ from the identity matrix are at the $0.1\%$ level, much
stronger than those obtained in the type-I seesaw scheme, just
because the flavor-changing processes with charged leptons are
allowed at the tree level in the type-III seesaw mechanism.

\subsection{Inverse and multiple seesaws}

Given the naturalness and testability as two prerequisites, the
double or inverse seesaw mechanism is another interesting
possibility of generating tiny neutrino masses at the TeV scale. The
idea of this seesaw picture is to add three heavy right-handed
neutrinos $N^{}_{\rm R}$, three SM gauge-singlet neutrinos
$S^{}_{\rm R}$ and one Higgs singlet $\Phi$ into the SM, such that
the gauge-invariant lepton mass terms can be written as
\begin{eqnarray}
-{\cal L}^{}_{\rm lepton} = \overline{l^{}_{\rm L}} Y^{}_l H
E^{}_{\rm R} + \overline{l^{}_{\rm L}} Y^{}_\nu \tilde{H} N^{}_{\rm
R} + \overline{N^{\rm c}_{\rm R}} Y^{}_S \Phi S^{}_{\rm R} +
\frac{1}{2} \overline{S^{\rm c}_{\rm R}} \mu S^{}_{\rm R} + {\rm
h.c.} \; ,
\end{eqnarray}
where the $\mu$-term is naturally small according to 't Hooft's
naturalness criterion, because it violates the lepton number. After
spontaneous gauge symmetry breaking, the overall neutrino mass term
turns out to be
\begin{eqnarray}
-{\cal L}^{}_{\rm mass} \; =\; \frac{1}{2}
~\overline{\left(\nu^{}_{\rm L} ~N^{\rm c}_{\rm R} ~S^{\rm c}_{\rm
R}\right)} \left( \begin{matrix} {\bf 0} & M^{}_{\rm D} & {\bf 0}
\cr M^T_{\rm D} & {\bf 0} & M^{}_S \cr {\bf 0} & M^T_S & \mu \cr
\end{matrix} \right) \left( \begin{matrix} \nu^{\rm c}_{\rm L} \cr N^{}_{\rm
R} \cr S^{}_{\rm R} \cr \end{matrix} \right) \; ,
\end{eqnarray}
where $M^{}_{\rm D} = Y^{}_\nu \langle H\rangle$ and $M^{}_S =
Y^{}_S \langle \Phi\rangle$. A diagonalization of the symmetric
$9\times 9$ matrix $\cal M$ leads us to the effective light neutrino
mass matrix
\begin{equation}
M^{}_\nu \approx M^{}_{\rm D} \frac{1}{M^T_S} \mu \frac{1}{M^{}_S}
M^T_{\rm D} \;
\end{equation}
in the leading-order approximation. Hence the smallness of
$M^{}_\nu$ can be attributed to both the smallness of $\mu$ itself
and the doubly-suppressed $M^{}_{\rm D}/M^{}_S$ term for $M^{}_{\rm
D} \ll M^{}_S$. For example, $\mu \sim {\cal O}(1)$ keV and
$M^{}_{\rm D}/M^{}_S \sim {\cal O}(10^{-2})$ naturally give rise to
a sub-eV $M^{}_\nu$. One has $M^{}_\nu = {\bf 0}$ in the limit $\mu
\rightarrow {\bf 0}$, which reflects the restoration of the
slightly-broken lepton number. The heavy sector consists of three
pairs of pseudo-Dirac neutrinos whose CP-conjugated Majorana
components have a tiny mass splitting characterized by the order of
$\mu$.

Going beyond the canonical (type-I) and inverse seesaw mechanisms,
one may build the so-called ``multiple" seesaw mechanisms to further
lower the seesaw scales.

\section{Non-unitary Neutrino Mixing}

It is worth remarking that the charged-current interactions of light
and heavy Majorana neutrinos are not completely independent in
either the type-I seesaw or the type-(I+II) seesaw. The standard
charged-current interactions of $\nu^{}_i$ and $N^{}_i$ are already
given in Eq. (34), where $V$ is just the light neutrino mixing
matrix responsible for neutrino oscillations, and $R$ describes the
strength of charged-current interactions between $(e, \mu, \tau)$
and $(N^{}_1, N^{}_2, N^{}_3)$. Since $V$ and $R$ belong to the same
unitary transformation done in Eq. (28) or Eq. (159), they must be
correlated with each other and their correlation signifies an
important relationship between neutrino physics and collider
physics.

It can be shown that $V$ and $R$ share nine rotation angles
($\theta^{}_{i4}$, $\theta^{}_{i5}$ and $\theta^{}_{i6}$ for $i=1$,
$2$ and $3$) and nine phase angles ($\delta^{}_{i4}$,
$\delta^{}_{i5}$ and $\delta^{}_{i6}$ for $i=1$, $2$ and $3$). To
see this point clearly, let us decompose $V$ into $V = A V^{}_0$,
where $V^{}_0$ is the standard (unitary) parametrization of the
$3\times 3$ PMNS matrix in which three CP-violating phases
$\delta^{}_{ij}$ (for $ij=12, 13, 23$) are associated with
$s^{}_{ij}$ (i.e., $c^{}_{ij} \equiv \cos\theta^{}_{ij}$ and
$\hat{s}^{}_{ij} \equiv e^{i\delta^{}_{ij}} \sin\theta^{}_{ij}$).
Because of $VV^\dagger = AA^\dagger = {\bf 1} - RR^\dagger$, it is
obvious that $V \rightarrow V^{}_0$ in the limit of $A \rightarrow
{\bf 1}$ (or equivalently, $R \rightarrow {\bf 0}$). Considering the
fact that the non-unitarity of $V$ must be a small effect (at most
at the percent level as constrained by current neutrino oscillation
data and precision electroweak data), we expect $s^{}_{ij} \leq
{\cal O}(0.1)$ (for $i=1,2,3$ and $j=4,5,6$) to hold. Then we obtain
\begin{eqnarray}
R = \left( \begin{matrix} \hat{s}^*_{14} & \hat{s}^*_{15} &
\hat{s}^*_{16} \cr \hat{s}^*_{24} & \hat{s}^*_{25} & \hat{s}^*_{26}
\cr \hat{s}^*_{34} & \hat{s}^*_{35} & \hat{s}^*_{36} \cr
\end{matrix} \right)
\end{eqnarray}
as an excellent approximations. A striking consequence of the
non-unitarity of $V$ is the loss of universality for the Jarlskog
invariants of CP violation, $J^{ij}_{\alpha\beta} \equiv {\rm
Im}(V^{}_{\alpha i} V^{}_{\beta j} V^*_{\alpha j} V^*_{\beta i})$,
where the Greek indices run over $(e, \mu, \tau)$ and the Latin
indices run over $(1,2,3$). For example, the extra CP-violating
phases of $V$ are possible to give rise to a significant asymmetry
between $\nu^{}_\mu \rightarrow \nu^{}_\tau$ and
$\overline{\nu}^{}_\mu \rightarrow \overline{\nu}^{}_\tau$
oscillations.

The probability of $\nu^{}_\alpha \rightarrow \nu^{}_\beta$
oscillations in vacuum, defined as $P^{}_{\alpha\beta}$, is given by
\begin{eqnarray}
P^{}_{\alpha\beta} = \frac{\displaystyle \sum^{}_i |V^{}_{\alpha
i}|^2 |V^{}_{\beta i}|^2 + 2 \sum^{}_{i<j} {\rm Re} \left(
V^{}_{\alpha i} V^{}_{\beta j} V^*_{\alpha j} V^*_{\beta i} \right)
\cos \Delta^{}_{ij} - \sum^{}_{i<j} J^{ij}_{\alpha\beta}
\sin\Delta^{}_{ij}} {\displaystyle \left(
VV^\dagger\right)^{}_{\alpha\alpha} \left(
VV^\dagger\right)^{}_{\beta\beta}} \; ,
\end{eqnarray}
where $\Delta^{}_{ij} \equiv \Delta m^2_{ij} L/(2E)$ with $\Delta
m^2_{ij} \equiv m^2_i - m^2_j$, $E$ being the neutrino beam energy
and $L$ being the baseline length. If $V$ is exactly unitary (i.e.,
$A = {\bf 1}$ and $V = V^{}_0$), the denominator of Eq. (168) will
become unity and the conventional formula of $P^{}_{\alpha\beta}$
will be reproduced. Note that $\nu^{}_\mu \rightarrow \nu^{}_\tau$
and $\overline{\nu}^{}_\mu \rightarrow \overline{\nu}^{}_\tau$
oscillations may serve as a good tool to probe possible signatures
of non-unitary CP violation. To illustrate this point, we consider a
short- or medium-baseline neutrino oscillation experiment with
$|\sin\Delta^{}_{13}| \sim |\sin\Delta^{}_{23}| \gg
|\sin\Delta^{}_{12}|$, in which the terrestrial matter effects are
expected to be insignificant or negligibly small. Then the dominant
CP-conserving and CP-violating terms of $P(\nu^{}_\mu \rightarrow
\nu^{}_\tau)$ and $P(\overline{\nu}^{}_\mu \rightarrow
\overline{\nu}^{}_\tau)$ are
\begin{eqnarray}
P(\nu^{}_\mu \rightarrow \nu^{}_\tau) \approx \sin^2 2\theta^{}_{23}
\sin^2 \frac{\Delta^{}_{23}}{2} - 2 \left( J^{23}_{\mu\tau} +
J^{13}_{\mu\tau} \right) \sin\Delta^{}_{23} \; ,
\nonumber \\
P(\overline{\nu}^{}_\mu \rightarrow \overline{\nu}^{}_\tau) \approx
\sin^2 2\theta^{}_{23} \sin^2 \frac{\Delta^{}_{23}}{2} + 2 \left(
J^{23}_{\mu\tau} + J^{13}_{\mu\tau} \right) \sin\Delta^{}_{23} \; ,
\end{eqnarray}
where the good approximation $\Delta^{}_{13} \approx \Delta^{}_{23}$
has been used in view of the experimental fact $|\Delta m^2_{13}|
\approx |\Delta m^2_{23}| \gg |\Delta m^2_{12}|$, and the
sub-leading and CP-conserving ``zero-distance" effect has been
omitted. For simplicity, I take $V^{}_0$ to be the exactly
tri-bimaximal mixing pattern (i.e., $\theta^{}_{12} =
\arctan(1/\sqrt{2})$, $\theta^{}_{13} =0$ and $\theta^{}_{23}
=\pi/4$ as well as $\delta^{}_{12} = \delta^{}_{13} = \delta^{}_{23}
=0$) and then arrive at
\begin{equation}
2\left( J^{23}_{\mu\tau} + J^{13}_{\mu\tau} \right) \; \approx \;
\sum^6_{l=4} s^{}_{2l} s^{}_{3l} \sin \left( \delta^{}_{2l} -
\delta^{}_{3l} \right) \; .
\end{equation}
Given $s^{}_{2l} \sim s^{}_{3l} \sim {\cal O}(0.1)$ and
$(\delta^{}_{2l} - \delta^{}_{3l}) \sim {\cal O}(1)$ (for
$l=4,5,6$), this non-trivial CP-violating quantity can reach the
percent level. When a long-baseline neutrino oscillation experiment
is concerned, however, the terrestrial matter effects must be taken
into account because they might fake the genuine CP-violating
signals. As for $\nu^{}_\mu \rightarrow \nu^{}_\tau$ and
$\overline{\nu}^{}_\mu \rightarrow \overline{\nu}^{}_\tau$
oscillations under discussion, the dominant matter effect results
from the neutral-current interactions and modifies the CP-violating
quantity of Eq. (170) in the following way:
\begin{eqnarray}
2\left( J^{23}_{\mu\tau} + J^{13}_{\mu\tau} \right) \Longrightarrow
\sum^6_{l=4} s^{}_{2l} s^{}_{3l} \left[ \sin \left( \delta^{}_{2l} -
\delta^{}_{3l} \right) + A^{}_{\rm NC} L \cos \left( \delta^{}_{2l}
- \delta^{}_{3l} \right) \right] \; ,
\end{eqnarray}
where $A^{}_{\rm NC} = G^{}_{\rm F} N^{}_n /\sqrt{2}~$ with $N^{}_n$
being the background density of neutrons, and $L$ is the baseline
length. It is easy to find $A^{}_{\rm NC} L \sim {\cal O}(1)$ for $L
\sim 4 \times 10^3$ km.

\section{Concluding Remarks}

I have briefly described some basic properties of massive neutrinos
in an essentially model-independent way in these lectures, which are
largely based on the book by Dr. Shun Zhou and myself \cite {XZ} and
on a few review articles or lectures \cite{ICHEP08}---\cite{Xing10}.
It is difficult to cite all the relevant references. I apologize for
missing other people's works due to the tight page limit of these
proceedings. For the same reason I am unable to write in the
cosmological matter-antimatter asymmetry and the leptogenesis
mechanism, although they were discussed in my lectures. Here let me
just give a few remarks on the naturalness and testability of TeV
seesaw mechanisms.

Although the seesaw ideas are elegant, they have to appeal for some
or many new degrees of freedom in order to interpret the observed
neutrino mass hierarchy and lepton flavor mixing. According to
Weinberg's {\it third law of progress in theoretical physics}, ``you
may use any degrees of freedom you like to describe a physical
system, but if you use the wrong ones, you will be sorry." What
could be better?

Anyway, we hope that the LHC might open a new window for us to
understand the origin of neutrino masses and the dynamics of lepton
number violation. A TeV seesaw might work ({\it naturalness}?) and
its heavy degrees of freedom might show up at the LHC ({\it
testability}?). A bridge between collider physics and neutrino
physics is highly anticipated and, if it exists, will lead to rich
phenomenology.

I am indebted to the organizers of AEPSHEP 2012 for their invitation
and hospitality. This work is supported in part by the National
Natural Science Foundation of China under grant No. 11135009.

\end{document}